\documentclass[reprint,
 superscriptaddress,
 amsmath,
 amssymb,
 aps,
 pra,
 floatfix,
 twocolumn]
 {revtex4-2}
\usepackage{etoolbox} 
\makeatletter
\patchcmd{\frontmatter@abstract@produce}
  {\vskip200\p@\@plus1fil
   \penalty-200\relax
   \vskip-200\p@\@plus-1fil}
  {}
  {}
  {}
\makeatother

\usepackage[x11names]{xcolor}
\usepackage{enumerate}
\usepackage[hang,flushmargin]{footmisc}
\usepackage{amsfonts}
\usepackage{amsthm}
\usepackage{graphicx} 
\usepackage{comment}
\usepackage{array}   
\usepackage{bm} 
\usepackage{dcolumn}
\usepackage{tikz}
\usetikzlibrary{calc}
\usepackage{upgreek}
\usepackage{isomath}
\usepackage{placeins}
\usepackage{multirow}
\usepackage{bigdelim}
\usepackage{natbib}
\usepackage{mathtools}
\usepackage{hyperref}
\usepackage{orcidlink}

\newcommand{\R}{\mathbb{R}}

\newcolumntype{L}{>{$}l<{$}} 

\newcommand{\bra}[1]{\langle #1\rvert}
\newcommand{\ket}[1]{\lvert #1\rangle}
\newcommand{\ip}[2]{\bra{#1} #2\rangle}

\newcommand{\rpd}[1]{\partial_t #1}

\DeclareMathOperator{\Tr}{Tr}

\begin{document}

\title{\mbox{\hspace{-.5pc}Engineering tunable decoherence-free subspaces with collective atom-cavity interactions}}

\author{Lyryl H. C. Vaecairn\orcidlink{0009-0000-5829-2243}}
\thanks{Corresponding author: Lyryl.Vaecairn@colorado.edu}
\affiliation{JILA and Department of Physics, University of Colorado, Boulder, Colorado, USA}

\author{Jarrod T. Reilly\orcidlink{0000-0001-5410-089X}}
\affiliation{JILA and Department of Physics, University of Colorado, Boulder, Colorado, USA}

\author{John Drew Wilson\orcidlink{0000-0001-6334-2460}}
\affiliation{JILA and Department of Physics, University of Colorado, Boulder, Colorado, USA}

\author{Simon B. J\"ager\orcidlink{0000-0002-2585-5246}}
\affiliation{Physikalisches Institut, Universit\"at Bonn, Bonn, Germany}

\author{Murray Holland\orcidlink{0000-0002-3778-1352}}
\affiliation{JILA and Department of Physics, University of Colorado, Boulder, Colorado, USA}

\date{31 March 2025}

\begin{abstract}
We propose schemes to design and control a time-dependent decoherence-free subspace (DFS) in a dissipative atom-cavity system.
These schemes use atoms with three internal energy levels, which allows for the DFS to be multi-dimensional--a condition important for quantum sensing, simulation, and computation.
We consider the use of tunable external driving lasers to transfer the system from a coherent spin state to a highly degenerate DFS.
We find that the typical state in the DFS is highly entangled.
Throughout evolution the state is kept in an instantaneous DFS, thereby allowing for pure states to be prepared.
We develop adiabatic shortcuts to carry out this evolution with higher purity and fidelity than standard adiabatic and dissipative methods.
\\\\
DOI: 10.1103/PhysRevA.111.033718
\end{abstract}

{
\let\clearpage\relax
\maketitle
}

\section{Introduction}

Quantum technologies come with the promise of great advancement in applications such as quantum metrology, communications, simulation, and computation.
In all of these fields, quantum properties can be controlled to out-perform their classical counterparts.
Most quantum technologies currently operate in Noisy Intermediate Scale Quantum regimes~\cite{Preskill2018}, where the control of quantum properties is limited by decoherence~\cite{Joos2003,Gardiner2010,Orszag2010,DiVincenzo1995,Unrah1995} and speed of state preparation~\cite{Aharonov2007,Albash2018,Venegas2020,Jin2021,Herb2024}.
In particular, dissipation prevents stable coherence over long timescales which, in turn, limits the generation of useful entanglement for quantum sensing~\cite{Reilly2023Speedingup,Wilson2024,Bohnet2016,Pellizzari1995} and quantum algorithms~\cite{Schulte1976}.
For near-term quantum advantage, removing dominant sources of error is an essential milestone.
One promising avenue is that of fault-tolerant error correcting schemes, which seek to eliminate all possible errors ~\cite{Zhou2018,Shor1995,Reichardt2024,Preskill1998,Arute2019}.
However, these schemes often require the control of individual qubits with a high degree of accuracy, thereby making their practical implementations challenging~\cite{Wang2024, Cong2022}.
In many cases, it would instead be ideal to use the innate properties of the physical system in order to remove unwanted noise sources without the need to correct errors based on measurement.

A decoherence-free subspace (DFS) allows for exactly this, by taking advantage of the structure of quantum decoherence in order to remove it completely in a quantum system~\cite{Wu2017,Karasik2008,Lidar2014,Reilly2022,RubiesBigorda2024}.
In a DFS, one engineers the dissipation to stabilize the desired quantum states against the dominant external noise sources.
Contrastingly, DFSs arise in quantum resetting schemes~\cite{Diniz2024} where this stabilization prohibits the system from returning to its original state.
A promising approach in engineering dissipation is the use of dissipative cavity modes, where one controls the coupling of an atomic ensemble to the cavity mode via a dynamical driving protocol~\cite{Ritsch2013,Reilly2022,Chitra2015,Damanet2019}.
These protocols have the advantage of removing the need for full control over all constituents of the quantum system, while the global coupling to the cavity mode creates a collective DFS.
Such implementations have been discussed in the context of quantum computation and quantum sensing~\cite{Lidar1998,Fortunato2002,Hamann2021,Horodecki2009,DallaTorre2013} where they allow stable coherence over longer timescales.
However, the reduction of the required control in such implementations comes at the cost of their universality because, in general, they only protect against a subset of all possible errors.
As a result, useful platforms and protocols which formulate a highly controllable DFS are extremely valuable.

In this paper, we describe the generation and control of a multi-dimensional DFS which is completely insensitive to decoherence from photon loss.
We consider a collection of atoms coupled to a cavity, each with three relevant internal levels, thereby leading to collective $\mathrm{SU}(3)$ dynamics.
This model has been studied in several theoretical works in the context of cavity-mediated transitions between hyperfine states~\cite{Masson2017,Masson2019,Grimsmo2013,Chen2012,Dimer2007,Zhang2018}, and has been realized in experiments using $\prescript{87}{}{\text{Rb}}$~\cite{Zhiqiang2017,Baden2014}.
Here, we develop a theory to analytically describe a DFS by exactly diagonalizing the non-Hermitian jump operators in the symmetric state manifold.

In particular, states can be prepared using time dependent dynamics where the state remains in an instantaneous DFS at every point in time.
This allows states to be prepared using adiabatic shortcuts~\cite{Guery2019,Wu2017} employing experimentally relevant driving schemes.
Within this system, we show that a transitionless quantum driving method is possible.
These methods allow one to prepare a decoherence-free state without any loss of purity and in timescales not limited by any error rate.
We contrast this with typical methods of preparing a DFS which require dissipative state preparation~\cite{DallaTorre2013} or adiabatic change to the system~\cite{Carollo2006,Parkins1995,Parkins1993}, and take orders of magnitude longer than our adiabatic shortcuts.
These shorter timescales mean there is far less opportunity to have errors against which the DFS does not protect.

We also provide a set of operators which drive between degenerate states in a decoherence-free qubit without causing any decoherence.
This provides a step towards realizing a controllable DFS for quantum simulation~\cite{Ritsch2013,Shankar2017} and sensing~\cite{Degan2017,Pezze2018,Reilly2023,Ye2024,Huang2024} in atom-cavity setups.

This article is organized as follows.
Sec.~\ref{sec:DFS} is an outline of the basic requirements for a subspace to be a stable DFS.
In Sec.~\ref{sec:model} we introduce the collective atom-cavity model which will be used to create DFSs.
From this, in Sec.~\ref{sec:findDFS}, we calculate the eigenstates of the decoherence operator to determine the dimension and structure of every DFS in the system.
Next, Sec.~\ref{sec:timeEvolve} is a discussion of various time evolution methods used to prepare a DFS.
Sec.~\ref{sec:qubit} defines and examines effective Pauli operators that drive within a two-dimensional DFS.
We conclude in Sec.~\ref{sec:conclusion} with a discussion of future work.
\vspace*{-1pc}
\section{Stable Decoherence-Free Subspaces} \label{sec:DFS}
\vspace*{-.5pc}
We begin with a brief introduction of the theoretical framework that we use to describe DFSs in open quantum systems.
We consider the situation where a system is coupled to a bath.
In the next section, we consider the system to be atoms within a cavity and the bath to be the electromagnetic field in- and outside of the cavity.
We describe the system after eliminating the bath degrees of freedom using a density matrix, $\hat{\rho}$.
The dynamics of this density operator is governed by a master equation under the Born-Markov approximations~\cite{Steck2007},
\begin{equation} \label{eq:MasterEquation}
    \frac{\partial\hat{\rho}}{\partial t} = \hat{\mathcal{L}} \hat{\rho} := \frac{1}{i \hbar} \left[ \hat{H}, \hat{\rho} \right] + \sum_i \hat{\mathcal{D}} \left[ \hat{L}_i \right] \hat{\rho},
\end{equation}
where $\hat{\mathcal{L}}$ is the Liouvillian superoperator.
Here, the system's unitary dynamics are governed by the Hamiltonian $\hat{H}$, while dissipation is described by the action of the jump operators $\hat{L}_i$ using the Lindblad superoperator,
\begin{equation}
   \hat{\mathcal{D}} \left[ \hat{L}_i \right] \hat{\rho} = \hat{L}_i \hat{\rho} \hat{L}_i^{\dagger} - \frac{1}{2} \left( \hat{L}_i^{\dagger} \hat{L}_i \hat{\rho} + \hat{\rho} \hat{L}_i^{\dagger} \hat{L}_i \right).
\end{equation}

\begin{figure}[t]
    \centering
    \includegraphics[width=\linewidth]{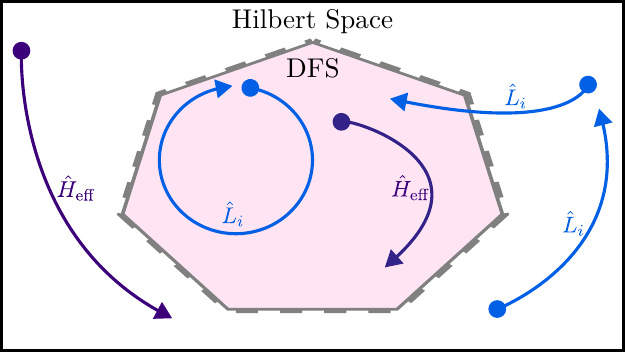}
    \caption{Action of $\hat{L}_i$ and $\hat{H}$ on the system.
    Within the DFS (pink region), states do not move under $\hat{L}_i$ because they are eigenstates.
    Outside the DFS, $\hat{L}_i$ moves certain states into the DFS and others remain in the general Hilbert space.
    $\hat{H}$ always acts unitarily, and cannot move a state into or out of the DFS.}
    \label{fig:explanation}
\end{figure}

Within the framework of the Lindblad master equation described by Eq.~\eqref{eq:MasterEquation} the formal definition of a DFS that we use is given by two criteria~\cite{Karasik2008,Wu2017,Reilly2022}:

(1) The first criterion is the general Lidar-Chuang-Whaley theorem~\cite{Lidar1998}.
For a DFS $\mathcal{H}_{\text{DFS}}=\text{span} \left[ \{\ket{\psi_k}:\ 1\le k\le m\} \right]$, all basis states must be degenerate eigenstates of all the jump operators $\hat{L}_i$,
\begin{equation}
    \hat{L}_i \ket{\psi_k} = \Lambda_i \ket{\psi_k}, \quad \text{for all } i,k.
\end{equation}
For a schematic of this we show in Fig.~\ref{fig:explanation} that the jump operator $\hat{L}_i$ does not couple to states outside of the DFS.
At the same time it can couple a generic state in the Hilbert space to a state inside of the DFS.
Note that this is a feature unique to non-Hermitian operators, as a Hermitian $\hat{L}_i$ only causes dephasing which cannot change the probabilities of measuring each eigenstate of $\hat{L}_i$.

(2) The DFS $\mathcal{H}_{\text{DFS}}$ must also be invariant under the effective Hamiltonian
\begin{equation} \label{H_eff}
    \hat{H}_{\text{eff}} = \hat{H} + \frac{i \hbar}{2} \sum_i \left( \Lambda_i^* \hat{L}_i - \Lambda_i \hat{L}_i^{\dagger}\right),
\end{equation}
such that the DFS basis states remain in the DFS upon action by $\hat{H}_{\text{eff}}$, i.e.,
\begin{equation} \label{DFScondition}
    \bra{\psi_n^{\perp}} \hat{H}_{\text{eff}} \ket{\psi_k} = 0.
\end{equation}
The above must hold for all basis states $\ket{\psi_n^{\perp}}$ of the complementary subspace $\mathcal{H}_{\text{CS}}$ and for all basis states $\ket{\psi_k}$ of the DFS $\mathcal{H}_{\text{DFS}}$.
The complementary subspace is the Hilbert space of states orthogonal to the DFS~\cite{Karasik2008}, $\mathcal{H}_{\text{CS}}=\text{span}\left[\left\{\ket{\psi_n^\perp}:\ \ip{\psi_n^\perp}{\psi_k}=0,\ \ket{\psi_k}\in\mathcal{H}_{\text{DFS}}\right\}\right]$.
In Fig.~\ref{fig:explanation} we have highlighted this by sketching that $\hat{H}_{\text{eff}}$ cannot couple any decoherence-free states outside of $\mathcal{H}_{\text{DFS}}$.

\begin{figure}[b]
    \centering
    \includegraphics[width=\linewidth]{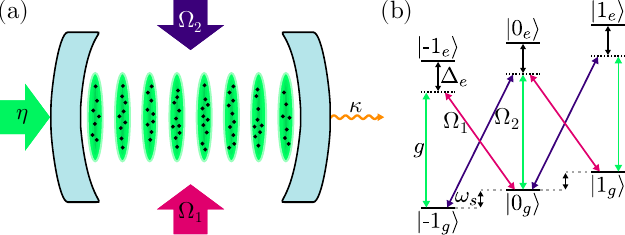}
    \caption{\textbf{(a)} Schematic of $N$ identical atoms trapped in the anti-nodes of a cavity \cite{Reilly2022}.
    \textbf{(b)} Internal level structure for a given atom.}
    \label{fig:Model}
\end{figure}

The system will typically damp into the DFS over time as the jump operators introduce stable decoherence-free population, whereas population in all other states will decay away.
The Hamiltonian cannot affect this behavior as it cannot move states into or out of the DFS.
We now manipulate the DFS via dynamical control over the jump operators $\hat{L}_i$ and $\hat{H}$.
In the next section, we discuss how this can be achieved for a realistic physical system.

\section{Model} \label{sec:model}

As shown in Fig.~\ref{fig:Model}(a), we consider a single mode of an optical cavity with frequency $\omega_c$ and decay rate $2 \kappa$.
We assume that $N$ atoms are trapped at the antinodes of the cavity mode, and thus couple to the cavity field with constant vacuum Rabi frequency $g$.
The internal structure of the atoms is shown in Fig.~\ref{fig:Model}(b).
Each atom has three ground states $\ket{0_g}$ and $\ket{\pm 1_g}$ and three excited states $\ket{0_e}$ and $\ket{\pm 1_e}$.
The ground states $\ket{\pm 1_g}$ are split from $\ket{0_g}$ by frequency $\pm \omega_s$, while each excited state has bare frequency $\omega_e\gg\omega_s$ with respect to its corresponding ground state.
The cavity mode drives the $\ket{i_g} \leftrightarrow \ket{i_e}$ transitions.

Dynamical control of the atom-cavity system is enabled by tunable external laser drives.
The cavity is driven through one of its mirrors by a laser with pump amplitude $\eta(t)$ and frequency $\omega_d$.
Our system also uses two external lasers perpendicular to the cavity axis which drive the atoms between the excited and ground state manifolds, while also changing their hyperfine state.
The first laser, shown in pink in Fig.~\ref{fig:Model}, has amplitude $\Omega_1(t)$, frequency $\omega_1$, and lowers the hyperfine state by one.
The second laser, shown in purple in Fig.~\ref{fig:Model}, has amplitude $\Omega_2(t)$, frequency $\omega_2$, and raises the hyperfine state by one. 
A key ingredient in this setup is that all amplitudes $\eta(t),\Omega_1(t),\Omega_2(t)$ can be tuned which gives us dynamical control over the atom-cavity system.

The scheme that is presented here has been realized in experiment~\cite{Zhiqiang2017} using a $F = 1 \leftrightarrow F' = 1$ hyperfine transition with a linearly polarized cavity and $\sigma_{\pm}$ light from the laser drives $\Omega_1$ and $\Omega_2$.
In this case, the $-1,0,1$ state labels would correspond to the magnetic quantum number, and the respective ground or excited state manifold can be mapped onto a spin-$1$ system.

We now define the Hamiltonian and jump operator in this model in order to engineer a tunable DFS for the ground states.
For this we eliminate the excited states and the cavity degrees of freedom which is done in detail in Appendix~\ref{appx:MasterEquationDerivation}.
For completeness we provide the parameter regimes where this is justified below.

(a) {\em Elimination of the excited states:}We work in the highly-detuned, weakly-driven limit where the detuning $\Delta_e = (\omega_1+\omega_2)/2-\omega_e$ is much larger than the coupling rates, $\lvert \Delta_e \rvert \gg g, \lvert \Omega_1 \rvert, \lvert \Omega_2 \rvert$.
This allows us to eliminate the excited states $\ket{i_e}$, $i\in\{-1,0,1\}$, so we are left with only the three ground states and the cavity.
We also assume that the cavity coupling is much larger than the drive, $g\gg\lvert\Omega_1\rvert,\lvert\Omega_2\rvert$, in order to eliminate Raman coupling between $\ket{\text{-}1_g}$ and $\ket{1_g}$.

(b){\em Elimination of the cavity:}
In addition, we assume that the lifetime of a cavity photon is much shorter than the typical timescale of dynamics within the atomic ground states.
This is based on $\kappa\gg|\sqrt{N}g\Omega_1/\Delta_e|,\ |\sqrt{N}g\Omega_2/\Delta_e|$, where the $\sqrt{N}$ scaling comes from the collective enhancement of the atom-cavity coupling.
To eliminate the cavity degrees of freedom we also need them to follow the dynamically modulated drives $\kappa\gg\rpd{\eta}/\eta,\rpd{\Omega}_1/\Omega_1,\rpd{\Omega}_2/\Omega_2$ with $\rpd{f}=\partial f/\partial t$.

Based on these assumptions the excited states and cavity degrees of freedom can be integrated out.
The effective master equation describing the density operator $\hat{\rho}_{\mathrm{at}}$ which now solely describes the ground states is given by
\begin{equation} 
    \frac{\partial\hat{\rho}_{\mathrm{at}}}{\partial t} = \hat{\mathcal{L}}_{\mathrm{at}} \hat{\rho}_{\mathrm{at}} := \frac{1}{i\hbar} \left[ \hat{H}_{\mathrm{at}}, \hat{\rho}_{\mathrm{at}} \right] + \hat{\mathcal{D}} \left[ \hat{L} \right] \hat{\rho}_{\mathrm{at}},\label{eq:masterequation2}
\end{equation}
with a single jump operator
\begin{equation} \label{eq:Ldef}
    \hat{L}=\sqrt{\Gamma_c}\left(\hat{J}_- + \mu^2 \hat{J}_+ + \chi\right),
\end{equation}
and the effective Hamiltonian
\begin{equation} \label{eq:H_at}
    \hat{H}_{\mathrm{at}} = \frac{\hbar \Delta_c'}{2 \kappa} \hat{L}^{\dagger} \hat{L},
\end{equation}
where we introduced the dressed cavity detuning $\Delta_c'=(\omega_1+\omega_2)/2-\omega_c-Ng^2/\Delta_e$.
We note that this detuning may be set to zero without fundamentally changing the system dynamics.

The jump operators are a superposition of the collective lowering and raising operators,
\begin{equation} \label{eq:J_pm def}
    \hat{J}_- = \sum_{j=1}^N \ket{- 1_g}_j \bra{0_g}_j + \ket{0_g}_j \bra{1_g}_j = \hat{J}_+^\dag.
\end{equation}
These are generated by cavity decay inducing collective emission which takes an atom from $\ket{1_g}$ to $\ket{0_g}$ or $\ket{0_g}$ to $\ket{\text{-}1_g}$.
While this process is in general collectively enhanced, we have introduced the single-particle decay rate
\begin{equation} \label{eq:gammaC}
    \Gamma_c=\frac{\kappa g^2|\Omega_1|^2}{2\Delta_e^2(\Delta_c'^2+\kappa^2)}.
\end{equation} 
Note that this rate can be controlled by the driving amplitude $\Omega_1$.
Throughout this paper we will, however, fix $\Omega_1$ and control the dynamics by changing the ratio of collective pumps described by
\begin{equation} 
    \mu=\sqrt{\frac{|\Omega_2|}{|\Omega_1|}}\label{eq:mu}
    \end{equation}
    and the cavity drive which is encoded in the quantity
\begin{equation}
    \chi=\frac{2\eta\Delta_e}{g|\Omega_1|}.\label{eq:chi}
\end{equation}

In the following section we will derive the possible DFSs in this system and discuss how they can be controlled.

\section{Engineering a Decoherence-Free Subspace} \label{sec:findDFS}
\vspace*{-.5pc}
Since the master equation in Eq.~\eqref{eq:masterequation2} has only a single jump operator $\hat{L}$ which also defines $\hat{H}_{\mathrm{at}}$, both criteria for stable DFS presented in Section~\ref{sec:DFS} are met when a state $\ket{\psi}$ satisfies
\begin{equation}
    \hat{L} \ket{\psi}=0.
\end{equation}
Therefore, a DFS is given by the span of all states which meet this criterion, $\mathcal{H}_{\text{DFS}}=\mathrm{span} \left[ \{\ket{\psi}: \, \hat{L} \ket{\psi}=0 \}\right]$, for given values of $\mu (t)$ and $\chi (t)$.
We now follow the methodology of Ref.~\cite{Reilly2022} to find analytical expressions for the DFS states.

We initialize our system at time $t=0$ in the coherent-spin state $\ket{\psi(0)} = \ket{- 1_g}^{\otimes N}$.
Both $\hat{L}$ and $\hat{H}_{\mathrm{at}}$ act symmetrically on all atoms and they conserve atom number, so we may restrict our computations to just the permutationally symmetric states.
This means that the Hermitian operators we consider belong to the generators of $\mathrm{SU}(3)$~\cite{Silva2022}, describing the collective 3-level dynamics~\cite{Mathur2010}.
This allows us to reduce the dimensionality of our Hilbert space from $3^N$ (which corresponds to the dimension of arbitrary many-body states) to~\cite{Silva2022,Stanley2011}
\begin{equation} \label{eq:HilbertDim}
    \binom{N+2}{2}=\frac{(N+2)(N+1)}{2}
\end{equation} 
(corresponding to symmetric states only).

We can represent both the states and operators using Schwinger bosons~\cite{Schwinger1952,Mathur2010}.
We define the annihilation (creation) operator $\hat{b}_i$ ($\hat{b}_i^\dag$) for $i\in\{-1,0,1\}$ which annihilates (creates) an atom in the state $\ket{i_g}$, and satisfy the canonical commutation relations $\left[\hat{b}_i,\hat{b}_j^\dag\right]=\delta_{i\,j}$.
This allows us to write 
\begin{equation} \label{eq:Jminus}
    \hat{J}_-=\hat{b}_{-1}^\dag \hat{b}_0+\hat{b}_0^\dag \hat{b}_1.
\end{equation}
Forming a vector of the system's Schwinger bosons, $\vec{b}=(\hat{b}_{-1}, \hat{b}_0, \hat{b}_1)^T$, we can therefore express the jump operator as
\begin{equation} \label{L as b combo}
    \hat{L}=\sqrt{\Gamma_c}\left(\vec{b}^\dag\mathbf{L}\vec{b}+\chi\right)
\end{equation}
with
\begin{equation}
    \mathbf{L}= 
        \begin{pmatrix}
            0&1&0
            \\\mu^2&0&1
            \\0&\mu^2&0
        \end{pmatrix}.
\end{equation}
This form allows us to diagonalize the jump operator $\hat{L}$ via a non-unitary transformation which we describe now.
\vspace*{-1pc}
\subsection{Jump operator eigenstates}
\vspace*{-.5pc}
For $\mu>0$, we may diagonalize $\mathbf{L}$ via the transformation $\mathbf{L}=\mathbf{VDV}^{-1}$.
This yields
\begin{align} 
    \mathbf{V}= &
    \frac{1}{2}\begin{pmatrix}
        \mu & \sqrt{2} \mu & \mu\\\sqrt{2} \mu^{2} & 0 & - \sqrt{2} \mu^{2}\\\mu^{3} & - \sqrt{2} \mu^{3} & \mu^{3}
    \end{pmatrix},\label{eq:V}\\
    \mathbf{D}= &
    \begin{pmatrix}
        \sqrt{2}\mu&0&0
        \\0&0&0
        \\0&0&-\sqrt{2}\mu
    \end{pmatrix},\label{eq:D}
\end{align}
which allows us to define new operators
\begin{align} 
    \vec{c}=&\begin{pmatrix}
        \hat{c}_1 \\ \hat{c}_2 \\ \hat{c}_3
    \end{pmatrix}
    = \mathbf{V}^\dag\vec{b} \label{eq:c def},\\
    \vec{d}=&\begin{pmatrix}
        \hat{d}_1 \\ \hat{d}_2 \\ \hat{d}_3
    \end{pmatrix} = \mathbf{V}^{-1}\vec{b}.\label{eq:d def}
\end{align}

Note that one consequence of ${\bf L}$ being non-Hermitian is that we have to define two types of operators $\hat{c}_i\neq\hat{d}_i$.
These operators agree only if ${\bf V}$ is unitary, ${\bf V}^\dag={\bf V}^{-1}$, which is true if ${\bf L}$ is Hermitian.
For the specific system at hand ${\bf L}$ is Hermitian if $\mu=1$.

For any other case the commutation relations read $[\hat{c}_i,\hat{d}_j^\dag]=\delta_{i\,j}$ while $[\hat{c}_i,\hat{c}_j^\dag]$ and $[\hat{d}_i,\hat{d}_j^\dag]$ do not obey canonical commutation relations.
Note that this implies that $\hat{c}_i$ and $\hat{d}_i$ are in general not bosonic annihilation operators.

The advantage of defining these operators is that we can express the jump operator as
\begin{equation} \label{eq:L as cd combo}
    \hat{L}=\sqrt{\Gamma_c}\left(\vec{c}^\dag\mathbf{D}\vec{d}+\chi\right)=\sqrt{\Gamma_c}\left[\sqrt{2}\mu\left(\hat{c}_{1}^\dag \hat{d}_{1}-\hat{c}_{3}^\dag \hat{d}_{3}\right)+\chi\right],
\end{equation}
which allows us to give a closed form for the general eigenstates
\begin{equation} \label{eq:LEigenstates}
    \ket{\psi_{\vec{k}}}=\mathcal{N}_{\vec{k}}(\hat{c}_{1}^\dag)^{k_1}(\hat{c}_2^\dag)^{k_2}(\hat{c}_3^\dag)^{k_3} \ket{0},
\end{equation}
where $\ket{0}$ is the bosonic vacuum state.
Here, $\mathcal{N}_{\vec{k}}$ is a normalization factor which is more complicated than the usual product of factorials and is explicitly calculated in Appendix~\ref{appx:overlap}.
The vector $\vec{k}=(k_1, k_2, k_3)$ defines the state where $0\le k_i\le N$ for $i\in\{1,2,3\}$ such that $k_1+k_2+k_3=N$.
As operators $\hat{c}_i^\dag$ are not true bosonic creation operators, states $\ket{\psi_{\vec{k}}}$, $\ket{\psi_{\vec{k}'}}$ with $\vec{k}'\neq\vec{k}$ are in general not orthogonal.
Still, the set $\{\ket{\psi_{\vec{k}}}\}$ for all valid $\vec{k}$ is linearly independent and thus forms a basis for the entire Hilbert space with dimension matching Eq.~\eqref{eq:HilbertDim}.

Using Eq.~\eqref{eq:L as cd combo} and the commutation relations of the $\hat{c}_i$ and $\hat{d}_j$ operators, we find that the eigenvalue corresponding to the eigenstate $\ket{\psi_k}$ is given by
\begin{equation} \label{eq:LEigenvalues}
    \hat{L} \ket{\psi_{\vec{k}}}=\sqrt{\Gamma_c}\left[\sqrt{2}\mu\left(k_1-k_3\right)+\chi\right]\ket{\psi_{\vec{k}}}.
\end{equation}
This result now allows us to derive the DFS analytically.
Since we require that $ \hat{L} \ket{\psi_{\vec{k}}}=0$, the states with
\begin{equation} \label{eq:chi as k}
    \chi=\sqrt{2}\mu\left(k_3-k_1\right)
\end{equation}
are in the DFS.
In particular, since we control $\chi$ and $\mu$ with the external driving lasers, we can engineer the DFS by tuning integer $C=\chi/\left(\sqrt{2}\mu\right)$ with $-N\leq C\leq N$.
Consequently all states with the same $k_3-k_1=C$ are in the DFS and we call them dark states.

There are a few considerations to note.

\begin{figure}[t]
    \centering
    \includegraphics[width=\linewidth]{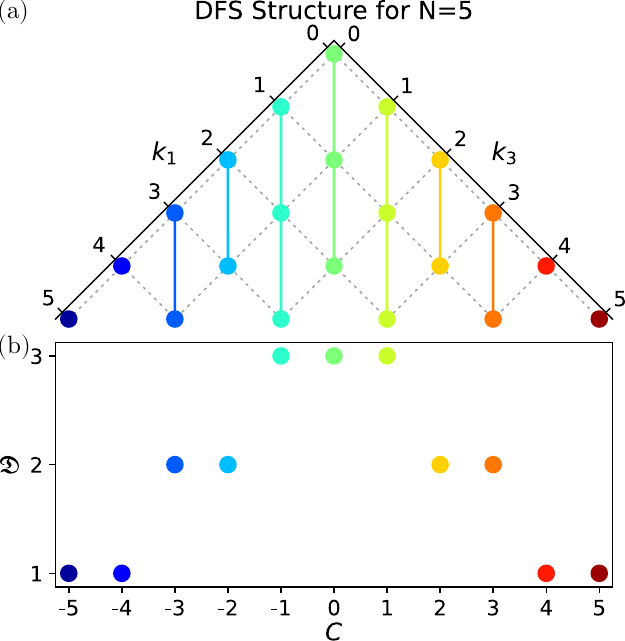}
    \caption{The DFS structure for $N=5$ atoms.
    \textbf{(a)} The states in each DFS.
    All $\vec{k}$ are represented as points at integer $(k_1,k_3)$.
    Lines connect $\vec{k}$ with constant $C=k_3-k_1$ corresponding to $\ket{\psi_{\vec{k}}}$ in the same subspace.
    \textbf{(b)} The dimension $\mathfrak{D}$ for each DFS $\mathcal{H}_{\text{DFS}}^{(C)}(\mu)$, labeled by $C$.}
    \label{fig:dims}
\end{figure}

(a) {\em Non-Integer $C$:}
In an experiment, $C$ may not be an exact integer due to fluctuations in $\chi$ and $\mu$.
In this case the system will not exactly reach a dark state but will approach a mixed state with a large overlap with the DFS and a small overlap with the orthogonal subspace of the DFS.
The small overlap will be determined by the magnitude of fluctuations and therefore we can assume it to be negligible provided that the fluctuations are perturbative.
Note that for the special case of the cavity-drive $\eta$ being turned off, then $\chi=0$ and $C=0$ is exactly an integer.
In this work, we will assume that $C$ is an integer for simplicity.

(b) {\em Zero Collective Pump Ratio:}
We will now briefly discuss the case of $\mu=0$.
This is a special case where $\mathbf{L}$ is not diagonalizable, so the above process cannot be applied.
Even so, for $\mu=0$ $\mathbf{L}$ generally has several dark states.
The most relevant in this work is the previously mentioned initial state 
\begin{equation} \label{eq:phi0 def}
    \ket{\psi(0)}=\ket{- 1_g}^{\otimes N}=\frac{(\hat{b}_{-1}^\dag)^N}{\sqrt{N!}}\ket{0},
\end{equation}
which is a dark state of $\hat{L}$ for $\chi=0$.
\vspace*{-1pc}
\subsection{Decoherence-free subspace structure} \label{sec:structure}
\vspace*{-.5pc}
In the previous section, we established that the DFS can be classified by an integer number $C =\chi/\left(\sqrt{2}\mu\right)$.
Therefore, the DFS defined by a specific $C$ will be spanned by all the eigenstates $\ket{\psi_{\vec{k}}}$ which satisfy $k_3-k_1=C$.
Hence, we may formally define the DFS for a specific choice of $C$ and $\mu$ to be
\begin{align}
    \mathcal{H}_{\text{DFS}}^{(C)}(\mu)=\text{span}\left[\left\{\ket{\psi_{\vec{k}}}:\ k_3-k_1=C\right\}\right].\label{eq:DFS}
\end{align} 
It is important to note that the states $\ket{\psi_{\vec{k}}}$ depend on $\mu$.
We can tune $\mu$ and $\chi$ while keeping $C$ constant to dynamically control these states.
This will ensure that at each instant of time, we are in a DFS with a fixed dimension but changing states.

Let us now examine the structure of these subspaces in more depth. In Fig.~\ref{fig:dims}(a), we show the structure of each possible DFS for $N=5$ atoms.
Although $\vec{k}$ is a three-dimensional object, the condition $k_1+k_2+k_3=N$ means there are only two independent degrees of freedom, so we may choose these to be $k_1,k_3$.
Lines of constant $k_3-k_1$ are shown connecting degenerate states together.
Consider, for example, the orange line going through $(k_1,k_3)=(0,3)$ and $(1,4)$.
This represents $\mathcal{H}_{\text{DFS}}^{(3)}(\mu)=\mathrm{span}\left[\left\{\ket{\psi_{0,2,3}}, \ket{\psi_{1,0,4}}\right\}\right]$ and corresponds to $C=3$ and the drive $\chi=3 \sqrt{2} \mu$.
There are only two eigenstates in this subspace because there are no other $\vec{k}$ which satisfy $k_3-k_1=3$ and $k_1+k_2+k_3=5$.
In general, we find that the dark states $\ket{\psi_{\vec{k}}}$ for a specific $C$ satisfy
\begin{equation}
    \begin{aligned}
        \vec{k}&\in\left\{(j,N-2j-|C|,j+|C|):\ 0\le j\le\mathfrak{D}-1, \, C \ge 0 \right\}, \\
        \vec{k}&\in\left\{(j+|C|,N-2j-|C|,j):\ 0\le j\le\mathfrak{D}-1, \, C \le 0 \right\}.
    \end{aligned}
\end{equation}
Here, the dimensionality of the DFS is given by 
\begin{equation} \label{eq:dim}
    \mathfrak{D}(N,C) =\left\lceil\frac{N+1-|C|}{2}\right\rceil,
\end{equation}
where $\lceil \bm{\cdot} \rceil$ is the ceiling function.
We plot the dimensionality of each DFS for the case of $N = 5$ in Fig.~\ref{fig:dims}(b), which demonstrates some general properties of Eq.~\eqref{eq:dim}.

Another natural question that arises is whether the eigenstates have interparticle entanglement.
In fact, we find that every eigenstate $\ket{\psi_{\vec{k}}}$ except for the product states $\vec{k}\in\{(N,0,0),\ (0,N,0),\ (0,0,N)\}$ has interparticle entanglement and metrological use~\cite{Reilly2023,Liu2020}.
Some of these states have a sensitivity to certain operators of half of the Heisenberg limit~\cite{Holland1993}.
While this is interesting, it is not fundamentally necessary for understanding the DFS, so we relegate these results to Appendix~\ref{appx:interparticleEntanglement}.
\vspace*{-1pc}
\subsection{Complementary subspace} \label{sec:complementarySubspace}
\vspace*{-.5pc}
When working with non-orthogonal basis states, it is challenging to find the orthogonal complement of a subspace.
As we have discussed in Sec.~\ref{sec:DFS}, this is particularly relevant for calculating the orthogonal complement of the DFS $\mathcal{H}_{\text{DFS}}^{(C)}(\mu)$ given in Eq.~\eqref{eq:DFS}.
Fortunately, by defining the operators $\hat{c}_i$ and $\hat{d}_i$ we have very natural access to the complementary Hilbert space which is based on the relation
\begin{equation} \label{eq:psi_n psi_k overlap}
    \ip{\psi_{\vec{n}}^\perp}{\psi_{\vec{k}}} =\mathcal{N}^\perp_{\vec{n}}\mathcal{N}_{\vec{k}}k_1!k_2!k_3!\delta_{\vec{n}\,\vec{k}}.
\end{equation}
with 
\begin{equation} \label{eq:psi_n}
    \ket{\psi_{\vec{n}}^\perp}=\mathcal{N}_{\vec{n}}^\perp(\hat{d}_{1}^\dag)^{n_1}(\hat{d}_2^\dag)^{n_2}(\hat{d}_3^\dag)^{n_3}\ket{0}.
\end{equation}
Here we introduced the new normalization $\mathcal{N}_{\vec{n}}^\perp$ which is explicitly calculated in Appendix~\ref{appx:overlap}, and $\vec{n}=(n_1, n_2, n_3)$ satisfies $0\le n_i\le N$ for $i\in\{1,2,3\}$ and $n_1+n_2+n_3=N$.

Using this definition and property we can write down the orthogonal complement of $\mathcal{H}_{\text{DFS}}^{(C)}(\mu)$ as
\begin{align}
    \mathcal{H}_{\text{CS}}^{(C)}(\mu) = \text{span} \left[ \{ \ket{\psi_{\vec{n}}^\perp}:\ n_3-n_1\neq C\} \right].
\end{align}

This result will be of importance in the next section where we will consider dynamically ramping $\mu$ to prepare the system in a given DFS.
When doing this there is typically leakage from the DFS to its complementary space.
The analytical result above does on one hand allow us to quantify this leakage and on the other hand to counter it and to create shortcuts.
\vspace*{-1pc}
\section{Time Evolution} \label{sec:timeEvolve}

We now consider dynamical control of $\mu$ and $\chi$ which is realized by changing the drive amplitude $\eta$ and the ratio of $\Omega_2/\Omega_1$.
Consequently, the DFS $\mathcal{H}_{\text{DFS}}^{(C)}(\mu(t))$ will change while we fix the integer $C=\chi/\left(\sqrt{2}\mu\right)$.
By keeping $\Omega_1$ constant, the cavity-induced spontaneous emission rate $\Gamma_c$ remains constant.
We initialize the system at $t=0$ in the ground state $\ket{\psi(0)}$ [see Eq.~\eqref{eq:phi0 def}] with $\mu=0=\chi$.
We then vary $\Omega_2(t)$ and $\eta(t)$ until at final time $t_f$ we reach $\Omega_2(t_f)=\Omega_1$, meaning that $\mu(t_f)=1$ and $\chi(t_f)=\sqrt{2}C$.
We target $\mu(t_f)=1$ so that $\hat{L}$ is Hermitian, whereupon we find its eigenstates are highly entangled (see Appendix~\ref{appx:interparticleEntanglement}).
However, we note that one may choose any $\mu\ge 0$ as the target DFS and our formalism will still hold.

As this paper only considers a single mode of dissipation, it is ideal to prepare the system in a DFS quickly so that other sources of unmitigated error have a negligible effect on the dynamics.
We look at evolution time in units of $1/\Gamma_c$, as this is the timescale corresponding to single-particle emission via the cavity.

We measure how well we have prepared the system in a pure state in the DFS at each time using the purity
\begin{equation}
    \mathcal{P}(t) = \Tr \left[ \hat{\rho}_{\mathrm{at}}(t)^2 \right]
\end{equation}
and the overlap with the DFS defined for the current $\mu(t)$, given by
\begin{equation}
    \mathcal{F}(\mu, t)=\sum_{i=1}^\mathfrak{D}\bra{\phi_{i}(\mu)}\hat{\rho}_{\mathrm{at}}(t)\ket{\phi_{i}(\mu)}
\end{equation}
where $\{\ket{\phi_i}\}$ is an orthogonal basis of $\mathcal{H}_{\text{DFS}}^{(C)}(\mu)$ with dimension $\mathfrak{D}$.
For a one-dimensional DFS, this overlap reduces to the Uhlmann-Jozsa fidelity~\cite{Jozsa1994}.
The basis is typically found by Gram-Schmidt orthonormalization of the basis vectors $\ket{\psi_{\vec{k}}}\in\mathcal{H}_{\text{DFS}}^{(C)}(\mu)$~\cite{Cheney2012}.
We also consider the overlap with the target DFS, $\mathcal{F}(1, t)$, which provides information about how close the system is to its desired final state for any time.
We set the threshold for a sufficiently good state preparation to be 
\begin{equation} \label{eq:threshold}
    \mathcal{P}(t_f)\ge 0.99,\ \mathcal{F}(1, t_f)\ge 0.99.
\end{equation}

We prepare the system in the target DFS using two standard protocols as well as a novel shortcut to adiabaticity protocol.
\vspace*{-.5pc}
\subsection{Instantaneous quench} \label{sec:quench}

In this scheme, we prepare our system in a DFS by quenching instantaneously into some $0\le\mu_q<1$ and waiting for some time, leaving the Hamiltonian and jump operator constant, as explored for a different setup in Ref.~\cite{DallaTorre2013}.
We implement this by increasing $\Omega_2$ instantaneously to the final quench value at $t=0$.
The goal is to ensure that $\hat{\rho}_{\mathrm{at}}(t_f)$ is pure and has a large overlap with $\mathcal{H}_{\text{DFS}}^{(C)}(1)$ defined by Eq.~\eqref{eq:threshold}.

We observe that for any $\mu_q<1$, the system will exponentially damp into the DFS.
This is the same logic as envisioned in Fig.~\ref{fig:explanation} that shows that decoherence-free population cannot exit the DFS whereas generic states can enter the DFS.
However, if $\hat{L}$ is Hermitian, there is no damping into the DFS.
This is why we choose $\mu_q<1$ to guarantee that the system damps into the DFS within a certain time.
This time increases when $\mu_q$ approaches 1, so in order to minimize $t_f$, we must choose the smallest possible $\mu_q$ such that $\mathcal{H}_{\text{DFS}}^{(C)}(\mu_q)$ has sufficient overlap with $\mathcal{H}_{\text{DFS}}^{(C)}(1)$.
We then let the system damp into the quench DFS until $\hat{\rho}_{\mathrm{at}}(t_f)$ meets the threshold given in Eq.~\eqref{eq:threshold}.

Figures~\ref{fig:N00} and~\ref{fig:0N0} show results for this quench scheme for $\mathcal{H}_{\text{DFS}}^{(-N)}(\mu)$ and $\mathcal{H}_{\text{DFS}}^{(0)}(\mu)$.
The threshold evolution times $t_f$ and associated quench values $\mu_q$ are given and the results are analyzed in Sec.~\ref{sec:results}.
We calculate $t_f$ numerically where we find it is of order several $100/\Gamma_c$.
This time is set by the smallest damping rate of the system at the value $\mu_q<1$, which is related to the difference of the two smallest eigenvalues of $\hat{\mathcal{D}}[\hat{L}(\mu_q)]$.
This very small dissipative gap is $\ll\Gamma_c$, which highlights that one needs to wait long times in order to achieve a high fidelity with the target state when considering such a quench protocol.
Therefore, it is worth considering more involved preparation schemes.
\vspace*{-.5pc}
\subsection{Adiabatic ramp} \label{sec:LinearRamp}

One alternative is an adiabatic ramp of $\mu$ from 0 to 1, changing $\hat{H}_{\mathrm{at}}(t)$ and $\hat{L}(t)$ adiabatically to drive the system into the desired DFS similarly to Ref.~\cite{Reilly2022}.
Adiabatic refers here to the fact that the ramp is slow enough that the system has sufficient time to damp into the instantaneous DFS.
For simplicity, we consider a linear ramp, where $\mu(t)=\beta t$ for constant $\beta>0$.
As $\mu=\sqrt{|\Omega_2|/|\Omega_1|}$ [see Eq.~\eqref{eq:mu}], this means $\Omega_2(t)$ has a quadratically increasing driving profile from 0 to $\Omega_1$.

At $t_f=1/\beta$, we reach $\mu(t_f)=\beta t_f=1$, so the Hamiltonian and jump operator of the system have reached their final values and the system should be ideally in some pure state within $\mathcal{H}_{\text{DFS}}^{(C)}(1)$.
This ideal result will never hold perfectly because the dynamics are not perfectly adiabatic unless the evolution takes infinite time, meaning that as $\mu$ changes, the changing Hamiltonian and jump operator will cause some component of the state $\hat{\rho}_{\mathrm{at}}(t)$ to enter the orthogonal complement $\mathcal{H}_{\text{CS}}(\mu)$.
To achieve a final state which meets the threshold for purity and target overlap as given in Eq.~\eqref{eq:threshold}, we numerically calculate the largest $\beta$ where the threshold is satisfied, as this gives the smallest still sufficient ramping time $t_f=1/\beta$.

Figures~\ref{fig:N00} and~\ref{fig:0N0} show simulation results for this ramp scheme in $\mathcal{H}_{\text{DFS}}^{(-N)}(\mu)$ and $\mathcal{H}_{\text{DFS}}^{(0)}(\mu)$.
We analyze these results and give the threshold $t_f$ values in Sec.~\ref{sec:results}.
We calculate $t_f$ numerically and find that it is similar to, though slightly shorter than, the times found for the quench protocol.
Intuitively, this can be understood by considering that the ramp needs to be adiabatic which becomes increasingly hard when $\mu$ approaches 1 and the damping rate vanishes.
Ideally one would should keep the ramping rate $\beta$ below the aforementioned dissipative gap $\ll \Gamma_c$.

The idea is now to engineer a protocol that corrects for the errors that happen during the adiabatic ramp.
This is realized in the next section by developing adiabatic shortcuts.
\vspace*{-1pc}
\subsection{Adiabatic shortcut} \label{sec:shortcut}

Here, we add an additional shortcut Hamiltonian $\hat{H}_s$ onto the system to shortcut adiabaticity and enter the target DFS in short timescales.
The purpose of this Hamiltonian is to remove all transitions between the DFS and the complementary subspace in order to drive directly from our initial state $\ket{\psi(0)}$ into the target DFS $\mathcal{H}_{\text{DFS}}^{(C)}(\mu)$ exactly and without any purity loss~\cite{Wu2017}.
Such a shortcut connects two specific states, $\ket{\psi(0)}$ and a specific dark state within $\mathcal{H}_{\text{DFS}}^{(C)}(1)$, and does not allow any other transitions.
This means that we no longer have to worry about letting the system damp into the target DFS or using adiabatic dynamics, so the speed of the time evolution is limited by the timescale separation that was imposed to eliminate the cavity degrees of freedom and atomic excited states.

As shown in Ref.~\cite{Wu2017} and explored in Appendix~\ref{appx:ShortcutDerivation}, this shortcut Hamiltonian must satisfy
\begin{equation}
    \bra{\psi_{\vec{n}}^\perp} \hat{H}_s \ket{\psi_{\vec{k}}} = i\hbar \ip{\psi_{\vec{n}}^\perp}{\rpd{\psi_{\vec{k}}}}.\label{eq:shortcutcrit}
\end{equation}
where $\ket{\psi_{\vec{n}}^\perp}$ is a complementary state defined in Eq.~\eqref{eq:psi_n} for which, as discussed in Sec.~\ref{sec:complementarySubspace}, $n_3-n_1\neq C$.
This means that $\ket{\psi_{\vec{k}}}$ has unitary behavior under $\hat{H}_{\mathrm{at}}(t)+\hat{H}_s(t)$, and is thus decoherence-free within a changing $\mathcal{H}_{\text{DFS}}^{(C)}(\mu(t))$~\cite{Karasik2008}.

In general one can choose an arbitrary Hamiltonian; however, in this work, we search for shortcut Hamiltonians which are quadratic in creation and annihilation operators
\begin{equation} \label{eq:H1_quadratic}
    \hat{H}_s=\sum_{i,j=-1}^1\alpha_{ij}\hat{b}_i^\dag\hat{b}_j
\end{equation}
for complex scalars $\alpha_{ij}=\alpha_{ji}^*$.
These quadratic terms can in practice be created from single-particle terms and one does not need to control atom-atom interactions.

The first question that arises is to which states one can shortcut with such a Hamiltonian.
It can be proven that we can only shortcut to states that do not possess interparticle entanglement.
This can be seen by matching the possible terms in Eq.~\eqref{eq:H1_quadratic} to the terms which appear in $\ip{\psi_{\vec{n}}^\perp}{\rpd{\psi_{\vec{k}}}}$ [see Eq.~\eqref{eq:timeDerivative} in the Appendix], but the proof does not appear in this work.
The intuitive explanation for this finding is that we have initialized the state with an unentangled product state $\ket{\psi(0)}$ [see Eq.~\eqref{eq:phi0 def}] and the shortcut Hamiltonian cannot create entanglement because it is made of single-particle terms.
This means we can shortcut exactly the states of the form $\mathcal{N}(\hat{c}_i^\dag)^N\ket{0}$ for $i\in\{1,2,3\}$ which describe unentangled states.

To be specific in the following, we distinguish shortcut Hamiltonians $\hat{H}_s$ that are used for the preparation of $\ket{\psi_{N,0,0}}$ and $\ket{\psi_{0,0,N}}$ which are each in a one-dimensional DFS from a Hamiltonian that shortcuts to $\ket{\psi_{0,N,0}}$ which resides in a DFS of maximum dimension given in Eq.~\eqref{eq:dim} for $C=0$.

We note that despite only being able to shortcut directly to unentangled states, the multi-dimensional $\mathcal{H}_{\text{DFS}}^{(0)}(\mu)$ contains highly entangled states which are degenerate with $\ket{\psi_{0,N,0}}$.
In Sec.~\ref{sec:qubit} we derive a driving scheme between states in a two-dimensional DFS which avoids decoherence.
For the cases of $N\in\{2,3\}$, $\mathcal{H}_{\text{DFS}}^{(0)}(\mu)$ is two-dimensional, meaning that we can drive between $\ket{\psi_{0,N,0}}$ and its degenerate states without introducing any decoherence.
\vspace*{-1pc}
\subsubsection{\textbf{Cavity drive modification}} \label{sec:CavityDrive}

We start by providing the shortcut Hamiltonian that is used for the preparation of $\ket{\psi_{N,0,0}}$ and $\ket{\psi_{0,0,N}}$, which are both in one-dimensional DFSs.
This Hamiltonian can be created by changing the driving amplitude of the injected cavity drive $\eta$ to $\eta+\eta_s$ so that $\chi$ transforms to $\chi+\chi_s$.
Remarkably, this does not require adding any new lasers to the system.
We can write this as effectively shifting the Hamiltonian $\hat{H}_{\mathrm{at}}$ to $\hat{H}_{\mathrm{at}}+\hat{H}_s$ where
\begin{equation} \label{eq:H1CavityDrive}
    \hat{H}_s=\frac{\hbar\sqrt{\Gamma_c}}{2}\left[\left(\frac{\Delta_c'}{\kappa}+i\right)\chi_s^*\hat{L}+\left(\frac{\Delta_c'}{\kappa}-i\right)\chi_s\hat{L}^\dag\right].
\end{equation}
We now need to find the appropriate value of $\chi_s$ such that $\hat{H}_s$ satisfies Eq.~\eqref{eq:shortcutcrit}.
This calculation is straight-forward but lengthy which is why we shifted it to Appendix~\ref{appx:CavityDrive}.
The final result reads
\begin{equation} \label{eq:chiS}
    \chi_s=\pm\frac{i2\sqrt{2}\rpd{\mu}}{\Gamma_c\left(\Delta_c'/\kappa-i\right)(\mu^4-1)}
\end{equation}
where the $+$ case will shortcut $\vec{k}=(0,0,N)$ and the $-$ case applies to $\vec{k}=(N,0,0)$.
We assume that $\mu(t)=\beta t$ so $\rpd{\mu}=\beta$ is constant.
In this case, the speed of the shortcut just depends on the maximum $\beta$.
As $\mu$ approaches 1, $\chi_s$ diverges, indicating that we cannot shortcut all the way to $\mu=1$, and we are limited by laser power.
Note that this divergence cannot be remedied by choosing a different $\mu(t)$ dependence.
To avoid the issue, we pick some maximum cutoff value for $|\chi+\chi_s|$ beyond which we continue ramping $\mu$ but keep $|\chi+\chi_s|$ constant.
This has a minimal effect on the characteristics of the evolution overall, and is further discussed in Appendix~\ref{appx:CavityDrive}.

This shortcut is simulated for $\ket{\psi_{N,0,0}}$ in Fig.~\ref{fig:N00}.
We set the cutoff to be $|\chi+\chi_s|\le 5\sqrt{2}N$.
The effect of this cutoff value is shown in Sec~\ref{sec:results}; the final purity and target overlap are not exactly 1 as they would be in a perfect shortcut.
\vspace*{-1pc}
\subsubsection{\textbf{Additional lasers}} \label{sec:CentralState}

We now discuss how to shortcut to the state $\ket{\psi_{0,N,0}}$ in the central DFS where $C=\chi=0$. This subspace has the maximal dimension.
To achieve this, consider a new shortcut Hamiltonian given by
\begin{equation} \label{eq:H1}
    \hat{H}_s=i\alpha\hat{b}_{-1}^\dag \hat{b}_1-i\alpha\hat{b}_1^\dag \hat{b}_{-1}
\end{equation}
which could be implemented experimentally by additional lasers coupling between the $\ket{1_g}$ and $\ket{\text{-}1_g}$ internal states of each atom.
The quantity $i\alpha$ is an imaginary free parameter which we need to find using Eq.~\eqref{eq:shortcutcrit}.
This derivation is shifted to Appendix~\ref{appx:CentralState} and the final result reads
\begin{align}
    \alpha=\frac{2\hbar\mu\rpd{\mu}}{\mu^4+1}.
\end{align}
Similarly to the shortcut given in Eq.~\eqref{eq:H1CavityDrive}, there is a trade-off between laser power and ramping speed imposed by $\rpd{\mu}$.
In contrast to Eq.~\eqref{eq:H1}, the value $\alpha$ and therefore the required laser power remains finite when one approaches $\mu=1$.
Thus we can ramp from the initial state exactly into $\mathcal{H}_{\text{DFS}}^{(0)}(1)$.
This shortcut is simulated in Fig.~\ref{fig:0N0} and up to numerical error which disappears with finer timesteps, the final state of the evolution is exactly $\ket{\psi_{0,N,0}}$.
\vspace*{-1pc}
\subsection{Results} \label{sec:results}

In this section we discuss all the numerical results obtained by simulating the protocols described in the previous sections.
To be specific, we simulated the time evolution of the system using a Runge–Kutta algorithm in Python.
The dimension of matrices representing operators and states is given by Eq.~\eqref{eq:HilbertDim}.
We compare our three methods of time evolution for a given DFS.
Figure~\ref{fig:N00} shows results for the single-dimensional DFS $\mathcal{H}_{\text{DFS}}^{(-N)}(\mu)$.
Figure~\ref{fig:0N0} uses the three-dimensional DFS $\mathcal{H}_{\text{DFS}}^{(0)}(\mu)$.
In both figures, the results of the shortcut method are in solid purple lines, the quench is in dashed blue, and the ramp is graphed in dotted orange.

\subsubsection{\textbf{Edge decoherence-free subspace}}

\begin{figure}[t]
    \centering
    \includegraphics[width=\linewidth]{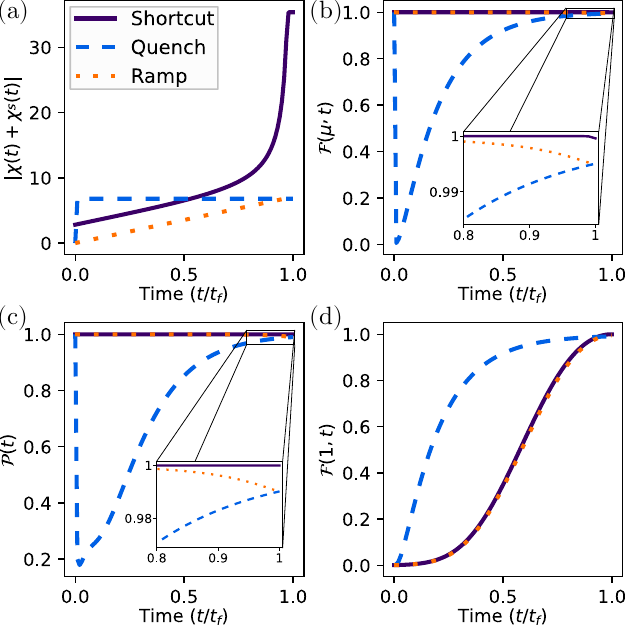}
    \caption{Time evolution of $\mathcal{H}_{\text{DFS}}^{(-N)}(\mu)$ in units of $t_f$ via shortcut (purple, solid), quench (blue, dashed), and ramp (orange, dotted) methods.
    We shortcut to $\ket{\psi_{N,0,0}}$.
    Here, $N=5$ and $\Delta_c'/\kappa=0.1$. The values of $t_f$ differ for each method and are found in Tab.~\ref{tab:N00}.
    \textbf{(a)} Cavity drive parameter $|\chi+\chi_s|$.
    The shortcut $\chi+\chi_s$ value is set at $25\sqrt{2}$.
    \textbf{(b)} Purity of $\hat{\rho}_{\mathrm{at}}(t)$.
    The inset shows behavior near $t=t_f$.
    \textbf{(c)} Overlap of $\hat{\rho}_{\mathrm{at}}(t)$ with $\mathcal{H}_{\text{DFS}}^{(-N)}(\mu(t))$ which measures how well the state matches the current DFS.
    The inset shows behavior near $t=t_f$.
    \textbf{(d)} Overlap of $\hat{\rho}_{\mathrm{at}}(t)$ with $\mathcal{H}_{\text{DFS}}^{(-N)}(1)$ which measures how well the state matches the target DFS.
    Note that $t_f$ is different for the different protocols and given in Tab.~\ref{tab:N00}.}
    \label{fig:N00}
\end{figure}

\begin{figure}[t]
    \centering
    \includegraphics[width=\linewidth]{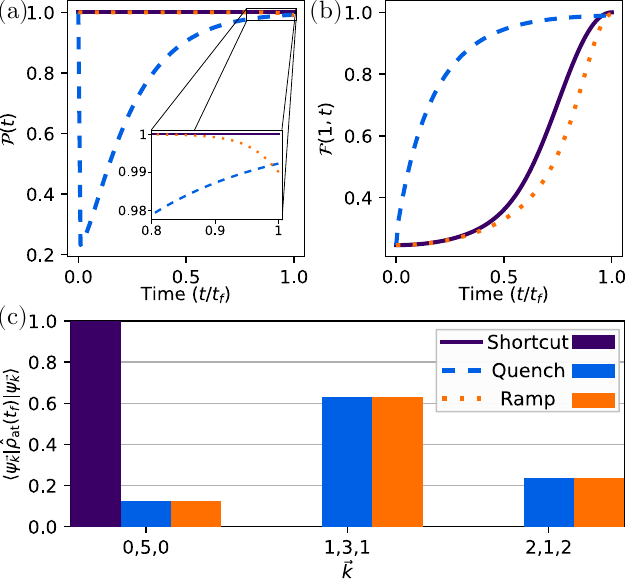}
    \caption{Time evolution of $\mathcal{H}_{\text{DFS}}^{(0)}(\mu)$ in units of $t_f$ via shortcut (purple, solid), quench (blue, dashed), and ramp (orange, dotted) methods.
    We shortcut to $\ket{\psi_{0,N,0}}$.
    Here, $N=5$ and $\Delta_c'/\kappa=0.1$. The values of $t_f$ differ for each method and are found in Tab.~\ref{tab:0N0}.
    \textbf{(a)} Purity of $\hat{\rho}_{\mathrm{at}}(t)$.
    The quench causes an immediate drop in purity similarly to Fig.~\ref{fig:N00}.
    The inset shows behavior near $t=t_f$.
    \textbf{(b)} Overlap of $\hat{\rho}_{\mathrm{at}}(t)$ with $\mathcal{H}_{\text{DFS}}^{(0)}(1)$ which measures how well the state matches the target DFS.
    \textbf{(c)} Probability distribution within $\mathcal{H}_{\text{DFS}}^{(0)}(1)$ for final state $\hat{\rho}_{\mathrm{at}}(t_f)$.
    The values of $t_f$ are different for the different protocols and given in Tab.~\ref{tab:0N0}.}
    \label{fig:0N0}
\end{figure}

First, we consider the extremal value of $C=-N$, i.e. the edge DFS $\mathcal{H}_{\text{DFS}}^{(-N)}(\mu)$.
This time-evolution is shown in Fig.~\ref{fig:N00}.
In Tab.~\ref{tab:N00} we show the numerical results for the evolution time $t_f$, the final purity $\mathcal{P}(t_f)$, and the final overlap $\mathcal{F}(1,t_f)$ with the $\mu=1$ target subspace.
For the quench, we also provide the final overlap $\mathcal{F}(\mu,t_f)$ with the $\mu(t_f)$ DFS because $\mu(t_f)=\mu_q=0.96$ in this case.
For the other methods, $\mu(t_f)=1$, so this final overlap is identical to $\mathcal{F}(1,t_f)$.

{\renewcommand{\arraystretch}{1.15}
\begin{table}[ht]
\center
\begin{tabular}{cLLLL}
     \hline\hline
     Method & t_f/\Gamma_c & \mathcal{P}(t_f) & \mathcal{F}(1,t_f) & \mathcal{F}(\mu(t_f), t_f)
     \\\hline
     Quench & 318 & .990 & .991 & .995
     \\
     Ramp & 137 & .990 & \multicolumn{2}{L}{.995}
     \\
     Shortcut & 1 & .999992 & \multicolumn{2}{L}{.9996} 
     \\\hline\hline
\end{tabular}
\caption{\label{tab:N00}
The different values for $t_f$ found and used to reach the state 
$\ket{\psi_{N,0,0}}$ and the values for the purities and overlaps reached after this time. The shortcut has unity purity and overlap up to numerical error.}
\end{table}}

Here, we can see that for both the quench and the ramp, the limiting factor for meeting the threshold given in Eq.~\eqref{eq:threshold} is purity, as $\mathcal{F}(1,t_f)$ for both methods is greater than $0.99$.
For the shortcut, despite having some small error due to the cutoff $\chi$ value, the result is far above the threshold for any evolution time sufficiently long for our approximations to hold.

In Fig.~\ref{fig:N00}\textbf{a}, for the quench and ramp, $\chi_s=0$, so $|\chi+\chi_s|=|\chi|=|\sqrt{2}\mu(-N)|$ which is proportional to $\mu$.
For the quench, $\chi$ increases from 0 to $\mu_q$ instantaneously.
For the shortcut, the cavity drive $\chi+\chi_s$ shows predominantly linear behavior at short times, as the dominant term is $\chi$, but as $\mu$ approaches 1 near the end of the evolution, $\chi_s$ given in Eq.~\eqref{eq:chiS} dominates and the power begins to diverge due to the $1/(\mu^4-1)$ dependence.

Figures~\ref{fig:N00}\textbf{b} and~\textbf{c} show the overlap $\mathcal{F}(\mu,t)$ with the current DFS $\mathcal{H}_{\text{DFS}}^{(-N)}(\mu)$, and the purity $\mathcal{P}(t)$, respectively.
For the quench, the overlap and purity both drop drastically when $\mu$ quenches because the state does not adapt immediately to the quick change in $\hat{H}_{\mathrm{at}}(t)$ and $\hat{L}(t)$.
After this, the purity increases exponentially to unity as $\hat{\rho}_{\mathrm{at}}(t)$ damps into the DFS.
For the other methods, the overlap and purity remain high throughout the evolution.
This is because the ramp is slow enough to ensure that $\hat{\rho}_{\mathrm{at}}(t)$ changes adiabatically, and the shortcut is defined to take initial pure dark state $\ket{\psi(0)}$ directly to final pure state $\ket{\psi_{5,0,0}}$ with unitary evolution.

Figure~\ref{fig:N00}\textbf{d} gives the overlap $\mathcal{F}(1,t)$ onto the target DFS $\mathcal{H}_{\text{DFS}}^{(-N)}(1)$.
This overlap starts at 0, as there is no overlap between $\ket{\psi(0)}$ and $\ket{\psi_{N,0,0}}$, which is the only state in $\mathcal{H}_{\text{DFS}}^{(-N)}(1)$.
As the state is driven towards its final value in each case, this target overlap increases.
As the system in both the ramp and the shortcut methods follows the current DFS $\mathcal{H}_{\text{DFS}}^{(-N)}(\mu)$ so closely throughout the evolution, their path through Hilbert space is similar, explaining why their target overlaps are similar throughout the evolution as well.
The key difference between these two is that the shortcut realizes these dynamics in a much shorter time.
\vspace*{-1pc}
\subsubsection{\textbf{Central decoherence-free subspace}}

Now, we consider time dynamics where we set $C=0$, corresponding to $\mathcal{H}_{\text{DFS}}^{(0)}(\mu)$, as shown in Fig.~\ref{fig:0N0}.
In Tab.~\ref{tab:0N0} we provide a similar set of numerical results for the previous state in Tab.~\ref{tab:N00}.
Here, the quench value for $\mu$ is $\mu(t_f)=\mu_q=0.98$.

{\renewcommand{\arraystretch}{1.15}
\begin{table}[ht]
\center
\begin{tabular}{cLLLL}
     \hline\hline
      Method & t_f/\Gamma_c & \mathcal{P}(t_f) & \mathcal{F}(1,t_f) & \mathcal{F}(\mu(t_f), t_f)
     \\\hline
     Quench & 396 & .992 & .990 & .996
     \\
     Ramp & 330 & .990 & \multicolumn{2}{L}{.995}
     \\
     Shortcut & 1 & 1 & \multicolumn{2}{L}{1} 
     \\\hline\hline
\end{tabular}
\caption{\label{tab:0N0}
The different values for $t_f$ used to reach the state $\ket{\psi_{0,N,0}}$ and the values for the purities and overlaps reached after this time.}
\end{table}}

The limiting factor for the quench scheme is the target overlap $\mathcal{F}(1,t_f)$.
We also had to pick a larger $\mu_q$ than in the $C=-N$ case, meaning that for $C=0$, there is a more significant difference between $\mathcal{H}_{\text{DFS}}^{(C)}(1-\epsilon)$ and $\mathcal{H}_{\text{DFS}}^{(C)}(1)$ for small $\epsilon$.
The chosen $t_f$ for the quench is slightly longer for $C=0$ than for $C=-N$.
Although the state damps more quickly into the $C=0$ DFS for a fixed $\mu_q$, at the same time we must increase $\mu_q$ to achieve the desired overlap, and this slows the damping.
From our numerics it appears that the latter effect is more significant in this case.

For $C=0$, $\chi=0$ always, so we omit the graph showing $\chi$.
Also, the overlap with the current DFS $\mathcal{F}(\mu,t)$ is similar to that of the $C=-N$ case shown in Fig.~\ref{fig:N00} for the different state, and is not used as a threshold for state preparation, so this is omitted as well.

Figure~\ref{fig:0N0}\textbf{a} shows the purity $\mathcal{P}(t)$ of each method.
This is similar to the corresponding graph in Fig.~\ref{fig:N00}, but the shape of the curves differ slightly.
We can see in the inset that the ramp purity stays near 1 longer, but decreases more quickly near the end of the evolution to reach the same $0.99$ threshold value.

Figure~\ref{fig:0N0}\textbf{b} gives the overlap with the target DFS $\mathcal{H}_{\text{DFS}}^{(0)}(1)$.
Here, even at $\mu=0$ there is non-zero overlap between $\ket{\psi(0)}$ and the target subspace.
We highlight that the shortcut and the ramp do not follow the same path through Hilbert space in this case because they end up at different states within the target DFS, and this is emphasized by the fact that they have different values for $\mathcal{F}(1,t)$ throughout the evolution.
\vspace*{-1pc}
\subsubsection{\textbf{Final state}} \label{sec:accessibleStates}
\vspace*{-.5pc}
As $\mathcal{H}_{\text{DFS}}^{(0)}(\mu)$ is multi-dimensional, determining the final state of the evolution analytically is nontrivial.
As the final purity $\mathcal{P}(t_f)$ is near 1 for all protocols, we can map $\hat{\rho}_{\mathrm{at}}(t_f)$ to a pure state $\ket{\psi(t_f)}$.
For the shortcut, we have $\ket{\psi(t_f)}=\ket{\psi_{0,5,0}}$ because our shortcuts are designed intentionally to drive between these states.
For the quench and the ramp, however, the purity ends near unity due to an additional underlying constraint.
The final state can be derived analytically using a procedure outlined below and derived in Appendix~\ref{appx:collectiveDipole}.

From Eq.~\eqref{eq:Ldef} we can see that $\hat{L}$ is composed of the operators $\hat{J}_\pm$.
These operators commute with the collective dipole operator, defined by
\begin{equation}~\label{eq:J2}
\hat{J}^2 = \hat{J}_-\hat{J}_++\hat{J}_+\hat{J}_-+\hat{J}_z^2,
\end{equation}
where $\hat{J}_z \equiv [ \hat{J}_+, \hat{J}_-]$ and the definition of $\hat{J}_\pm$ is given by Eq.~\eqref{eq:Jminus}.
In collections of two-level atoms, $\hat{J}^2$ is constant for any permutationally symmetric ensemble.
Collections of three-level atoms differ because, even in permutationally symmetric ensembles, $\hat{J}^2$ is not constant.
This can be seen from the fact that, starting with the state $(\hat{b}_{1}^\dag)^{N}\ket{0}/\sqrt{N!}$, repeated applications of $\hat{J}_-$ could never reach a state like $(\hat{b}_{0}^\dag)^{N}\ket{0}/\sqrt{N!}$ and, therefore, $\hat{J}^2$ acts differently on these two states.
Following convention, we will label the eigenvalues of $\hat{J}^2$ by a quantum number $j(j+1)$ where, for a three level system, $j$ is an integer between $0$ and $N$.

This imposes a constraint on the state in $\mathcal{H}_{\text{DFS}}^{(C)}(1)$ realized at the end of the evolution.
The initial state $\ket{\psi(0)}$ defined in Eq.~\eqref{eq:phi0 def} is an $\hat{J}^2$ eigenstate with $j=N$.
Both $\hat{H}_{\mathrm{at}}$ and $\hat{L}$ commute with $\hat{J}^2$ for all $\mu$, so $j=N$ is fixed throughout the evolution.
This constraint, along with the original constraint that $k_3-k_1=C$, means that the final state for both of these schemes must be 
\begin{equation}
    \ket{\psi(t_f)}=\ket{j=N,k_3-k_1=C}.
\end{equation}
In general, a state belonging to the symmetric ensemble of three-level atoms only needs to two quantum numbers to specify it~\cite{Mathur2010}.

This state can be calculated algorithmically in terms of the original basis using Clebsch-Gordon coefficients with the shells of this $\hat{J}^2$~\cite{Alex2011,Sakurai2020}.
Doing so for $C=0$ and $N=5$, we find
\begin{equation} \label{eq:psi final}
\begin{aligned}
    \ket{\psi(t_f)} = \frac{1}{\sqrt{63}}\left(\sqrt{8}\ket{\psi_{0,5,0}}+\sqrt{40}\ket{\psi_{1,3,1}}+\sqrt{15}\ket{\psi_{2,1,2}}\right).
\end{aligned}
\end{equation}
These populations are exactly what is found from the numerical calculations, which are shown in Fig.~\ref{fig:0N0}\textbf{c}.
There, we plot the overlaps between $\hat{\rho}_{\mathrm{at}}(t_f)$ and the three states included in Eq.~\eqref{eq:psi final}.
We note that for any $N$, the final state will be pure.

From these results, and the fact that the final overlap with the DFS is nearly unity, i.e. $\mathcal{F}(1,t_f)\approx1$, all three protocols are capable of preparing a pure state in the desired DFS.
This can be read off from Figs.~\ref{fig:N00} and~\ref{fig:0N0}.
The shortcut is clearly the fastest protocol, meaning it also has the opportunity to prepare the system in the target DFS while avoiding errors occurring on longer timescales which are not accounted for here.
\vspace*{-1pc}
\section{Realizing a Qubit} \label{sec:qubit}
\vspace*{-.5pc}
Once we have prepared a DFS, we may drive between decoherence-free states without causing any decoherence. This allows us to store quantum information and perform quantum information tasks by building a logical qubit which is stabilized by cavity-photon loss. Here, we focus on a two-dimensional DFS at $\mu=1$ for simplicity, but these results may be generalized.

A DFS will satisfy $\mathfrak{D}(N,C)=2$ exactly when
\begin{equation} \label{eq:qubit_constraint}
    C\in\{\pm(N-3),\pm(N-2)\}
\end{equation}
for $N\ge 3$, and $C=0$ for $N=2$. This can be derived using Eq.~\eqref{eq:dim}. Denote the two states in this subspace as
\begin{equation}
    \begin{aligned}
        |g\rangle= & |\psi_{k_1,k_2,k_3}\rangle
        \\|e\rangle= & |\psi_{k_1+1,k_2-2,k_3+1}\rangle
    \end{aligned}
\end{equation}
for fixed $k_1,k_2,k_3$ such that $C=k_3-k_1$ satisfies Eq.~\eqref{eq:qubit_constraint}. As each multi-dimensional DFS in our system contains entangled states, the operators which drive between $|g\rangle$ and $|e\rangle$ must contain multi-particle interaction terms. Any single-particle term such as $\hat{c}_i^\dag\hat{c}_j$ for $i\neq j$ will necessarily move some population outside of the DFS by changing $k_3-k_1$. For an operator to drive between states within a fixed DFS without causing decoherence, a sufficient condition is to enforce that the operator commutes with $\hat{L}$~\cite{Knill2000}. In this case, any time-evolution caused by such an operator will commute with $\hat{L}$, meaning that if the initial state of the system is a dark state of $\hat{L}$, the final state must be as well.

To drive between qubit states, we define effective Pauli operators~\cite{Sakurai2020} as follows:
\begin{equation}
    \begin{aligned}
        \hat{\sigma}_x= & \frac{1}{\mathcal{N}(N,C)}(\hat{c}_2^\dag\hat{c}_3\hat{c}_2^\dag\hat{c}_1+\hat{c}_1^\dag\hat{c}_2\hat{c}_3^\dag\hat{c}_2)
        \\\hat{\sigma}_y= & \frac{i}{\mathcal{N}(N,C)}(\hat{c}_2^\dag\hat{c}_3\hat{c}_2^\dag\hat{c}_1-\hat{c}_1^\dag\hat{c}_2\hat{c}_3^\dag\hat{c}_2)
        \\\hat{\sigma}_z= & \frac{-i}{2}\left[\hat{\sigma}_x,\hat{\sigma}_y\right]
        \\= & \frac{1}{\mathcal{N}^2(N,C)}(4\hat{n}_1\hat{n}_2\hat{n}_3-\hat{n}_1\hat{n}_2^2-\hat{n}_2^2\hat{n}_3
        \\ & -\hat{n}_2^2+2\hat{n}_1\hat{n}_3+\hat{n}_1\hat{n}_2+\hat{n}_2\hat{n}_3+\hat{n}_2)
    \end{aligned}
\end{equation}
where $\hat{n}_i=\hat{c}_i^\dag\hat{c}_i$ for $i\in\{1,2,3\}$ and
\begin{equation}
    \mathcal{N}(N,C)=\begin{cases}
        \sqrt{6N-12} & |C|=N-3,
        \\\sqrt{2N-2} & |C|=N-2,
    \end{cases}
\end{equation}
is a normalization dependent on the DFS. Note that these operators are Hermitian only at $\mu=1$ where the $\hat{c}_i$ operators satisfy canonical commutation relations.

The above operators behave identically to typical Pauli operators when acted on states within the chosen DFS. 
For complex coefficients $\alpha, \beta$, one can verify 
\begin{equation}
    \begin{aligned}
        \hat{\sigma}_x\left(\alpha|g\rangle+\beta|e\rangle\right)= & \beta|g\rangle+\alpha|e\rangle,
        \\\hat{\sigma}_y\left(\alpha|g\rangle+\beta|e\rangle\right)= & i\beta|g\rangle-i\alpha|e\rangle,
        \\\hat{\sigma}_z\left(\alpha|g\rangle+\beta|e\rangle\right)= & -\alpha|g\rangle+\beta|e\rangle,
    \end{aligned}
\end{equation}
which follows the usual operation of Pauli matrices.

Note that these operators behave differently from Pauli operators when the entire Hilbert space is considered. In particular, Pauli operators satisfy
\begin{equation}
    \left[\hat{\sigma}_i,\hat{\sigma}_j\right]=2i\epsilon_{i\,j\,k}\hat{\sigma}_k
\end{equation}
for $i,j,k\in\{x,y,z\}$ where $\epsilon_{i\,j\,k}$ is the Levi-Civita symbol~\cite{Sakurai2020}. In this system, this relation holds only when acting on states within the DFS and is not true over the entire Hilbert space because in general, $\left[\hat{\sigma}_i,\hat{\sigma}_j\right]$ contains higher orders of the $\hat{c}_i$ operators.

The entangling operations described by the $\hat{\sigma}_i$ must be engineered, but can in principle be realized in experiment~\cite{Bohnet2016}. Moreover, this complexity comes with a great advantage: the very same complexity that is needed to control the qubit also hinders flips between the states $\ket{g}$ and $\ket{e}$. The reason for this is that the qubit is protected by the DFS, so such flips require two-particle terms which makes the qubit naturally robust against single-particle decoherence.
\vspace*{-1pc}
\section{Conclusion and Outlook} \label{sec:conclusion}
\vspace*{-1pc}
We have analyzed an experimentally relevant atom-cavity system wherein one may create and control a DFS.
The DFS is in general multi-dimensional, containing degenerate states which do not decohere under photon loss.
This DFS comes from diagonalizing the non-Hermitian jump operator via a non-unitary transformation.
This diagonalization yields degenerate eigenspaces of the jump operator, where we can tune the external drive to make any of these eigenspaces into a DFS.
Within this DFS we show several protocols to prepare pure states by quenching, adiabatically changing the DFS in time, or by employing adiabatic shortcuts.
These adiabatic shortcuts allow states to be prepared on rapid timescales only limited by the lifetime of photons in the cavity--i.e. the validity of the model.
We also provide operators which fully control a decoherence-free qubit without causing any decoherence.

Our work provides an important stepping stone towards controlling multi-dimensional DFSs in an atom-cavity system, and opens the door to several interesting questions.
Future work will explore the use of this system in experimentally relevant parameter regimes, where we expect to see protection against experimentally relevant noise sources.
Of particular interest is the control of the dimensionality of the DFS in order to create qudits which are protected against photon decay.
These protected qudits may be relevant for quantum computing and quantum metrology~\cite{Ostermann2013,Roos2006}.

Lastly, we plan to extend this formalism to arbitrarily-many ground states where we expect more complicated dynamics to play out.
This is inspired in large part by atom-optical lattice systems~\cite{Kollath2016,Colella2019,Sheikhan2016}, which can be used to realize complex and controllable physics~\cite{Ledesma2024,Ledesma2024GateSet}.
There, one may find that more than the symmetric state manifold, and more than one jump operator become relevant.
\vspace*{-1pc}
\section{Acknowledgments}
\vspace*{-.5pc}
We would like to thank Shah Saad Alam and Nicholas T. Weir for useful discussions.
This work is supported by the Department of Energy Quantum Systems Accelerator (Grant No.\ 7565477) and the National Science Foundation (Grants~No.~PHY-2317149 and No.\ OSI-2016244).

\appendix
\vspace*{-1pc}
\section{Effective Master Equation} \label{appx:MasterEquationDerivation}
\vspace*{-.5pc}
In this appendix we provide the details for the derivation of the master equation given in Eq.~\eqref{eq:masterequation2} and the subsequent equations in that section.
Begin with the system shown in Fig.~\ref{fig:Model}.
The coherent dynamics are governed by the Hamiltonian

\begin{equation}
    \begin{aligned}
        \hat{H_0}=&\hbar\omega_c \hat{a}^\dag \hat{a}+\hbar\left(\eta \hat{a}^\dag e^{-i\omega_d t}+\text{H.c.}\right)
        \\&+\sum_{j=1}^N\hbar\omega_s\left(\ket{1_g}_j\bra{1_g}-\ket{\text{-}1_g}_j\bra{\text{-}1_g}\right) + \hbar\omega_e\ket{0_e}_j\bra{0_e}
        \\& + \hbar\left(\omega_e-\omega_s\right)\ket{\text{-}1_e}_j\bra{\text{-}1_e} + \hbar\left(\omega_e+\omega_s\right)\ket{1_e}_j\bra{1_e}
        \\&+\hbar g\hat{a}^\dag\left(\ket{\text{-}1_g}_j\bra{\text{-}1_e}+\ket{0_g}_j\bra{0_e}+\ket{1_g}_j\bra{1_e}\right)+\text{H.c.}
        \\&+\frac{\hbar\Omega_1}{2}\left(\ket{\text{-}1_e}_j\bra{0_g}e^{-i\omega_1 t} + \ket{0_e}_j\bra{1_g}e^{-i\omega_1 t}\right)+\text{H.c.}
        \\&+\frac{\hbar\Omega_2}{2}\left(\ket{1_e}_j\bra{0_g}e^{-i\omega_2 t}+\ket{0_e}_j\bra{\text{-}1_g}e^{-i\omega_2 t}\right)+\text{H.c.}
    \end{aligned}
\end{equation}
where $j$ denotes the $j$-th atom in the cavity and $\omega_s, \omega_e$ are the bare frequencies of states $\ket{1_g}_j,\ \ket{0_e}_j,$, respectively.
We have defined $\hat{a}$ ($\hat{a}^\dag$) as the annihilation (creation) operator for a photon in the cavity mode with frequency $\omega_c$.
The amplitude of the laser field driving the cavity is given by $\eta$, and it has frequency $\omega_d$.
This assumes no cross-coupling between the $\Omega_1,\ \Omega_2$ drives, and that spontaneous decay from the excited states is negligible.
\vspace*{-1pc}
\subsection{Interaction picture}
\vspace*{-.5pc}
In the first step we transform the Hamiltonian into the interaction picture using $\hat{\rho} \to \hat{U} \hat{\rho} \hat{U}^\dag$ with $\hat{U} = \exp [ i \hat{H}' t/ \hbar ]$ with
\begin{equation}
    \begin{aligned}
        \hat{H}'= & \hbar\frac{\omega_1+\omega_2}{2}\hat{a}^\dag\hat{a}+\sum_{j=1}^N\frac{-\omega_1+\omega_2}{2}\hbar(\ket{1_g}_j\bra{1_g}-\ket{\text{-}1_g}_j\bra{\text{-}1_g})
        \\& +\hbar\frac{\omega_1+\omega_2}{2}\ket{0_e}_j\bra{0_e}+\hbar\omega_1\ket{\text{-}1_e}_j\bra{\text{-}1_e}+\hbar\omega_2\ket{1_e}_j\bra{1_e}
    \end{aligned}
\end{equation}
so that our Hamiltonian becomes
\begin{equation}
    \begin{aligned}
        \hat{H}_I=&-\hbar\Delta_c \hat{a}^\dag \hat{a}+\hbar\left(\eta \hat{a}^\dag e^{i\Delta_d t}+\text{H.c.}\right)
        \\& +\hbar\sum_{j=1}^N-\Delta_s\left(\ket{1_g}_j\bra{1_g}-\ket{\text{-}1_g}_j\bra{\text{-}1_g}\right)-\Delta_e\ket{0_e}_j\bra{0_e}
        \\& -\left(\Delta_e-\Delta_s\right)\ket{\text{-}1_e}_j\bra{\text{-}1_e}-\left(\Delta_e+\Delta_s\right)\ket{1_e}_j\bra{1_e}
        \\& + g\hat{a}^\dag\left(\ket{\text{-}1_g}_j\bra{\text{-}1_e}+\ket{0_g}_j\bra{0_e}+\ket{1_g}_j\bra{1_e}\right) + \text{H.c.}
        \\& +\frac{\Omega_1}{2}\left(\ket{\text{-}1_e}_j\bra{0_g} + \ket{0_e}_j\bra{1_g}+\text{H.c.}\right)
        \\& +\frac{\Omega_2}{2}\left(\ket{1_e}_j\bra{0_g}+\ket{0_e}_j\bra{\text{-}1_g}+\text{H.c.}\right)
    \end{aligned}
\end{equation}
where we have introduced the detunings $\Delta_c=(\omega_1+\omega_2)/2-\omega_c,\ \Delta_e=(\omega_1+\omega_2)/2-\omega_e,\ \Delta_s=(-\omega_1+\omega_2)/2-\omega_s,\ \Delta_d=(\omega_1+\omega_2)/2-\omega_d$.
\vspace*{-1pc}
\subsection{Excited-state elimination}
\vspace*{-.5pc}
We now eliminate the cavity degrees of freedom.
This explicitly assumes the highly-detuned limit, $\left|\Delta_e\right|\gg g,\ \left|\Delta_s\right|,\ \left|\Omega_1\right|,\ \left|\Omega_2\right|$.
This allows us to adiabatically eliminate the excited states $\ket{\pm 1_e},\ \ket{0_e}$ which evolve on a faster timescale than the ground states~\cite{Steck2007}.
After this elimination we may write the Hamiltonian in terms of the ground states and the cavity.
To simplify the result, we set $\Delta_s=\Delta_d=0$ and take the weak-drive limit $g\gg \left|\Omega_1\right|,\ \left|\Omega_2\right|$ to suppress undesired coupling between $\ket{1_g}$ and $\ket{-1_g}$.
The resulting Hamiltonian is
\begin{equation}
    \hat{H}= -\hbar\Delta_c' \hat{a}^\dag \hat{a} + \hbar\left\{ \hat{a}^\dag\left[\eta+\frac{g}{2\Delta_e}\left(\Omega_1\hat{J}_- + \Omega_2 \hat{J}_+\right)\right]+\text{H.c.}\right\}
\end{equation}
where $\Delta_c'=\Delta_c -N g^2/\Delta_e$ is the dressed cavity detuning.
The reduced density matrix describing just the ground states and the cavity field, $\hat{\rho}_{\text{at-cav}}$, can be described by the master equation
\begin{equation}
    \hat{\mathcal{L}}_{\text{diss}}\hat{\rho}_{\text{at-cav}} = \frac{1}{i \hbar} \left[ \hat{H}, \hat{\rho}_{\text{at-cav}} \right] + 2\kappa\hat{\mathcal{D}}\left[\hat{a}\right]\hat{\rho}_{\text{at-cav}},
\end{equation}
using the above Hamiltonian and dissipation coming from cavity photon losses.
\vspace*{-1pc}
\subsection{Cavity-mode elimination}

We now eliminate the cavity field to derive a time evolution that only involves the ground states.
For this we assume that the cavity photon lifetime is much shorter than collectively-enhanced processes, $\kappa/\sqrt{N}\gg|g\Omega_1/\Delta_e|,\ |g\Omega_2/\Delta_e|$.
This allows us to adiabatically eliminate the cavity mode and determine our final effective master equation over the atomic degrees of freedom similarly to Refs.~\cite{Jager2022,Reilly2022}.
The final master equation for the reduced density matrix $\hat{\rho}_{\mathrm{at}}$ calculated by taking the partial trace of $\hat{\rho}_{\text{at-cav}}$ over the cavity mode is given by Eq.~\eqref{eq:masterequation2}.
\vspace*{-1pc}
\section{States} \label{appx:overlap}
\vspace*{-.5pc}
The purpose of this appendix is to provide more information and details about the eigenstates of the jump operator $\hat{L}$.
\vspace*{-1pc}
\subsection{Eigenvalue}
\vspace*{-.5pc}
In this first part we discuss some technical details on the derivation of the eigenvalues.
For this we use the form of Eq.~\eqref{eq:L as cd combo} and apply it onto the eigenstate $\ket{\psi_{\vec{k}}}$ given by Eq.~\eqref{eq:LEigenstates}.

We will commute the ladder operators in $\hat{L}$ through $\ket{\psi_{\vec{k}}}$ which means the goal is to simplify
\begin{equation}
    \left(\hat{c}_1^\dag\hat{d}_1-\hat{c}_3^\dag\hat{d}_3\right)(\hat{c}_1^\dag)^{k_1}(\hat{c}_2^\dag)^{k_2}(\hat{c}_3^\dag)^{k_3}\ket{0}
\end{equation}
by commuting the first term through the second. As $[\hat{c}_i,\hat{d}_j^\dag]=\delta_{i\,j}$ and $[\hat{c}_i^\dag,\hat{c}_j^\dag]=0$, the only terms which do not commute are
\begin{equation} \label{eq:dc commutated}
    \hat{d}_i(\hat{c}_i^\dag)^{k_i}=(\hat{c}_{i}^\dag)^{k_i}\hat{d}_{i}+ k_i(\hat{c}_i^\dag)^{k_i-1}
\end{equation}
for $i\in\{1,3\}$.
The first term of Eq.~\eqref{eq:dc commutated} disappears when applied on the vacuum state.
This is used to show the relation
\begin{equation} \label{eq:cd action}
    \hat{c}_i^\dag\hat{d}_i\ket{\psi_{\vec{k}}}=k_i\ket{\psi_{\vec{k}}}
\end{equation}
and leads to the result in Eq.~\eqref{eq:LEigenvalues}.
\vspace*{-1pc}
\subsection{Eigenstate overlap}
\vspace*{-.5pc}
We now calculate the overlap between $\hat{L}$ eigenstates to determine the normalization $\mathcal{N}_{\vec{k}}$.

For $\mu=1$, the $\hat{c}_i$ operators satisfy canonical commutation relations $[\hat{c}_i,\hat{c}_j^\dag]=\delta_{i\,j}$, so their overlap is $\ip{\psi_{\vec{k}'}}{\psi_{\vec{k}}}=\delta_{\vec{k}'\,\vec{k}}$, and their normalization takes the standard form
\begin{equation}
    \frac{1}{\mathcal{N}_{\vec{k}}^2}=k_1!k_2!k_3!.
\end{equation}

For $0<\mu<1$, this expression becomes more involved.
We will ``orthogonalize'' the $\hat{c}_i$ operators similarly to Ref.~\cite{Reilly2022}.
We write $\hat{c}_2=a_1\hat{c}_{1}+\hat{c}_{\beta}$ and $\hat{c}_3=a_2\hat{c}_{1}+a_3\hat{c}_{\beta}+\hat{c}_{\gamma}$ assuming $a_i\in\R$ for $i\in\{1,2,3\}$ and $[\hat{c}_1,\hat{c}_{\beta}^\dag]=[\hat{c}_1,\hat{c}_{\gamma}^\dag]=[\hat{c}_{\beta},\hat{c}_{\gamma}^\dag]=0$.
We can then use $a_1=[\hat{c}_2,\hat{c}_1^\dag]/[\hat{c}_1,\hat{c}_1^\dag]$, $a_2=[\hat{c}_3,\hat{c}_1^\dag]/[\hat{c}_1,\hat{c}_1^\dag]$, $a_3=[\hat{c}_3,\hat{c}_{\beta}^\dag]/[\hat{c}_{\beta},\hat{c}_{\beta}^\dag]$.
Using this approach we find
\begin{align}
    a_1=a_3= & \sqrt{2}\frac{1 - \mu^2}{\mu^2 + 1},\\
     a_2= & \frac{(\mu - 1)^2(\mu + 1)^2}{(\mu^2 + 1)^2},
    \\
    \hat{c}_{\beta} = & \frac{\mu^{2}}{\mu^{2} + 1}(\sqrt{2} \mu \hat{b}_{-1} + \left(\mu^{2} - 1\right) \hat{b}_{0} - \sqrt{2} \mu \hat{b}_{1}),\\
    \hat{c}_{\gamma} = & \frac{2 \mu^{3}}{\left(\mu^{2} + 1\right)^{2}}\left(\mu^{2} \hat{b}_{-1} - \sqrt{2} \mu \hat{b}_{0} + \hat{b}_{1}\right).
\end{align}

We use these relations and then solve for the corresponding operators.
Now calculate the overlap between two $\hat{L}$ eigenstates in terms of these new operators.
\begin{equation} \label{eq:overlap initial}
    \begin{aligned}
        \ip{\psi_{\vec{k}'}}{\psi_{\vec{k}}}=&\mathcal{N}_{\vec{k}'} \mathcal{N}_{\vec{k}} \bra{0} \hat{c}_1^{k_1'} \hat{c}_2^{k_2'} \hat{c}_3^{k_3'} (\hat{c}_{3}^\dag)^{k_3} (\hat{c}_2^\dag)^{k_2} (\hat{c}_1^\dag)^{k_1} \ket{0}
        \\= & \mathcal{N}_{\vec{k}'} \mathcal{N}_{\vec{k}} \bra{0} \hat{c}_1^{k_1'} \left(a_1\hat{c}_{1}+\hat{c}_{\beta}\right)^{k_2'}\left(a_2\hat{c}_{1}+a_3\hat{c}_{\beta}+\hat{c}_{\gamma}\right)^{k_3'}
        \\ & \times(a_2\hat{c}_{1}^\dag+a_3\hat{c}_{\beta}^\dag+\hat{c}_{\gamma}^\dag)^{k_3}(a_1\hat{c}_{1}^\dag+\hat{c}_{\beta}^\dag)^{k_2} (\hat{c}_1^\dag)^{k_1} \ket{0}
    \end{aligned}
\end{equation}
We may apply the multinomial theorem to each term above, as the operators in each term commute with each other~\cite{Stanley2011},
\begin{equation}
    \left(x_1+...+x_m\right)^n=\sum_{\substack{q_1,...,q_m\ge 0 \\ q_1+...+q_m=n}}\frac{n!}{q_1!...q_m!}x_1^{q_1}...x_m^{q_m}.
\end{equation}
This turns Eq.~\eqref{eq:overlap initial} into a sum of terms.
Since $\hat{c}_1,\ \hat{c}_{\beta}$, and $\hat{c}_{\gamma}$ commute by construction, we may rearrange the result and apply $\bra{0}\hat{c}_i^k (\hat{c}_i^\dag)^l\ket{0}=k![\hat{c}_i,\hat{c}_i^\dag]^k\delta_{k\,l}$.
We achieve

\begin{widetext}
    \begin{equation}
        \begin{aligned}
            \ip{\psi_{\vec{k}'}}{\psi_{\vec{k}}}=&\mathcal{N}_{\vec{k}'} \mathcal{N}_{\vec{k}} \sum_{q=0}^{k_2'}\sum_{s=0}^{k_2}\sum_{r_1=\max(0, s-q)}^{\min(k_3', k_3+s-q)}\sum_{r_2=0}^{\min(k_3'-r_1, k_3-q+s-r_1)}\binom{k_2'}{q}\binom{k_2}{s}\binom{k_3'}{r_1}\binom{k_3'-r_1}{r_2}\binom{k_3}{q-s+r_1}\binom{k_3-q+s-r_1}{r_2}
            \\&\times  \left(N-q-r_1-r_2\right)! \left(q+r_1\right)! r_2! a_1^{k_2'+k_2-q-s} a_2^{k_3'+k_3-2r_1-2r_2-q+s} a_3^{q-s+2r_1}[\hat{c}_{1},\hat{c}_{1}^\dag]^{N-q-r_1-r_2} [\hat{c}_{\beta},\hat{c}_{\beta}^{\dag}]^{q+r_1} [\hat{c}_{\gamma},\hat{c}_{\gamma}^{\dag}]^{r_2}
        \end{aligned}
    \end{equation}
    Using the explicit results for $a_i$ and the operators $\hat{c}_1$, $\hat{c}_{\beta}$, and $\hat{c}_{\gamma}$ we get
    \begin{equation}
        \begin{aligned}
            \ip{\psi_{\vec{k}'}}{\psi_{\vec{k}}}=&\mathcal{N}_{\vec{k}'} \mathcal{N}_{\vec{k}} \left(-1\right)^{k_{2} + k'_2}\sum_{q=0}^{k_2'}\sum_{s=0}^{k_2}\sum_{r_1=\max(0, s-q)}^{\min(k_3', k_3+s-q)}\sum_{r_2=0}^{\min(k_3'-r_1, k_3-q+s-r_1)} 2^{- 2 N + k_2/2 + k'_2/2 + 2 q + 3 r_{1} + 4 r_{2} - s} 
            \\&\times\binom{k_2'}{q}\binom{k_2}{s}\binom{k_3'}{r_1}\binom{k_3'-r_1}{r_2}\binom{k_3}{q-s+r_1}\binom{k_3-q+s-r_1}{r_2} \left(N-q-r_1-r_2\right)! \left(q+r_1\right)! r_2!
            \\&\times\mu^{2 N + 2 q + 2 r_{1} + 4 r_{2}} \left[\left(\mu - 1\right) \left(\mu + 1\right)\right]^{k_{2} + k'_2 + 2 k_{3} + 2 k'_3 - 2 q - 2 r_{1} - 4 r_{2}} \left(\mu^{2} + 1\right)^{2 N - k_{2} - k'_2 - 2 k_{3} - 2 k'_3}.
        \end{aligned}
    \end{equation}
    To calculate the normalization, we set $\vec{k}'=\vec{k}$ and use that $\ip{\psi_{\vec{k}'}}{\psi_{\vec{k}}}=1$.
    The final result is
    \begin{equation}
        \begin{aligned}
            \frac{1}{\mathcal{N}_{\vec{k}}^2}=&\sum_{q,s=0}^{k_2}\sum_{r_1=\max(0, s-q)}^{\min(k_3, k_3+s-q)}\sum_{r_2=0}^{\min(k_3-r_1, k_3-q+s-r_1)}2^{- 2 N + k_{2} + 2 q + 3 r_{1} + 4 r_{2} - s} 
            \\&\times \binom{k_2}{q}\binom{k_2}{s}\binom{k_3}{r_1}\binom{k_3-r_1}{r_2}\binom{k_3}{q-s+r_1}\binom{k_3-q+s-r_1}{r_2} (N-q-r_1-r_2)! (q+r_1)! r_2!
            \\&\times \mu^{2 N + 2 q + 2 r_{1} + 4 r_{2}} \left(\left(\mu - 1\right) \left(\mu + 1\right)\right)^{2 k_{2} + 4 k_{3} - 2 q - 2 r_{1} - 4 r_{2}} \left(\mu^{2} + 1\right)^{2 N - 2 k_{2} - 4 k_{3}}.
        \end{aligned}
    \end{equation}
\end{widetext}

\subsection{Complementary states}

We now discuss properties of the orthogonal complement of the eigenspaces.

The overlap and normalization calculations for complementary states 
\begin{equation}
    \ket{\psi_{\vec{n}}^\perp}=\mathcal{N}_{\vec{n}}^\perp(\hat{d}_{1}^\dag)^{n_1}(\hat{d}_2^\dag)^{n_2}(\hat{d}_3^\dag)^{n_3}\ket{0}.
\end{equation}
are similar to those of $\hat{L}$ eigenstates, but we must expand the $\hat{d}_i$ operators and use their commutation relations.
These states are eigenstates of $\hat{L}^\dag$ satisfying
\begin{equation} \label{Ldag psi^perp}
    \hat{L}^\dag\ket{\psi_{\vec{n}}^\perp}=\sqrt{\Gamma_c}\left[\sqrt{2}\mu\left(n_1-n_3\right)+\chi\right]\ket{\psi_{\vec{n}}^\perp}.
\end{equation}
The overlap is given by
\begin{equation}
    \ip{\psi_{\vec{n}'}^\perp}{\psi_{\vec{n}}^\perp}=\left(-1\right)^{n_{2} + n'_2}\ip{\psi_{\vec{n}'}}{\psi_{\vec{n}}}
\end{equation}
and the state normalization is
\begin{equation}
    \mathcal{N}^\perp_{\vec{n}}=\mu^{4N}\mathcal{N}_{\vec{n}}
\end{equation}
for $0<\mu\le 1$.

These states are orthogonal to the $\hat{L}$ eigenstates.
To calculate this we use
\begin{equation}
    \ip{\psi^\perp_{\vec{n}}}{\psi_{\vec{k}}} =\mathcal{N}^\perp_{\vec{n}}\mathcal{N}_{\vec{k}}\bra{0}\hat{d}_{1}^{n_1}\hat{d}_2^{n_2}\hat{d}_3^{n_3}(\hat{c}_{1}^\dag)^{k_1}(\hat{c}_2^\dag)^{k_2}(\hat{c}_3^\dag)^{k_3}\ket{0}
\end{equation}
The nontrivial commutation relations here are
\begin{equation}
    \bra{0}\hat{d}_i^{n_i}(\hat{c}_i^\dag)^{k_i}\ket{0}=k_i!\delta_{n_i\,k_i}
\end{equation}
for $i\in\{1,2,3\}$, which follows from Eq.~\eqref{eq:dc commutated}.
The result is
\begin{equation}
    \ip{\psi^\perp_{\vec{n}}}{\psi_{\vec{k}}} =\mathcal{N}^\perp_{\vec{n}}\mathcal{N}_{\vec{k}}k_1!k_2!k_3!\delta_{\vec{n}\,\vec{k}}.
\end{equation}

\section{Interparticle Entanglement} \label{appx:interparticleEntanglement}

We will now briefly discuss the entanglement and its usefulness in this appendix.

We start with a general statement about which eigenstates are entangled.
We derive eigenstates of the jump operators
\begin{align}
\hat{L}=\sqrt{\Gamma_c}\left(\sum_{j=1}^N\hat{\ell}_j+\chi\right)
\end{align}
with single-particle operators
\begin{align}
\hat{\ell}_j=(\ket{- 1_g}_j \bra{0_g} + \ket{0_g}_j \bra{1_g})+\mu^2(\ket{0_g}_j \bra{-1_g} + \ket{1_g}_j \bra{0_g}).
\end{align}
This means that each symmetric unentangled pure state is uniquely defined by the eigenstates of $\hat{\ell}_j$.
This is true because if one fixes the state for a single $j$, one fixes all other states because the state needs to be a symmetric tensor product of single-particle states.
There are three eigenstates of $\hat{\ell}_j$ which correspond to the three unentangled eigenstates of $\hat{L}$ which satisfy $\vec{k}\in\{(N,0,0),(0,N,0),(0,0,N)\}$.
All other eigenstates are consequently entangled.

This general statement does not quantify the amount of entanglement in the other eigenstates.
Below we will provide an example of an eigenstate whose entanglement results in metrological use.
This state is given by $\ket{\psi_{N/2,0,N/2}}$ for even $N$ and is reminiscent of a ring state when $\mu \approx 1$~\cite{DallaTorre2013,Reilly2022}.
This state is degenerate with the product state $\ket{\psi_{0,N,0}}$ as both are dark when $\chi=0$.
This means that we may shortcut to $\mathcal{H}_{\text{DFS}}^{(0)}(\mu=1)$ and in principle drive into this highly sensitive and entangled state without leaving the DFS.

To determine if the interparticle entanglement in the state is metrologically useful, we can find the quantum Fisher information (QFI)~\cite{Liu2020} with respect to a given parameter.
The QFI is a sufficient entanglement witness when it is larger than $N$~\cite{Reilly2023}, and it is capped by the Heisenberg limit of $N^2$~\cite{Holland1993}.
In the effective two-level system formed at $\mu=1$ by basis states $\hat{c}_1^\dag\ket{0}$ and $\hat{c}_3^\dag\ket{0}$, we can use the methodology of Ref.~\cite{Reilly2023} to find that $\ket{\psi_{N/2,0,N/2}}$ is highly sensitive to action of
\begin{equation} \label{eq:Keffx}
    \hat{K}_{\mathrm{eff}}^x=\frac{\hat{c}_1^\dag\hat{c}_3+\hat{c}_3^\dag\hat{c}_1}{2}.
\end{equation}
The QFI of $\ket{\psi_{N/2,0,N/2}}$ with respect to this operator can be found analytically,
\begin{equation}
    4 \left[ \left\langle\left(\hat{K}_{\mathrm{eff}}^x\right)^2\right\rangle-\left\langle \hat{K}_{\mathrm{eff}}^x\right\rangle^2 \right]=\frac{N^2}{2}+N,
\end{equation}
which has Heisenberg-limit scaling.

Meanwhile, when $N$ is odd, the eigenstates with the highest degree of interparticle entanglement satisfy $\vec{k}=((N\pm 1)/2,0,(N\mp 1)/2)$.
For the same operator, the QFI at $\mu=1$ is given by
\begin{equation}
    4 \left[ \left\langle\left(\hat{K}_{\mathrm{eff}}^x\right)^2\right\rangle-\left\langle \hat{K}_{\mathrm{eff}}^x\right\rangle^2 \right]=\frac{N^2 - 1}{2}+N,
\end{equation}
which also exhibits Heisenberg-limit scaling.

\section{The Collective Dipole Operator in SU(3)} \label{appx:collectiveDipole}

We will now consider the collective dipole operator~\cite{Griffiths2017} given by Eq.~\eqref{eq:J2}.
We can write this as
\begin{equation}
    \begin{aligned}
        \hat{J}^2= & \hat{K}_-\hat{K}_++\hat{K}_+\hat{K}_-+\hat{K}_z^2,
        \\\hat{K}_-= & \hat{c}_1^\dag\hat{c}_2+\hat{c}_2^\dag\hat{c}_3=\hat{K}_+^\dag,
        \\\hat{K}_z= & \hat{c}_3^\dag\hat{c}_3-\hat{c}_1^\dag\hat{c}_1.
    \end{aligned}
\end{equation}
Here, the $\hat{c}_i$ operators are defined at $\mu=1$, meaning that $\vec{c}=\vec{d}$ and the jump operator can be written as $\hat{L}(\mu=1)=\sqrt{\Gamma_c}\left(\sqrt{2}\hat{K}_z+\chi\right)$.
In the $\mu=1$ limit, finding the eigenstates of $\hat{K}_z$ also finds the eigenstates of $\hat{L}$.
Now, we rephrase the time evolution of the system as rotating the states from a simultaneous $\hat{J}^2$, $\hat{J}_z$ eigenstate into a simultaneous $\hat{J}^2$, $\hat{K}_z$, eigenstate.
The degeneracy of the DFS is then found from the fact that one may find states sharing the same eigenvalue of $\hat{K}_z$, but living on different $\hat{J}^2$ shells.

Now, because these atoms are fundamentally three-level systems we can imagine that the allowed eigenvalues of $\hat{J}^2$ should match those of symmetrically adding up collections of spin-$1$ particles.
We let $j(j+1)$ label the eigenvalues of $\hat{J}^2$ where $j$ is a whole number between $0$ and $N$ and $m_j$ label the eigenvalues of $\hat{J}_z$.
The allowed $j$ and $m_j$ values in the system are
\begin{equation} \label{eq:allowed j}
    \begin{aligned}
        &j\in\left\{N,N-2,N-4,...,0\right\}, \quad \mathrm{for\ even\ }N\\
        &j\in\left\{N,N-2,N-4,...,1\right\}, \quad \mathrm{for\ odd\ }N
    \end{aligned}
\end{equation}
For each of these $j$, the allowed $m_j$ values are
\begin{equation}
    m_j\in\{-j,-j+1,...,j\}.
\end{equation}
For the operator $\hat{K}_z$ we similarly define the eigenvalue $m_k$ which has allowed values running over the same range as $m_j$.
This defines the $(N+1)(N+2)/2$ linearly independent states which match Eq.~\eqref{eq:HilbertDim}.
Later in this section, we also show these are the only accessible states by repeated application of $\hat{J}_-$ and orthogonalization.

The initial state $\ket{\psi(0)}$ defined in Eq.~\eqref{eq:phi0 def} is a simultaneous eigenstate of $\hat{J}^2$ and $\hat{J}_z$, where $\hat{J}^2\ket{\psi(0)}=N(N+1)\ket{\psi(0)}$ and $\hat{J}_z\ket{\psi(0)}=-N\ket{\psi(0)}$, meaning that $j=N$ and $m_j = -N$.
At the end of the evolution when $\mu=1$, we have $\hat{K}_z\ket{\psi(t_f)}=C\ket{\psi(t_f)}$, meaning that $m_k=C$.
Both $\hat{H}_{\mathrm{at}}$ and $\hat{L}$ commute with $\hat{J}^2$ at all $\mu$, so for the quench and ramp methods discussed in Secs.~\ref{sec:quench} and \ref{sec:LinearRamp}, $j=N$ is conserved throughout the evolution.
Thus the final state for both of these schemes is
\begin{equation} \label{eq:psi final as jm}
    \ket{\psi(t_f)}=\ket{j=N,m_k=C}
\end{equation}
By examination, the final state $\ket{\psi(t_f)}$ given in Eq.~\eqref{eq:psi final} satisfies
\begin{equation}
    \begin{aligned}
        \hat{J}^2\ket{\psi(t_f)}= & 5(5+1)\ket{\psi(t_f)}
        \\\hat{K}_z\ket{\psi(t_f)}= & 0\ket{\psi(t_f)}
    \end{aligned}
\end{equation}
meaning that this state can be equivalently written as $\ket{j=5,m_k=0}$, which is consistent with Eq.~\eqref{eq:psi final as jm}.
As $j$ and $m_k$ are both constrained at the end of the evolution, they define the only accessible state in the target DFS, so the final purity must be 1.

We calculate this state algorithmically using the same methods as one does in finding the usual Clebsch-Gordon coefficients~\cite{Alex2011,Sakurai2020}.
In order to do so, we use the following helpful expression for $\hat{J}_{\pm}$~\cite{Griffiths2017}
\begin{equation} \label{eq:J_pm action} 
    \hat{J}_{\pm}\ket{j,m_j}=\sqrt{[j(j+1)-m(m\pm 1)]/2}\ket{j,m_j\pm 1}.
\end{equation}
We then write the accessible $\ket{j,m}$ in terms of the original basis represented by Schwinger bosons $\hat{b}_i^\dag$ or $\hat{c}_i^\dag$.
We define
\begin{equation}
    \ket{i,j,k}\equiv\frac{1}{\sqrt{i!j!k!}}(\hat{b}_{-1}^\dag)^i(\hat{b}_0^\dag)^j(\hat{b}_1^\dag)^k\ket{0}
\end{equation}
for $0\le i,j,k\le N$ and $i+j+k=N$.
The following process and thus the coefficients are identical if we instead define $\ket{i,j,k}$ using $\hat{c}_i$ operators and apply $\hat{K}_-$ instead of $\hat{J}_-$.

We calculate coefficients as follows.
Begin with 
\begin{equation} \label{eq:N N}
    \ket{j=N,m_j=N}=\ket{0,0,N},
\end{equation}
and apply $\hat{J}_-$.
This yields $\hat{J}_-\ket{0,0,N}=\sqrt{N}\ket{0,1,N-1}$ per Eq.~\eqref{eq:J_pm action}, meaning that 
\begin{equation} \label{eq:N N-1}
    \ket{N,N-1}=\ket{0,1,N-1}.
\end{equation}
We then apply $\hat{J}_-$ again to this state, which yields
\begin{equation} \label{eq:N N-2}
    \ket{N,N-2}=\frac{\ket{1,0,N-1}+\sqrt{2N-2}\ket{0,2,N-2}}{\sqrt{2N-1}}.
\end{equation}
We can continue applying $\hat{J}_-$ until we reach the state
\begin{equation}
    \ket{N,-N}=\ket{N,0,0},
\end{equation}
after which point subsequent applications of $\hat{J}_-$ give 0.
We can alternatively orthogonalize the state found in Eq.~\eqref{eq:N N-2}, yielding
\begin{equation}
    \ket{N-2,N-2}=\frac{\sqrt{2N-2}\ket{1,0,N-1}-\ket{0,2,N-2}}{\sqrt{2N-1}}.
\end{equation}
This means that $j=N$ is the largest $\hat{J}^2$ shell, and the next smaller shell is that of $j=N-2$.
The state $\ket{N-2,N-2}$ as defined above is in fact an eigenstate of $\hat{J}^2$ with eigenvalue $N-2$, and an eigenstate of $\hat{J}_z$ with eigenvalue $N-2$.

Repeating the above steps, we may apply $\hat{J}_-$ to $\ket{N-2,N-2}$ repeatedly to calculate $\ket{N-2,m_j}$ for $-(N-2)\le m_j\le N-2$.
If we repeat the above orthogonalization we would find a new $j=N-4$ shell where the state $\ket{N-4,N-4}$ would have the largest $m_j$ eigenvalue.
This pattern continues until $j=1$ or $0$ for odd and even $N$ respectively.
This leads to the earlier statement in Eq.~\eqref{eq:allowed j} about the range of values $j$ may take on.
We use this procedure to generate the analytical expression for the final evolution state for $\mathcal{H}_{\text{DFS}}^{(0)}(1)$ as defined in Eq.~\eqref{eq:psi final}. There we only need the first shell.

\section{Shortcut Derivation} \label{appx:ShortcutDerivation}

In this appendix we provide additional steps used for deriving the adiabatic shortcuts.

As we point out in the main text [see Eq.~\eqref{eq:shortcutcrit}], a shortcut Hamiltonian $\hat{H}_s$ to some state $\ket{\psi_{\vec{k}}}$ must satisfy
\begin{equation} \label{eq:shortcutCriterion}
    \bra{\psi_{\vec{n}}^{\perp}} \hat{H}_s \ket{\psi_{\vec{k}}} = i\hbar \ip{\psi_{\vec{n}}^{\perp}}{\rpd{\psi_{\vec{k}}}}
\end{equation}
\cite{Wu2017} for any $\vec{n}$ such that $n_3-n_1\neq C=k_3-k_1$, meaning all $\ket{\psi_{\vec{n}}^\perp}\in\mathcal{H}_{\text{CS}}^{(C)}(\mu)$.
We may apply this additional Hamiltonian in concert with the original $\hat{H}_{\mathrm{at}}$ in order to drive directly from the initial state $\ket{\psi(0)}$ into the target state.

Below will will provide an exact calculation of the right-hand side of Eq.~\eqref{eq:shortcutCriterion}.
We begin by calculating the time derivative, using the eigenstate form given in Eq.~\eqref{eq:LEigenstates}.
\begin{equation}
    \begin{aligned}
        \ket{\rpd{\psi_{\vec{k}}}} = & \rpd{\mathcal{N}}_{\vec{k}}(\hat{c}_1^\dag)^{k_1}(\hat{c}_2^\dag)^{k_2}(\hat{c}_3^\dag)^{k_3}\ket{0}
        \\ & +\mathcal{N}_{\vec{k}}[k_1(\hat{c}_1^\dag)^{k_1-1}\rpd{\hat{c}}_1^\dag](\hat{c}_2^\dag)^{k_2}(\hat{c}_3^\dag)^{k_3}\ket{0}
        \\& +\mathcal{N}_{\vec{k}}(\hat{c}_1^\dag)^{k_1}[k_2(\hat{c}_2^\dag)^{k_2-1}\rpd{\hat{c}}_2^\dag](\hat{c}_3^\dag)^{k_3}\ket{0}
        \\& +\mathcal{N}_{\vec{k}}(\hat{c}_1^\dag)^{k_1}(\hat{c}_2^\dag)^{k_2}[k_3(\hat{c}_3^\dag)^{k_3-1}\rpd{\hat{c}}_3^\dag]\ket{0}.
    \end{aligned}
\end{equation}
Note that the vacuum state $\ket{0}$ is static because the Hamiltonian and jump operator do not create particles.
The time dependence in $\hat{c}_i^\dag$ comes exclusively from $\mu(t)$.
By solving for these time derivatives and acting the resulting operators on $\ket{\psi_{\vec{k}}}$, this becomes
\begin{equation}
    \begin{aligned}
        \ket{\rpd{\psi_{\vec{k}}}}= & \left(\frac{\rpd{\mathcal{N}}_{\vec{k}}}{\mathcal{N}_{\vec{k}}}+\frac{2N\rpd{\mu}}{\mu}\right)\ket{\psi_{\vec{k}}}
        \\&-\frac{\rpd{\mu}}{\sqrt{2}\mu}\mathcal{N}_{\vec{k}}\Bigg(\frac{k_1\ket{\psi_{k_1-1,k_2+1,k_3}}}{\mathcal{N}_{k_1-1,k_2+1,k_3}} + \frac{k_2\ket{\psi_{k_1+1,k_2-1,k_3}}}{\mathcal{N}_{k_1+1,k_2-1,k_3}} 
        \\&+ \frac{k_2\ket{\psi_{k_1,k_2-1,k_3+1}}}{\mathcal{N}_{k_1,k_2-1,k_3+1}} + \frac{k_3\ket{\psi_{k_1,k_2+1,k_3-1}}}{\mathcal{N}_{k_1,k_2+1,k_3-1}}\Bigg).
    \end{aligned}
\end{equation}

Now, we take the inner product with some complementary state $\langle\psi_{\vec{n}}^\perp|$ where $n_3-n_1\neq C$.
By the properties of the complementary space, there are exactly four non-zero terms that correspond to
\begin{equation} \label{eq:allowed n}
    \begin{aligned}
        \vec{n}=(k_1-1,k_2+1,k_3),(k_1+1,k_2-1,k_3),
        \\(k_1,k_2-1,k_3+1),(k_1,k_2+1,k_3-1).
    \end{aligned}
\end{equation}
For any of these $\vec{n}$,
\begin{equation} \label{eq:timeDerivative}
    i\hbar\ip{\psi_{\vec{n}}^{\perp}}{\rpd{\psi_{\vec{k}}}}= \frac{-i\hbar\rpd{\mu}}{\sqrt{2}\mu}\mathcal{N}_{\vec{k}}\mathcal{N}^\perp_{\vec{n}}k_1!k_2!k_3!n_p
\end{equation}
where $p\in\{1,2,3\}$ is such that $n_p=k_p+1$.
Therefore, a shortcut Hamiltonian must couple $|\psi_{\vec{k}}\rangle$ to exactly the $|\psi_{\vec{n}}\rangle$ where $\vec{n}$ satisfies Eq.~\eqref{eq:allowed n}.

\subsection{Cavity-drive modification} \label{appx:CavityDrive}

We now discuss details for the implementation of the shortcut Hamiltonian for $\vec{k}\in\{(0,0,N),\ (N,0,0)\}$.
This shortcut is achieved by modifying the amplitude and phase of the cavity-mode drive $\eta$.
Keeping $\Delta_e,\ g,\ \Omega_1$ constant, we can write this as shifting $\chi$ to $\chi+\chi_s$.
This modifies the jump operator and the Hamiltonian of the system, but we may express the entire transformation in terms of an additional shortcut Hamiltonian.
This results in the shortcut Hamiltonian given by Eq.~\eqref{eq:H1CavityDrive},
\begin{equation}
    \hat{H}_s=\frac{\hbar\sqrt{\Gamma_c}}{2}\left(\left(\frac{\Delta_c'}{\kappa}+i\right)\chi_s^*\hat{L}+\left(\frac{\Delta_c'}{\kappa}-i\right)\chi_s\hat{L}^\dag\right).
\end{equation}
where we have substracted a constant energy offset.

We now must solve for the $\chi_s$ which allows this $\hat{H}_s$ to fulfill Eq.~\eqref{eq:shortcutCriterion}.
The relevant expression becomes
\begin{equation}
    \bra{\psi_{\vec{n}}^\perp}\hat{H}_s\ket{\psi_{\vec{k}}}=\frac{\hbar\sqrt{\Gamma_c}}{2}\left(\frac{\Delta_c'}{\kappa}-i\right)\chi_s\bra{\psi_{\vec{n}}^\perp}\hat{L}^\dag\ket{\psi_{\vec{k}}}
\end{equation}
because $\hat{L}\ket{\psi_{\vec{k}}}=0$ for any dark state.
Rewrite $\hat{L}^\dag$ as
\begin{equation}
    \begin{aligned}
        \hat{L}^\dag= & \frac{\sqrt{\Gamma_c}}{2\mu}\Big[\sqrt{2}\left(1+\mu^4\right)\left(\hat{c}_1^\dag \hat{d}_1-\hat{c}_3^\dag \hat{d}_3\right)
        \\& +\left(1-\mu^4\right)\left(\hat{c}_1^\dag \hat{d}_2-\hat{c}_2^\dag \hat{d}_1+\hat{c}_2^\dag \hat{d}_3-\hat{c}_3^\dag \hat{d}_2\right)\Big]+\sqrt{\Gamma_c}\chi.
    \end{aligned}
\end{equation}
We can apply this to $\ket{\psi_{\vec{k}}}$ and find that $\bra{\psi_{\vec{n}}^\perp}\hat{H}_s\ket{\psi_{\vec{k}}}$ is nonzero for exactly the $\vec{n}$ given in Eq.~\eqref{eq:allowed n}. We again define $p\in\{1,2,3\}$ such that $n_p=k_p+1$.
The result for these vectors reads
\begin{equation} \label{eq:H1 ip}
    \bra{\psi_{\vec{n}}^\perp}\hat{H}_s\ket{\psi_{\vec{k}}}= \pm\hbar\Gamma_c\chi_s\frac{\mu^4-1}{4\mu}\left(\frac{\Delta_c'}{\kappa}-i\right)\mathcal{N}_{\vec{k}}\mathcal{N}_{\vec{n}}^\perp k_1!k_2!k_3!n_p.
\end{equation}
The $+$ sign applies to $\vec{k}=(0,0,N)$ and the $-$ sign to $\vec{k}=(N,0,0)$.

We can now use Eq.~\eqref{eq:timeDerivative} to obtain 
\begin{equation}
    \chi_s=\pm\frac{i2\sqrt{2}\rpd{\mu}}{\Gamma_c\left(\Delta_c'/\kappa-i\right)\left(\mu^4-1\right)}
\end{equation}
where again, the $+$ sign will shortcut $\ket{\psi_{0,0,N}}$ and the $-$ sign will shortcut $\ket{\psi_{N,0,0}}$.

We consider that $\chi_s$ will diverge at $\mu=1$ unless $\rpd{\mu}$ were to approach 0 as $\mu\to 1$, slowing the evolution.
To avoid this issue, we instead set some maximum cutoff $\chi+\chi_s$, reached at $\mu=1-\epsilon$ for some small $\epsilon$.
We apply the exact shortcut to drive into $\mathcal{H}_{\text{DFS}}^{(\pm N)}(1-\epsilon)$, then for $1-\epsilon<\mu\le 1$ we apply the cutoff value of $\chi+\chi_s$ to the system rather than the ideal value.
The minimum laser amplitude required to prepare a state in this system without any shortcut is $\sqrt{2}N$, calculated by setting $\mu=1$ and $k_3-k_1=C=N$ in Eq.~\eqref{eq:chi as k}.
In simulation, we set the cutoff to be $|\chi+\chi_s|\le 5\sqrt{2}N$ where the scalar five is arbitrary and minimally affects the results.
This cutoff scheme causes some error: of order $10^{-4}$ on $\mathcal{F}(1,t)$ and $10^{-6}$ on $\mathcal{P}(t)$ for simulated values.
The error is small because the cutoff is relevant only for a short period of the evolution when the system already has high overlap with the target DFS.

\subsection{Additional lasers} \label{appx:CentralState}

In this appendix we provide additional details for the shortcut to the $\vec{k}=(0,N,0)$ state.
The shortcut Hamiltonian that we use is
\begin{equation}
     \hat{H}_s= i\alpha\left(\hat{b}_{-1}^\dag \hat{b}_1-\hat{b}_1^\dag \hat{b}_{-1}\right)
\end{equation}
for some real scalar $\alpha$, which would be implemented by additional lasers coupling between $\ket{1_g}$ and $\ket{\text{-}1_g}$.
We can write this as 
\begin{equation} \label{eq:H1_central_init}
    \begin{aligned}
        \hat{H}_s= & \frac{i\alpha}{4}\Bigg[\left(\mu^2-\mu^{-2}\right)\left(\hat{c}_1^\dag \hat{d}_1- 2 \hat{c}_2^\dag \hat{d}_2+\hat{c}_3^\dag \hat{d}_3+\hat{c}_1^\dag \hat{d}_3+\hat{c}_3^\dag \hat{d}_1\right)
        \\& + \sqrt{2}\left(\mu^2+\mu^{-2}\right)\left(\hat{c}_2^\dag \hat{d}_1-\hat{c}_1^\dag \hat{d}_2 + \hat{c}_2^\dag \hat{d}_3 - \hat{c}_3^\dag \hat{d}_2\right) \Bigg].
    \end{aligned}
\end{equation}

When we calculate $\bra{\psi_{\vec{n}}^\perp}\hat{H}_s\ket{\psi_{\vec{k}}}$, we need the first line of Eq.~\eqref{eq:H1_central_init} to disappear in order to solve Eq.~\eqref{eq:shortcutCriterion}.
The result takes the form
\begin{equation} \label{eq:H1transition}
    \bra{\psi_{\vec{n}}^{\perp}} \hat{H}_s \ket{\psi_{\vec{k}}} = \frac{-i\alpha}{4}\sqrt{2}\left(\mu^2+\mu^{-2}\right)\mathcal{N}_{\vec{k}}\mathcal{N}_{\vec{n}}^\perp k_1!k_2!k_3!N
\end{equation}
for $\vec{n}\in\{(1,N-1,0), (0,N-1,1)\}$, and is zero otherwise.
Using Eq.~\eqref{eq:shortcutCriterion} with Eqs.~\eqref{eq:H1transition} and~\eqref{eq:timeDerivative} we can find a consistent $\alpha$ which solves both cases at once, leading to the shortcut Hamiltonian
\begin{equation}
    \hat{H}_s=\frac{i2\hbar\mu\rpd{\mu}}{\mu^4+1}\left(b_{-1}^\dag b_1-b_1^\dag b_{-1}\right)
\end{equation}
which will perfectly shortcut $\ket{\psi_{0,N,0}}$.
The power of this shortcut is finite for all $\mu$, and is proportional to $\rpd{\mu}$, so we will need a stronger Hamiltonian if we ramp $\mu$ from 0 to 1 more quickly.


\begin{thebibliography}{76}%
\makeatletter
\providecommand \@ifxundefined [1]{%
 \@ifx{#1\undefined}
}%
\providecommand \@ifnum [1]{%
 \ifnum #1\expandafter \@firstoftwo
 \else \expandafter \@secondoftwo
 \fi
}%
\providecommand \@ifx [1]{%
 \ifx #1\expandafter \@firstoftwo
 \else \expandafter \@secondoftwo
 \fi
}%
\providecommand \natexlab [1]{#1}%
\providecommand \enquote  [1]{``#1''}%
\providecommand \bibnamefont  [1]{#1}%
\providecommand \bibfnamefont [1]{#1}%
\providecommand \citenamefont [1]{#1}%
\providecommand \href@noop [0]{\@secondoftwo}%
\providecommand \href [0]{\begingroup \@sanitize@url \@href}%
\providecommand \@href[1]{\@@startlink{#1}\@@href}%
\providecommand \@@href[1]{\endgroup#1\@@endlink}%
\providecommand \@sanitize@url [0]{\catcode `\\12\catcode `\$12\catcode `\&12\catcode `\#12\catcode `\^12\catcode `\_12\catcode `\%12\relax}%
\providecommand \@@startlink[1]{}%
\providecommand \@@endlink[0]{}%
\providecommand \url  [0]{\begingroup\@sanitize@url \@url }%
\providecommand \@url [1]{\endgroup\@href {#1}{\urlprefix }}%
\providecommand \urlprefix  [0]{URL }%
\providecommand \Eprint [0]{\href }%
\providecommand \doibase [0]{https://doi.org/}%
\providecommand \selectlanguage [0]{\@gobble}%
\providecommand \bibinfo  [0]{\@secondoftwo}%
\providecommand \bibfield  [0]{\@secondoftwo}%
\providecommand \translation [1]{[#1]}%
\providecommand \BibitemOpen [0]{}%
\providecommand \bibitemStop [0]{}%
\providecommand \bibitemNoStop [0]{.\EOS\space}%
\providecommand \EOS [0]{\spacefactor3000\relax}%
\providecommand \BibitemShut  [1]{\csname bibitem#1\endcsname}%
\let\auto@bib@innerbib\@empty
\bibitem [{\citenamefont {Preskill}(2018)}]{Preskill2018}%
  \BibitemOpen
  \bibfield  {author} {\bibinfo {author} {\bibfnamefont {J.}~\bibnamefont {Preskill}},\ }\bibfield  {title} {\bibinfo {title} {Quantum {C}omputing in the {NISQ} era and beyond},\ }\href {https://doi.org/10.22331/q-2018-08-06-79} {\bibfield  {journal} {\bibinfo  {journal} {{Quantum}}\ }\textbf {\bibinfo {volume} {2}},\ \bibinfo {pages} {79} (\bibinfo {year} {2018})}\BibitemShut {NoStop}%
\bibitem [{\citenamefont {Joos}\ \emph {et~al.}(2003)\citenamefont {Joos}, \citenamefont {Zeh}, \citenamefont {Kiefer}, \citenamefont {Giulini}, \citenamefont {Kupsch},\ and\ \citenamefont {Stamatescu}}]{Joos2003}%
  \BibitemOpen
  \bibfield  {author} {\bibinfo {author} {\bibfnamefont {E.}~\bibnamefont {Joos}}, \bibinfo {author} {\bibfnamefont {H.~D.}\ \bibnamefont {Zeh}}, \bibinfo {author} {\bibfnamefont {C.}~\bibnamefont {Kiefer}}, \bibinfo {author} {\bibfnamefont {D.}~\bibnamefont {Giulini}}, \bibinfo {author} {\bibfnamefont {J.}~\bibnamefont {Kupsch}},\ and\ \bibinfo {author} {\bibfnamefont {I.-O.}\ \bibnamefont {Stamatescu}},\ }\href {https://link.springer.com/book/10.1007/978-3-662-05328-7} {\emph {\bibinfo {title} {Decoherence and the Appearance of a Classical World in Quantum Theory}}},\ \bibinfo {edition} {2nd}\ ed.\ (\bibinfo  {publisher} {Springer-Verlag},\ \bibinfo {address} {Berlin, Germany},\ \bibinfo {year} {2003})\BibitemShut {NoStop}%
\bibitem [{\citenamefont {Gardiner}\ and\ \citenamefont {Zoller}(2010)}]{Gardiner2010}%
  \BibitemOpen
  \bibfield  {author} {\bibinfo {author} {\bibfnamefont {C.~W.}\ \bibnamefont {Gardiner}}\ and\ \bibinfo {author} {\bibfnamefont {P.}~\bibnamefont {Zoller}},\ }\href {https://link.springer.com/book/9783540223016} {\emph {\bibinfo {title} {Quantum Noise: a Handbook of Markovian and non-Markovian Quantum Stochastic Methods with Applications to Quantum Optics}}},\ \bibinfo {edition} {3rd}\ ed.\ (\bibinfo  {publisher} {Springer},\ \bibinfo {address} {Berlin, Germany},\ \bibinfo {year} {2010})\BibitemShut {NoStop}%
\bibitem [{\citenamefont {Orszag}(2010)}]{Orszag2010}%
  \BibitemOpen
  \bibfield  {author} {\bibinfo {author} {\bibfnamefont {M.}~\bibnamefont {Orszag}},\ }\href {https://link.springer.com/book/10.1007/978-3-031-54853-6} {\emph {\bibinfo {title} {Quantum Optics: Including Noise Reduction, Trapped Ions, Quantum Trajectories, and Decoherence}}},\ \bibinfo {edition} {2nd}\ ed.\ (\bibinfo  {publisher} {Springer-Verlag},\ \bibinfo {address} {Berlin, Germany},\ \bibinfo {year} {2010})\BibitemShut {NoStop}%
\bibitem [{\citenamefont {DiVincenzo}(1995)}]{DiVincenzo1995}%
  \BibitemOpen
  \bibfield  {author} {\bibinfo {author} {\bibfnamefont {D.~P.}\ \bibnamefont {DiVincenzo}},\ }\bibfield  {title} {\bibinfo {title} {Quantum computation},\ }\href {https://doi.org/10.1126/science.270.5234.255} {\bibfield  {journal} {\bibinfo  {journal} {Science}\ }\textbf {\bibinfo {volume} {270}},\ \bibinfo {pages} {255} (\bibinfo {year} {1995})}\BibitemShut {NoStop}%
\bibitem [{\citenamefont {Unruh}(1995)}]{Unrah1995}%
  \BibitemOpen
  \bibfield  {author} {\bibinfo {author} {\bibfnamefont {W.~G.}\ \bibnamefont {Unruh}},\ }\bibfield  {title} {\bibinfo {title} {Maintaining coherence in quantum computers},\ }\href {https://doi.org/10.1103/PhysRevA.51.992} {\bibfield  {journal} {\bibinfo  {journal} {Phys. Rev. A}\ }\textbf {\bibinfo {volume} {51}},\ \bibinfo {pages} {992} (\bibinfo {year} {1995})}\BibitemShut {NoStop}%
\bibitem [{\citenamefont {Aharonov}\ \emph {et~al.}(2007)\citenamefont {Aharonov}, \citenamefont {Dam}, \citenamefont {Kempe}, \citenamefont {Landau}, \citenamefont {Lloyd},\ and\ \citenamefont {Regev}}]{Aharonov2007}%
  \BibitemOpen
  \bibfield  {author} {\bibinfo {author} {\bibfnamefont {D.}~\bibnamefont {Aharonov}}, \bibinfo {author} {\bibfnamefont {W.~V.}\ \bibnamefont {Dam}}, \bibinfo {author} {\bibfnamefont {J.}~\bibnamefont {Kempe}}, \bibinfo {author} {\bibfnamefont {Z.}~\bibnamefont {Landau}}, \bibinfo {author} {\bibfnamefont {S.}~\bibnamefont {Lloyd}},\ and\ \bibinfo {author} {\bibfnamefont {O.}~\bibnamefont {Regev}},\ }\bibfield  {title} {\bibinfo {title} {Adiabatic quantum computation is equivalent to standard quantum computation},\ }\href {https://doi.org/10.1137/s0097539705447323} {\bibfield  {journal} {\bibinfo  {journal} {SIAM Journal on Computing}\ }\textbf {\bibinfo {volume} {37}},\ \bibinfo {pages} {166–194} (\bibinfo {year} {2007})}\BibitemShut {NoStop}%
\bibitem [{\citenamefont {Albash}\ and\ \citenamefont {Lidar}(2018)}]{Albash2018}%
  \BibitemOpen
  \bibfield  {author} {\bibinfo {author} {\bibfnamefont {T.}~\bibnamefont {Albash}}\ and\ \bibinfo {author} {\bibfnamefont {D.~A.}\ \bibnamefont {Lidar}},\ }\bibfield  {title} {\bibinfo {title} {Adiabatic quantum computation},\ }\href {https://doi.org/10.1103/RevModPhys.90.015002} {\bibfield  {journal} {\bibinfo  {journal} {Rev. Mod. Phys.}\ }\textbf {\bibinfo {volume} {90}},\ \bibinfo {pages} {015002} (\bibinfo {year} {2018})}\BibitemShut {NoStop}%
\bibitem [{\citenamefont {Venegas-Gomez}\ \emph {et~al.}(2020)\citenamefont {Venegas-Gomez}, \citenamefont {Schachenmayer}, \citenamefont {Buyskikh}, \citenamefont {Ketterle}, \citenamefont {Chiofalo},\ and\ \citenamefont {Daley}}]{Venegas2020}%
  \BibitemOpen
  \bibfield  {author} {\bibinfo {author} {\bibfnamefont {A.}~\bibnamefont {Venegas-Gomez}}, \bibinfo {author} {\bibfnamefont {J.}~\bibnamefont {Schachenmayer}}, \bibinfo {author} {\bibfnamefont {A.~S.}\ \bibnamefont {Buyskikh}}, \bibinfo {author} {\bibfnamefont {W.}~\bibnamefont {Ketterle}}, \bibinfo {author} {\bibfnamefont {M.~L.}\ \bibnamefont {Chiofalo}},\ and\ \bibinfo {author} {\bibfnamefont {A.~J.}\ \bibnamefont {Daley}},\ }\bibfield  {title} {\bibinfo {title} {Adiabatic preparation of entangled, magnetically ordered states with cold bosons in optical lattices},\ }\href {https://doi.org/10.1088/2058-9565/abb004} {\bibfield  {journal} {\bibinfo  {journal} {Quantum Science and Technology}\ }\textbf {\bibinfo {volume} {5}},\ \bibinfo {pages} {045013} (\bibinfo {year} {2020})}\BibitemShut {NoStop}%
\bibitem [{\citenamefont {Jin}\ \emph {et~al.}(2021)\citenamefont {Jin}, \citenamefont {Bao}, \citenamefont {Duan}, \citenamefont {Lu}, \citenamefont {Wang}, \citenamefont {Zhao}, \citenamefont {Shen},\ and\ \citenamefont {Xiao}}]{Jin2021}%
  \BibitemOpen
  \bibfield  {author} {\bibinfo {author} {\bibfnamefont {S.}~\bibnamefont {Jin}}, \bibinfo {author} {\bibfnamefont {H.}~\bibnamefont {Bao}}, \bibinfo {author} {\bibfnamefont {J.}~\bibnamefont {Duan}}, \bibinfo {author} {\bibfnamefont {X.}~\bibnamefont {Lu}}, \bibinfo {author} {\bibfnamefont {M.}~\bibnamefont {Wang}}, \bibinfo {author} {\bibfnamefont {K.-F.}\ \bibnamefont {Zhao}}, \bibinfo {author} {\bibfnamefont {H.}~\bibnamefont {Shen}},\ and\ \bibinfo {author} {\bibfnamefont {Y.}~\bibnamefont {Xiao}},\ }\bibfield  {title} {\bibinfo {title} {Adiabaticity in state preparation for spin squeezing of large atom ensembles},\ }\href {https://doi.org/10.1364/PRJ.413288} {\bibfield  {journal} {\bibinfo  {journal} {Photon. Res.}\ }\textbf {\bibinfo {volume} {9}},\ \bibinfo {pages} {2296} (\bibinfo {year} {2021})}\BibitemShut {NoStop}%
\bibitem [{\citenamefont {Herb}\ and\ \citenamefont {Degen}(2024)}]{Herb2024}%
  \BibitemOpen
  \bibfield  {author} {\bibinfo {author} {\bibfnamefont {K.}~\bibnamefont {Herb}}\ and\ \bibinfo {author} {\bibfnamefont {C.~L.}\ \bibnamefont {Degen}},\ }\bibfield  {title} {\bibinfo {title} {Quantum speed limit in quantum sensing},\ }\href {https://doi.org/10.1103/PhysRevLett.133.210802} {\bibfield  {journal} {\bibinfo  {journal} {Phys. Rev. Lett.}\ }\textbf {\bibinfo {volume} {133}},\ \bibinfo {pages} {210802} (\bibinfo {year} {2024})}\BibitemShut {NoStop}%
\bibitem [{\citenamefont {Reilly}\ \emph {et~al.}(2024)\citenamefont {Reilly}, \citenamefont {J\"ager}, \citenamefont {Wilson}, \citenamefont {Cooper}, \citenamefont {Eggert},\ and\ \citenamefont {Holland}}]{Reilly2023Speedingup}%
  \BibitemOpen
  \bibfield  {author} {\bibinfo {author} {\bibfnamefont {J.~T.}\ \bibnamefont {Reilly}}, \bibinfo {author} {\bibfnamefont {S.~B.}\ \bibnamefont {J\"ager}}, \bibinfo {author} {\bibfnamefont {J.~D.}\ \bibnamefont {Wilson}}, \bibinfo {author} {\bibfnamefont {J.}~\bibnamefont {Cooper}}, \bibinfo {author} {\bibfnamefont {S.}~\bibnamefont {Eggert}},\ and\ \bibinfo {author} {\bibfnamefont {M.~J.}\ \bibnamefont {Holland}},\ }\bibfield  {title} {\bibinfo {title} {Speeding up squeezing with a periodically driven dicke model},\ }\href {https://doi.org/10.1103/PhysRevResearch.6.033090} {\bibfield  {journal} {\bibinfo  {journal} {Phys. Rev. Res.}\ }\textbf {\bibinfo {volume} {6}},\ \bibinfo {pages} {033090} (\bibinfo {year} {2024})}\BibitemShut {NoStop}%
\bibitem [{\citenamefont {Wilson}\ \emph {et~al.}(2024)\citenamefont {Wilson}, \citenamefont {Reilly}, \citenamefont {Zhang}, \citenamefont {Luo}, \citenamefont {Chu}, \citenamefont {Thompson}, \citenamefont {Rey},\ and\ \citenamefont {Holland}}]{Wilson2024}%
  \BibitemOpen
  \bibfield  {author} {\bibinfo {author} {\bibfnamefont {J.~D.}\ \bibnamefont {Wilson}}, \bibinfo {author} {\bibfnamefont {J.~T.}\ \bibnamefont {Reilly}}, \bibinfo {author} {\bibfnamefont {H.}~\bibnamefont {Zhang}}, \bibinfo {author} {\bibfnamefont {C.}~\bibnamefont {Luo}}, \bibinfo {author} {\bibfnamefont {A.}~\bibnamefont {Chu}}, \bibinfo {author} {\bibfnamefont {J.~K.}\ \bibnamefont {Thompson}}, \bibinfo {author} {\bibfnamefont {A.~M.}\ \bibnamefont {Rey}},\ and\ \bibinfo {author} {\bibfnamefont {M.~J.}\ \bibnamefont {Holland}},\ }\bibfield  {title} {\bibinfo {title} {Entangled matter waves for quantum enhanced sensing},\ }\href {https://doi.org/10.1103/PhysRevA.110.L041301} {\bibfield  {journal} {\bibinfo  {journal} {Phys. Rev. A}\ }\textbf {\bibinfo {volume} {110}},\ \bibinfo {pages} {L041301} (\bibinfo {year} {2024})}\BibitemShut {NoStop}%
\bibitem [{\citenamefont {Bohnet}\ \emph {et~al.}(2016)\citenamefont {Bohnet}, \citenamefont {Sawyer}, \citenamefont {Britton}, \citenamefont {Wall}, \citenamefont {Rey}, \citenamefont {Foss-Feig},\ and\ \citenamefont {Bollinger}}]{Bohnet2016}%
  \BibitemOpen
  \bibfield  {author} {\bibinfo {author} {\bibfnamefont {J.~G.}\ \bibnamefont {Bohnet}}, \bibinfo {author} {\bibfnamefont {B.~C.}\ \bibnamefont {Sawyer}}, \bibinfo {author} {\bibfnamefont {J.~W.}\ \bibnamefont {Britton}}, \bibinfo {author} {\bibfnamefont {M.~L.}\ \bibnamefont {Wall}}, \bibinfo {author} {\bibfnamefont {A.~M.}\ \bibnamefont {Rey}}, \bibinfo {author} {\bibfnamefont {M.}~\bibnamefont {Foss-Feig}},\ and\ \bibinfo {author} {\bibfnamefont {J.~J.}\ \bibnamefont {Bollinger}},\ }\bibfield  {title} {\bibinfo {title} {Quantum spin dynamics and entanglement generation with hundreds of trapped ions},\ }\href {https://doi.org/10.1126/science.aad9958} {\bibfield  {journal} {\bibinfo  {journal} {Science}\ }\textbf {\bibinfo {volume} {352}},\ \bibinfo {pages} {1297} (\bibinfo {year} {2016})}\BibitemShut {NoStop}%
\bibitem [{\citenamefont {Pellizzari}\ \emph {et~al.}(1995)\citenamefont {Pellizzari}, \citenamefont {Gardiner}, \citenamefont {Cirac},\ and\ \citenamefont {Zoller}}]{Pellizzari1995}%
  \BibitemOpen
  \bibfield  {author} {\bibinfo {author} {\bibfnamefont {T.}~\bibnamefont {Pellizzari}}, \bibinfo {author} {\bibfnamefont {S.~A.}\ \bibnamefont {Gardiner}}, \bibinfo {author} {\bibfnamefont {J.~I.}\ \bibnamefont {Cirac}},\ and\ \bibinfo {author} {\bibfnamefont {P.}~\bibnamefont {Zoller}},\ }\bibfield  {title} {\bibinfo {title} {Decoherence, continuous observation, and quantum computing: A cavity qed model},\ }\href {https://doi.org/10.1103/PhysRevLett.75.3788} {\bibfield  {journal} {\bibinfo  {journal} {Phys. Rev. Lett.}\ }\textbf {\bibinfo {volume} {75}},\ \bibinfo {pages} {3788} (\bibinfo {year} {1995})}\BibitemShut {NoStop}%
\bibitem [{\citenamefont {Schulte-Herbr\"uggen}\ \emph {et~al.}(2012)\citenamefont {Schulte-Herbr\"uggen}, \citenamefont {Marx}, \citenamefont {Fahmy}, \citenamefont {Kauffman}, \citenamefont {Lomonaco}, \citenamefont {Khaneja},\ and\ \citenamefont {Glaser}}]{Schulte1976}%
  \BibitemOpen
  \bibfield  {author} {\bibinfo {author} {\bibfnamefont {T.}~\bibnamefont {Schulte-Herbr\"uggen}}, \bibinfo {author} {\bibfnamefont {R.}~\bibnamefont {Marx}}, \bibinfo {author} {\bibfnamefont {A.}~\bibnamefont {Fahmy}}, \bibinfo {author} {\bibfnamefont {L.}~\bibnamefont {Kauffman}}, \bibinfo {author} {\bibfnamefont {S.}~\bibnamefont {Lomonaco}}, \bibinfo {author} {\bibfnamefont {N.}~\bibnamefont {Khaneja}},\ and\ \bibinfo {author} {\bibfnamefont {S.~J.}\ \bibnamefont {Glaser}},\ }\bibfield  {title} {\bibinfo {title} {Control aspects of quantum computing using pure and mixed states},\ }\href {https://doi.org/10.1098/rsta.2011.0513} {\bibfield  {journal} {\bibinfo  {journal} {Philosophical Transactions of the Royal Society A: Mathematical, Physical and Engineering Sciences}\ }\textbf {\bibinfo {volume} {370}},\ \bibinfo {pages} {4651–4670} (\bibinfo {year} {2012})}\BibitemShut {NoStop}%
\bibitem [{\citenamefont {Zhou}\ \emph {et~al.}(2018)\citenamefont {Zhou}, \citenamefont {Zhang}, \citenamefont {Preskill},\ and\ \citenamefont {Jiang}}]{Zhou2018}%
  \BibitemOpen
  \bibfield  {author} {\bibinfo {author} {\bibfnamefont {S.}~\bibnamefont {Zhou}}, \bibinfo {author} {\bibfnamefont {M.}~\bibnamefont {Zhang}}, \bibinfo {author} {\bibfnamefont {J.}~\bibnamefont {Preskill}},\ and\ \bibinfo {author} {\bibfnamefont {L.}~\bibnamefont {Jiang}},\ }\bibfield  {title} {\bibinfo {title} {Achieving the heisenberg limit in quantum metrology using quantum error correction},\ }\bibfield  {journal} {\bibinfo  {journal} {Nature Communications}\ }\textbf {\bibinfo {volume} {9}},\ \href {https://doi.org/10.1038/s41467-017-02510-3} {10.1038/s41467-017-02510-3} (\bibinfo {year} {2018})\BibitemShut {NoStop}%
\bibitem [{\citenamefont {Shor}(1995)}]{Shor1995}%
  \BibitemOpen
  \bibfield  {author} {\bibinfo {author} {\bibfnamefont {P.~W.}\ \bibnamefont {Shor}},\ }\bibfield  {title} {\bibinfo {title} {Scheme for reducing decoherence in quantum computer memory},\ }\href {https://doi.org/10.1103/PhysRevA.52.R2493} {\bibfield  {journal} {\bibinfo  {journal} {Phys. Rev. A}\ }\textbf {\bibinfo {volume} {52}},\ \bibinfo {pages} {R2493} (\bibinfo {year} {1995})}\BibitemShut {NoStop}%
\bibitem [{\citenamefont {Reichardt}\ \emph {et~al.}(2024)\citenamefont {Reichardt}, \citenamefont {Aasen}, \citenamefont {Chao}, \citenamefont {Chernoguzov}, \citenamefont {van Dam}, \citenamefont {Gaebler}, \citenamefont {Gresh}, \citenamefont {Lucchetti}, \citenamefont {Mills}, \citenamefont {Moses}, \citenamefont {Neyenhuis}, \citenamefont {Paetznick}, \citenamefont {Paz}, \citenamefont {Siegfried}, \citenamefont {da~Silva}, \citenamefont {Svore}, \citenamefont {Wang},\ and\ \citenamefont {Zanner}}]{Reichardt2024}%
  \BibitemOpen
  \bibfield  {author} {\bibinfo {author} {\bibfnamefont {B.~W.}\ \bibnamefont {Reichardt}}, \bibinfo {author} {\bibfnamefont {D.}~\bibnamefont {Aasen}}, \bibinfo {author} {\bibfnamefont {R.}~\bibnamefont {Chao}}, \bibinfo {author} {\bibfnamefont {A.}~\bibnamefont {Chernoguzov}}, \bibinfo {author} {\bibfnamefont {W.}~\bibnamefont {van Dam}}, \bibinfo {author} {\bibfnamefont {J.~P.}\ \bibnamefont {Gaebler}}, \bibinfo {author} {\bibfnamefont {D.}~\bibnamefont {Gresh}}, \bibinfo {author} {\bibfnamefont {D.}~\bibnamefont {Lucchetti}}, \bibinfo {author} {\bibfnamefont {M.}~\bibnamefont {Mills}}, \bibinfo {author} {\bibfnamefont {S.~A.}\ \bibnamefont {Moses}}, \bibinfo {author} {\bibfnamefont {B.}~\bibnamefont {Neyenhuis}}, \bibinfo {author} {\bibfnamefont {A.}~\bibnamefont {Paetznick}}, \bibinfo {author} {\bibfnamefont {A.}~\bibnamefont {Paz}}, \bibinfo {author} {\bibfnamefont {P.~E.}\ \bibnamefont {Siegfried}}, \bibinfo {author} {\bibfnamefont {M.~P.}\ \bibnamefont {da~Silva}}, \bibinfo {author} {\bibfnamefont
  {K.~M.}\ \bibnamefont {Svore}}, \bibinfo {author} {\bibfnamefont {Z.}~\bibnamefont {Wang}},\ and\ \bibinfo {author} {\bibfnamefont {M.}~\bibnamefont {Zanner}},\ }\bibfield  {title} {\bibinfo {title} {Demonstration of quantum computation and error correction with a tesseract code},\ }\href {https://arxiv.org/abs/2409.04628} {\bibfield  {journal} {\bibinfo  {journal} {arXiv preprint arXiv:2409.04628}\ } (\bibinfo {year} {2024})},\ \Eprint {https://arxiv.org/abs/2409.04628} {2409.04628 [quant-ph]} \BibitemShut {NoStop}%
\bibitem [{\citenamefont {Preskill}(1998)}]{Preskill1998}%
  \BibitemOpen
  \bibfield  {author} {\bibinfo {author} {\bibfnamefont {J.}~\bibnamefont {Preskill}},\ }\bibinfo {title} {Fault-tolerant quantum computation},\ in\ \href {https://doi.org/10.1142/9789812385253_0008} {\emph {\bibinfo {booktitle} {Introduction to Quantum Computation and Information}}}\ (\bibinfo  {publisher} {World Scientific},\ \bibinfo {address} {Singapore},\ \bibinfo {year} {1998})\ pp.\ \bibinfo {pages} {213--269}\BibitemShut {NoStop}%
\bibitem [{\citenamefont {Arute}\ \emph {et~al.}(2019)\citenamefont {Arute} \emph {et~al.}}]{Arute2019}%
  \BibitemOpen
  \bibfield  {author} {\bibinfo {author} {\bibfnamefont {F.}~\bibnamefont {Arute}} \emph {et~al.},\ }\bibfield  {title} {\bibinfo {title} {Quantum supremacy using a programmable superconducting processor},\ }\href {https://doi.org/10.1038/s41586-019-1666-5} {\bibfield  {journal} {\bibinfo  {journal} {Nature}\ }\textbf {\bibinfo {volume} {574}},\ \bibinfo {pages} {505} (\bibinfo {year} {2019})}\BibitemShut {NoStop}%
\bibitem [{\citenamefont {Wang}\ and\ \citenamefont {Liu}(2024)}]{Wang2024}%
  \BibitemOpen
  \bibfield  {author} {\bibinfo {author} {\bibfnamefont {Y.}~\bibnamefont {Wang}}\ and\ \bibinfo {author} {\bibfnamefont {J.}~\bibnamefont {Liu}},\ }\bibfield  {title} {\bibinfo {title} {A comprehensive review of quantum machine learning: from nisq to fault tolerance},\ }\href {https://doi.org/10.1088/1361-6633/ad7f69} {\bibfield  {journal} {\bibinfo  {journal} {Reports on Progress in Physics}\ }\textbf {\bibinfo {volume} {87}},\ \bibinfo {pages} {116402} (\bibinfo {year} {2024})}\BibitemShut {NoStop}%
\bibitem [{\citenamefont {Cong}\ \emph {et~al.}(2022)\citenamefont {Cong}, \citenamefont {Levine}, \citenamefont {Keesling}, \citenamefont {Bluvstein}, \citenamefont {Wang},\ and\ \citenamefont {Lukin}}]{Cong2022}%
  \BibitemOpen
  \bibfield  {author} {\bibinfo {author} {\bibfnamefont {I.}~\bibnamefont {Cong}}, \bibinfo {author} {\bibfnamefont {H.}~\bibnamefont {Levine}}, \bibinfo {author} {\bibfnamefont {A.}~\bibnamefont {Keesling}}, \bibinfo {author} {\bibfnamefont {D.}~\bibnamefont {Bluvstein}}, \bibinfo {author} {\bibfnamefont {S.-T.}\ \bibnamefont {Wang}},\ and\ \bibinfo {author} {\bibfnamefont {M.~D.}\ \bibnamefont {Lukin}},\ }\bibfield  {title} {\bibinfo {title} {Hardware-efficient, fault-tolerant quantum computation with rydberg atoms},\ }\href {https://doi.org/10.1103/PhysRevX.12.021049} {\bibfield  {journal} {\bibinfo  {journal} {Phys. Rev. X}\ }\textbf {\bibinfo {volume} {12}},\ \bibinfo {pages} {021049} (\bibinfo {year} {2022})}\BibitemShut {NoStop}%
\bibitem [{\citenamefont {Wu}\ \emph {et~al.}(2017)\citenamefont {Wu}, \citenamefont {Huang}, \citenamefont {Li},\ and\ \citenamefont {Yi}}]{Wu2017}%
  \BibitemOpen
  \bibfield  {author} {\bibinfo {author} {\bibfnamefont {S.~L.}\ \bibnamefont {Wu}}, \bibinfo {author} {\bibfnamefont {X.~L.}\ \bibnamefont {Huang}}, \bibinfo {author} {\bibfnamefont {H.}~\bibnamefont {Li}},\ and\ \bibinfo {author} {\bibfnamefont {X.~X.}\ \bibnamefont {Yi}},\ }\bibfield  {title} {\bibinfo {title} {Adiabatic evolution of decoherence-free subspaces and its shortcuts},\ }\href {https://doi.org/10.1103/PhysRevA.96.042104} {\bibfield  {journal} {\bibinfo  {journal} {Phys. Rev. A}\ }\textbf {\bibinfo {volume} {96}},\ \bibinfo {pages} {042104} (\bibinfo {year} {2017})}\BibitemShut {NoStop}%
\bibitem [{\citenamefont {Karasik}\ \emph {et~al.}(2008)\citenamefont {Karasik}, \citenamefont {Marzlin}, \citenamefont {Sanders},\ and\ \citenamefont {Whaley}}]{Karasik2008}%
  \BibitemOpen
  \bibfield  {author} {\bibinfo {author} {\bibfnamefont {R.~I.}\ \bibnamefont {Karasik}}, \bibinfo {author} {\bibfnamefont {K.-P.}\ \bibnamefont {Marzlin}}, \bibinfo {author} {\bibfnamefont {B.~C.}\ \bibnamefont {Sanders}},\ and\ \bibinfo {author} {\bibfnamefont {K.~B.}\ \bibnamefont {Whaley}},\ }\bibfield  {title} {\bibinfo {title} {Criteria for dynamically stable decoherence-free subspaces and incoherently generated coherences},\ }\href {https://doi.org/10.1103/PhysRevA.77.052301} {\bibfield  {journal} {\bibinfo  {journal} {Phys. Rev. A}\ }\textbf {\bibinfo {volume} {77}},\ \bibinfo {pages} {052301} (\bibinfo {year} {2008})}\BibitemShut {NoStop}%
\bibitem [{\citenamefont {Lidar}(2014)}]{Lidar2014}%
  \BibitemOpen
  \bibfield  {author} {\bibinfo {author} {\bibfnamefont {D.~A.}\ \bibnamefont {Lidar}},\ }\bibinfo {title} {Review of decoherence-free subspaces, noiseless subsystems, and dynamical decoupling},\ in\ \href {https://doi.org/https://doi.org/10.1002/9781118742631.ch11} {\emph {\bibinfo {booktitle} {Quantum Information and Computation for Chemistry}}}\ (\bibinfo  {publisher} {John Wiley \& Sons, Ltd},\ \bibinfo {address} {Hoboken, NJ},\ \bibinfo {year} {2014})\ pp.\ \bibinfo {pages} {295--354}\BibitemShut {NoStop}%
\bibitem [{\citenamefont {Reilly}\ \emph {et~al.}(2022)\citenamefont {Reilly}, \citenamefont {J\"ager}, \citenamefont {Cooper},\ and\ \citenamefont {Holland}}]{Reilly2022}%
  \BibitemOpen
  \bibfield  {author} {\bibinfo {author} {\bibfnamefont {J.~T.}\ \bibnamefont {Reilly}}, \bibinfo {author} {\bibfnamefont {S.~B.}\ \bibnamefont {J\"ager}}, \bibinfo {author} {\bibfnamefont {J.}~\bibnamefont {Cooper}},\ and\ \bibinfo {author} {\bibfnamefont {M.~J.}\ \bibnamefont {Holland}},\ }\bibfield  {title} {\bibinfo {title} {Adiabatic control of decoherence-free subspaces in an open collective system},\ }\href {https://doi.org/10.1103/PhysRevA.106.023703} {\bibfield  {journal} {\bibinfo  {journal} {Phys. Rev. A}\ }\textbf {\bibinfo {volume} {106}},\ \bibinfo {pages} {023703} (\bibinfo {year} {2022})}\BibitemShut {NoStop}%
\bibitem [{\citenamefont {Rubies-Bigorda}\ \emph {et~al.}(2024)\citenamefont {Rubies-Bigorda}, \citenamefont {Masson}, \citenamefont {Yelin},\ and\ \citenamefont {Asenjo-Garcia}}]{RubiesBigorda2024}%
  \BibitemOpen
  \bibfield  {author} {\bibinfo {author} {\bibfnamefont {O.}~\bibnamefont {Rubies-Bigorda}}, \bibinfo {author} {\bibfnamefont {S.~J.}\ \bibnamefont {Masson}}, \bibinfo {author} {\bibfnamefont {S.~F.}\ \bibnamefont {Yelin}},\ and\ \bibinfo {author} {\bibfnamefont {A.}~\bibnamefont {Asenjo-Garcia}},\ }\bibfield  {title} {\bibinfo {title} {Deterministic generation of photonic entangled states using decoherence-free subspaces},\ }\href {https://arxiv.org/abs/2410.03325} {\bibfield  {journal} {\bibinfo  {journal} {arXiv preprint arXiv:2410.03325}\ } (\bibinfo {year} {2024})},\ \Eprint {https://arxiv.org/abs/2410.03325} {2410.03325 [quant-ph]} \BibitemShut {NoStop}%
\bibitem [{\citenamefont {Diniz}\ \emph {et~al.}(2025)\citenamefont {Diniz}, \citenamefont {Villas-Boas},\ and\ \citenamefont {Santos}}]{Diniz2024}%
  \BibitemOpen
  \bibfield  {author} {\bibinfo {author} {\bibfnamefont {C.~M.}\ \bibnamefont {Diniz}}, \bibinfo {author} {\bibfnamefont {C.~J.}\ \bibnamefont {Villas-Boas}},\ and\ \bibinfo {author} {\bibfnamefont {A.~C.}\ \bibnamefont {Santos}},\ }\bibfield  {title} {\bibinfo {title} {Scalable quantum eraser with superconducting integrated circuits},\ }\href {https://doi.org/10.1088/2058-9565/adbded} {\bibfield  {journal} {\bibinfo  {journal} {Quantum Science and Technology}\ }\textbf {\bibinfo {volume} {10}},\ \bibinfo {pages} {025039} (\bibinfo {year} {2025})}\BibitemShut {NoStop}%
\bibitem [{\citenamefont {Ritsch}\ \emph {et~al.}(2013)\citenamefont {Ritsch}, \citenamefont {Domokos}, \citenamefont {Brennecke},\ and\ \citenamefont {Esslinger}}]{Ritsch2013}%
  \BibitemOpen
  \bibfield  {author} {\bibinfo {author} {\bibfnamefont {H.}~\bibnamefont {Ritsch}}, \bibinfo {author} {\bibfnamefont {P.}~\bibnamefont {Domokos}}, \bibinfo {author} {\bibfnamefont {F.}~\bibnamefont {Brennecke}},\ and\ \bibinfo {author} {\bibfnamefont {T.}~\bibnamefont {Esslinger}},\ }\bibfield  {title} {\bibinfo {title} {Cold atoms in cavity-generated dynamical optical potentials},\ }\href {https://doi.org/10.1103/RevModPhys.85.553} {\bibfield  {journal} {\bibinfo  {journal} {Rev. Mod. Phys.}\ }\textbf {\bibinfo {volume} {85}},\ \bibinfo {pages} {553} (\bibinfo {year} {2013})}\BibitemShut {NoStop}%
\bibitem [{\citenamefont {Chitra}\ and\ \citenamefont {Zilberberg}(2015)}]{Chitra2015}%
  \BibitemOpen
  \bibfield  {author} {\bibinfo {author} {\bibfnamefont {R.}~\bibnamefont {Chitra}}\ and\ \bibinfo {author} {\bibfnamefont {O.}~\bibnamefont {Zilberberg}},\ }\bibfield  {title} {\bibinfo {title} {Dynamical many-body phases of the parametrically driven, dissipative dicke model},\ }\href {https://doi.org/10.1103/PhysRevA.92.023815} {\bibfield  {journal} {\bibinfo  {journal} {Phys. Rev. A}\ }\textbf {\bibinfo {volume} {92}},\ \bibinfo {pages} {023815} (\bibinfo {year} {2015})}\BibitemShut {NoStop}%
\bibitem [{\citenamefont {Damanet}\ \emph {et~al.}(2019)\citenamefont {Damanet}, \citenamefont {Daley},\ and\ \citenamefont {Keeling}}]{Damanet2019}%
  \BibitemOpen
  \bibfield  {author} {\bibinfo {author} {\bibfnamefont {F.~m.~c.}\ \bibnamefont {Damanet}}, \bibinfo {author} {\bibfnamefont {A.~J.}\ \bibnamefont {Daley}},\ and\ \bibinfo {author} {\bibfnamefont {J.}~\bibnamefont {Keeling}},\ }\bibfield  {title} {\bibinfo {title} {Atom-only descriptions of the driven-dissipative dicke model},\ }\href {https://doi.org/10.1103/PhysRevA.99.033845} {\bibfield  {journal} {\bibinfo  {journal} {Phys. Rev. A}\ }\textbf {\bibinfo {volume} {99}},\ \bibinfo {pages} {033845} (\bibinfo {year} {2019})}\BibitemShut {NoStop}%
\bibitem [{\citenamefont {Lidar}\ \emph {et~al.}(1998)\citenamefont {Lidar}, \citenamefont {Chuang},\ and\ \citenamefont {Whaley}}]{Lidar1998}%
  \BibitemOpen
  \bibfield  {author} {\bibinfo {author} {\bibfnamefont {D.~A.}\ \bibnamefont {Lidar}}, \bibinfo {author} {\bibfnamefont {I.~L.}\ \bibnamefont {Chuang}},\ and\ \bibinfo {author} {\bibfnamefont {K.~B.}\ \bibnamefont {Whaley}},\ }\bibfield  {title} {\bibinfo {title} {Decoherence-free subspaces for quantum computation},\ }\href {https://doi.org/10.1103/PhysRevLett.81.2594} {\bibfield  {journal} {\bibinfo  {journal} {Phys. Rev. Lett.}\ }\textbf {\bibinfo {volume} {81}},\ \bibinfo {pages} {2594} (\bibinfo {year} {1998})}\BibitemShut {NoStop}%
\bibitem [{\citenamefont {Fortunato}\ \emph {et~al.}(2002)\citenamefont {Fortunato}, \citenamefont {Viola}, \citenamefont {Hodges}, \citenamefont {Teklemariam},\ and\ \citenamefont {Cory}}]{Fortunato2002}%
  \BibitemOpen
  \bibfield  {author} {\bibinfo {author} {\bibfnamefont {E.~M.}\ \bibnamefont {Fortunato}}, \bibinfo {author} {\bibfnamefont {L.}~\bibnamefont {Viola}}, \bibinfo {author} {\bibfnamefont {J.}~\bibnamefont {Hodges}}, \bibinfo {author} {\bibfnamefont {G.}~\bibnamefont {Teklemariam}},\ and\ \bibinfo {author} {\bibfnamefont {D.~G.}\ \bibnamefont {Cory}},\ }\bibfield  {title} {\bibinfo {title} {Implementation of universal control on a decoherence-free qubit},\ }\href {https://doi.org/10.1088/1367-2630/4/1/305} {\bibfield  {journal} {\bibinfo  {journal} {New Journal of Physics}\ }\textbf {\bibinfo {volume} {4}},\ \bibinfo {pages} {5} (\bibinfo {year} {2002})}\BibitemShut {NoStop}%
\bibitem [{\citenamefont {Hamann}\ \emph {et~al.}(2022)\citenamefont {Hamann}, \citenamefont {Sekatski},\ and\ \citenamefont {D{\"u}r}}]{Hamann2021}%
  \BibitemOpen
  \bibfield  {author} {\bibinfo {author} {\bibfnamefont {A.}~\bibnamefont {Hamann}}, \bibinfo {author} {\bibfnamefont {P.}~\bibnamefont {Sekatski}},\ and\ \bibinfo {author} {\bibfnamefont {W.}~\bibnamefont {D{\"u}r}},\ }\bibfield  {title} {\bibinfo {title} {Approximate decoherence free subspaces for distributed sensing},\ }\href {https://doi.org/10.1088/2058-9565/ac44de} {\bibfield  {journal} {\bibinfo  {journal} {Quantum Science and Technology}\ }\textbf {\bibinfo {volume} {7}},\ \bibinfo {pages} {025003} (\bibinfo {year} {2022})}\BibitemShut {NoStop}%
\bibitem [{\citenamefont {Horodecki}\ \emph {et~al.}(2009)\citenamefont {Horodecki}, \citenamefont {Horodecki}, \citenamefont {Horodecki},\ and\ \citenamefont {Horodecki}}]{Horodecki2009}%
  \BibitemOpen
  \bibfield  {author} {\bibinfo {author} {\bibfnamefont {R.}~\bibnamefont {Horodecki}}, \bibinfo {author} {\bibfnamefont {P.}~\bibnamefont {Horodecki}}, \bibinfo {author} {\bibfnamefont {M.}~\bibnamefont {Horodecki}},\ and\ \bibinfo {author} {\bibfnamefont {K.}~\bibnamefont {Horodecki}},\ }\bibfield  {title} {\bibinfo {title} {Quantum entanglement},\ }\href {https://doi.org/10.1103/RevModPhys.81.865} {\bibfield  {journal} {\bibinfo  {journal} {Rev. Mod. Phys.}\ }\textbf {\bibinfo {volume} {81}},\ \bibinfo {pages} {865} (\bibinfo {year} {2009})}\BibitemShut {NoStop}%
\bibitem [{\citenamefont {Dalla~Torre}\ \emph {et~al.}(2013)\citenamefont {Dalla~Torre}, \citenamefont {Otterbach}, \citenamefont {Demler}, \citenamefont {Vuletic},\ and\ \citenamefont {Lukin}}]{DallaTorre2013}%
  \BibitemOpen
  \bibfield  {author} {\bibinfo {author} {\bibfnamefont {E.~G.}\ \bibnamefont {Dalla~Torre}}, \bibinfo {author} {\bibfnamefont {J.}~\bibnamefont {Otterbach}}, \bibinfo {author} {\bibfnamefont {E.}~\bibnamefont {Demler}}, \bibinfo {author} {\bibfnamefont {V.}~\bibnamefont {Vuletic}},\ and\ \bibinfo {author} {\bibfnamefont {M.~D.}\ \bibnamefont {Lukin}},\ }\bibfield  {title} {\bibinfo {title} {Dissipative preparation of spin squeezed atomic ensembles in a steady state},\ }\href {https://doi.org/10.1103/PhysRevLett.110.120402} {\bibfield  {journal} {\bibinfo  {journal} {Phys. Rev. Lett.}\ }\textbf {\bibinfo {volume} {110}},\ \bibinfo {pages} {120402} (\bibinfo {year} {2013})}\BibitemShut {NoStop}%
\bibitem [{\citenamefont {Masson}\ \emph {et~al.}(2017)\citenamefont {Masson}, \citenamefont {Barrett},\ and\ \citenamefont {Parkins}}]{Masson2017}%
  \BibitemOpen
  \bibfield  {author} {\bibinfo {author} {\bibfnamefont {S.~J.}\ \bibnamefont {Masson}}, \bibinfo {author} {\bibfnamefont {M.~D.}\ \bibnamefont {Barrett}},\ and\ \bibinfo {author} {\bibfnamefont {S.}~\bibnamefont {Parkins}},\ }\bibfield  {title} {\bibinfo {title} {Cavity qed engineering of spin dynamics and squeezing in a spinor gas},\ }\href {https://doi.org/10.1103/PhysRevLett.119.213601} {\bibfield  {journal} {\bibinfo  {journal} {Phys. Rev. Lett.}\ }\textbf {\bibinfo {volume} {119}},\ \bibinfo {pages} {213601} (\bibinfo {year} {2017})}\BibitemShut {NoStop}%
\bibitem [{\citenamefont {Masson}\ and\ \citenamefont {Parkins}(2019)}]{Masson2019}%
  \BibitemOpen
  \bibfield  {author} {\bibinfo {author} {\bibfnamefont {S.~J.}\ \bibnamefont {Masson}}\ and\ \bibinfo {author} {\bibfnamefont {S.}~\bibnamefont {Parkins}},\ }\bibfield  {title} {\bibinfo {title} {Extreme spin squeezing in the steady state of a generalized dicke model},\ }\href {https://doi.org/10.1103/PhysRevA.99.023822} {\bibfield  {journal} {\bibinfo  {journal} {Phys. Rev. A}\ }\textbf {\bibinfo {volume} {99}},\ \bibinfo {pages} {023822} (\bibinfo {year} {2019})}\BibitemShut {NoStop}%
\bibitem [{\citenamefont {Grimsmo}\ and\ \citenamefont {Parkins}(2013)}]{Grimsmo2013}%
  \BibitemOpen
  \bibfield  {author} {\bibinfo {author} {\bibfnamefont {A.~L.}\ \bibnamefont {Grimsmo}}\ and\ \bibinfo {author} {\bibfnamefont {S.}~\bibnamefont {Parkins}},\ }\bibfield  {title} {\bibinfo {title} {Cavity-qed simulation of qubit-oscillator dynamics in the ultrastrong-coupling regime},\ }\href {https://doi.org/10.1103/PhysRevA.87.033814} {\bibfield  {journal} {\bibinfo  {journal} {Phys. Rev. A}\ }\textbf {\bibinfo {volume} {87}},\ \bibinfo {pages} {033814} (\bibinfo {year} {2013})}\BibitemShut {NoStop}%
\bibitem [{\citenamefont {Chen}\ \emph {et~al.}(2012)\citenamefont {Chen}, \citenamefont {Shi}, \citenamefont {Zheng},\ and\ \citenamefont {Gu}}]{Chen2012}%
  \BibitemOpen
  \bibfield  {author} {\bibinfo {author} {\bibfnamefont {L.-B.}\ \bibnamefont {Chen}}, \bibinfo {author} {\bibfnamefont {P.}~\bibnamefont {Shi}}, \bibinfo {author} {\bibfnamefont {C.-H.}\ \bibnamefont {Zheng}},\ and\ \bibinfo {author} {\bibfnamefont {Y.-J.}\ \bibnamefont {Gu}},\ }\bibfield  {title} {\bibinfo {title} {Generation of three-dimensional entangled state between a single atom and a bose-einstein condensate via adiabatic passage},\ }\href {https://doi.org/10.1364/oe.20.014547} {\bibfield  {journal} {\bibinfo  {journal} {Optics Express}\ }\textbf {\bibinfo {volume} {20}},\ \bibinfo {pages} {14547} (\bibinfo {year} {2012})}\BibitemShut {NoStop}%
\bibitem [{\citenamefont {Dimer}\ \emph {et~al.}(2007)\citenamefont {Dimer}, \citenamefont {Estienne}, \citenamefont {Parkins},\ and\ \citenamefont {Carmichael}}]{Dimer2007}%
  \BibitemOpen
  \bibfield  {author} {\bibinfo {author} {\bibfnamefont {F.}~\bibnamefont {Dimer}}, \bibinfo {author} {\bibfnamefont {B.}~\bibnamefont {Estienne}}, \bibinfo {author} {\bibfnamefont {A.~S.}\ \bibnamefont {Parkins}},\ and\ \bibinfo {author} {\bibfnamefont {H.~J.}\ \bibnamefont {Carmichael}},\ }\bibfield  {title} {\bibinfo {title} {Proposed realization of the dicke-model quantum phase transition in an optical cavity qed system},\ }\href {https://doi.org/10.1103/PhysRevA.75.013804} {\bibfield  {journal} {\bibinfo  {journal} {Phys. Rev. A}\ }\textbf {\bibinfo {volume} {75}},\ \bibinfo {pages} {013804} (\bibinfo {year} {2007})}\BibitemShut {NoStop}%
\bibitem [{\citenamefont {Zhang}\ \emph {et~al.}(2018)\citenamefont {Zhang}, \citenamefont {Lee}, \citenamefont {Kumar}, \citenamefont {Arnold}, \citenamefont {Masson}, \citenamefont {Grimsmo}, \citenamefont {Parkins},\ and\ \citenamefont {Barrett}}]{Zhang2018}%
  \BibitemOpen
  \bibfield  {author} {\bibinfo {author} {\bibfnamefont {Z.}~\bibnamefont {Zhang}}, \bibinfo {author} {\bibfnamefont {C.~H.}\ \bibnamefont {Lee}}, \bibinfo {author} {\bibfnamefont {R.}~\bibnamefont {Kumar}}, \bibinfo {author} {\bibfnamefont {K.~J.}\ \bibnamefont {Arnold}}, \bibinfo {author} {\bibfnamefont {S.~J.}\ \bibnamefont {Masson}}, \bibinfo {author} {\bibfnamefont {A.~L.}\ \bibnamefont {Grimsmo}}, \bibinfo {author} {\bibfnamefont {A.~S.}\ \bibnamefont {Parkins}},\ and\ \bibinfo {author} {\bibfnamefont {M.~D.}\ \bibnamefont {Barrett}},\ }\bibfield  {title} {\bibinfo {title} {Dicke-model simulation via cavity-assisted raman transitions},\ }\href {https://doi.org/10.1103/PhysRevA.97.043858} {\bibfield  {journal} {\bibinfo  {journal} {Phys. Rev. A}\ }\textbf {\bibinfo {volume} {97}},\ \bibinfo {pages} {043858} (\bibinfo {year} {2018})}\BibitemShut {NoStop}%
\bibitem [{\citenamefont {Zhiqiang}\ \emph {et~al.}(2017)\citenamefont {Zhiqiang}, \citenamefont {Lee}, \citenamefont {Kumar}, \citenamefont {Arnold}, \citenamefont {Masson}, \citenamefont {Parkins},\ and\ \citenamefont {Barrett}}]{Zhiqiang2017}%
  \BibitemOpen
  \bibfield  {author} {\bibinfo {author} {\bibfnamefont {Z.}~\bibnamefont {Zhiqiang}}, \bibinfo {author} {\bibfnamefont {C.~H.}\ \bibnamefont {Lee}}, \bibinfo {author} {\bibfnamefont {R.}~\bibnamefont {Kumar}}, \bibinfo {author} {\bibfnamefont {K.~J.}\ \bibnamefont {Arnold}}, \bibinfo {author} {\bibfnamefont {S.~J.}\ \bibnamefont {Masson}}, \bibinfo {author} {\bibfnamefont {A.~S.}\ \bibnamefont {Parkins}},\ and\ \bibinfo {author} {\bibfnamefont {M.~D.}\ \bibnamefont {Barrett}},\ }\bibfield  {title} {\bibinfo {title} {Nonequilibrium phase transition in a spin-1 dicke model},\ }\href {https://doi.org/10.1364/OPTICA.4.000424} {\bibfield  {journal} {\bibinfo  {journal} {Optica}\ }\textbf {\bibinfo {volume} {4}},\ \bibinfo {pages} {424} (\bibinfo {year} {2017})}\BibitemShut {NoStop}%
\bibitem [{\citenamefont {Baden}\ \emph {et~al.}(2014)\citenamefont {Baden}, \citenamefont {Arnold}, \citenamefont {Grimsmo}, \citenamefont {Parkins},\ and\ \citenamefont {Barrett}}]{Baden2014}%
  \BibitemOpen
  \bibfield  {author} {\bibinfo {author} {\bibfnamefont {M.~P.}\ \bibnamefont {Baden}}, \bibinfo {author} {\bibfnamefont {K.~J.}\ \bibnamefont {Arnold}}, \bibinfo {author} {\bibfnamefont {A.~L.}\ \bibnamefont {Grimsmo}}, \bibinfo {author} {\bibfnamefont {S.}~\bibnamefont {Parkins}},\ and\ \bibinfo {author} {\bibfnamefont {M.~D.}\ \bibnamefont {Barrett}},\ }\bibfield  {title} {\bibinfo {title} {Realization of the dicke model using cavity-assisted raman transitions},\ }\href {https://doi.org/10.1103/PhysRevLett.113.020408} {\bibfield  {journal} {\bibinfo  {journal} {Phys. Rev. Lett.}\ }\textbf {\bibinfo {volume} {113}},\ \bibinfo {pages} {020408} (\bibinfo {year} {2014})}\BibitemShut {NoStop}%
\bibitem [{\citenamefont {Gu\'ery-Odelin}\ \emph {et~al.}(2019)\citenamefont {Gu\'ery-Odelin}, \citenamefont {Ruschhaupt}, \citenamefont {Kiely}, \citenamefont {Torrontegui}, \citenamefont {Mart\'{\i}nez-Garaot},\ and\ \citenamefont {Muga}}]{Guery2019}%
  \BibitemOpen
  \bibfield  {author} {\bibinfo {author} {\bibfnamefont {D.}~\bibnamefont {Gu\'ery-Odelin}}, \bibinfo {author} {\bibfnamefont {A.}~\bibnamefont {Ruschhaupt}}, \bibinfo {author} {\bibfnamefont {A.}~\bibnamefont {Kiely}}, \bibinfo {author} {\bibfnamefont {E.}~\bibnamefont {Torrontegui}}, \bibinfo {author} {\bibfnamefont {S.}~\bibnamefont {Mart\'{\i}nez-Garaot}},\ and\ \bibinfo {author} {\bibfnamefont {J.~G.}\ \bibnamefont {Muga}},\ }\bibfield  {title} {\bibinfo {title} {Shortcuts to adiabaticity: Concepts, methods, and applications},\ }\href {https://doi.org/10.1103/RevModPhys.91.045001} {\bibfield  {journal} {\bibinfo  {journal} {Rev. Mod. Phys.}\ }\textbf {\bibinfo {volume} {91}},\ \bibinfo {pages} {045001} (\bibinfo {year} {2019})}\BibitemShut {NoStop}%
\bibitem [{\citenamefont {Carollo}\ \emph {et~al.}(2006)\citenamefont {Carollo}, \citenamefont {Massimo~Palma}, \citenamefont {\L{}ozinski}, \citenamefont {Santos},\ and\ \citenamefont {Vedral}}]{Carollo2006}%
  \BibitemOpen
  \bibfield  {author} {\bibinfo {author} {\bibfnamefont {A.}~\bibnamefont {Carollo}}, \bibinfo {author} {\bibfnamefont {G.}~\bibnamefont {Massimo~Palma}}, \bibinfo {author} {\bibfnamefont {A.}~\bibnamefont {\L{}ozinski}}, \bibinfo {author} {\bibfnamefont {M.~F. m.~c.}\ \bibnamefont {Santos}},\ and\ \bibinfo {author} {\bibfnamefont {V.}~\bibnamefont {Vedral}},\ }\bibfield  {title} {\bibinfo {title} {Geometric phase induced by a cyclically evolving squeezed vacuum reservoir},\ }\href {https://doi.org/10.1103/PhysRevLett.96.150403} {\bibfield  {journal} {\bibinfo  {journal} {Phys. Rev. Lett.}\ }\textbf {\bibinfo {volume} {96}},\ \bibinfo {pages} {150403} (\bibinfo {year} {2006})}\BibitemShut {NoStop}%
\bibitem [{\citenamefont {Parkins}\ \emph {et~al.}(1995)\citenamefont {Parkins}, \citenamefont {Marte}, \citenamefont {Zoller}, \citenamefont {Carnal},\ and\ \citenamefont {Kimble}}]{Parkins1995}%
  \BibitemOpen
  \bibfield  {author} {\bibinfo {author} {\bibfnamefont {A.~S.}\ \bibnamefont {Parkins}}, \bibinfo {author} {\bibfnamefont {P.}~\bibnamefont {Marte}}, \bibinfo {author} {\bibfnamefont {P.}~\bibnamefont {Zoller}}, \bibinfo {author} {\bibfnamefont {O.}~\bibnamefont {Carnal}},\ and\ \bibinfo {author} {\bibfnamefont {H.~J.}\ \bibnamefont {Kimble}},\ }\bibfield  {title} {\bibinfo {title} {Quantum-state mapping between multilevel atoms and cavity light fields},\ }\href {https://doi.org/10.1103/PhysRevA.51.1578} {\bibfield  {journal} {\bibinfo  {journal} {Phys. Rev. A}\ }\textbf {\bibinfo {volume} {51}},\ \bibinfo {pages} {1578} (\bibinfo {year} {1995})}\BibitemShut {NoStop}%
\bibitem [{\citenamefont {Parkins}\ \emph {et~al.}(1993)\citenamefont {Parkins}, \citenamefont {Marte}, \citenamefont {Zoller},\ and\ \citenamefont {Kimble}}]{Parkins1993}%
  \BibitemOpen
  \bibfield  {author} {\bibinfo {author} {\bibfnamefont {A.~S.}\ \bibnamefont {Parkins}}, \bibinfo {author} {\bibfnamefont {P.}~\bibnamefont {Marte}}, \bibinfo {author} {\bibfnamefont {P.}~\bibnamefont {Zoller}},\ and\ \bibinfo {author} {\bibfnamefont {H.~J.}\ \bibnamefont {Kimble}},\ }\bibfield  {title} {\bibinfo {title} {Synthesis of arbitrary quantum states via adiabatic transfer of zeeman coherence},\ }\href {https://doi.org/10.1103/PhysRevLett.71.3095} {\bibfield  {journal} {\bibinfo  {journal} {Phys. Rev. Lett.}\ }\textbf {\bibinfo {volume} {71}},\ \bibinfo {pages} {3095} (\bibinfo {year} {1993})}\BibitemShut {NoStop}%
\bibitem [{\citenamefont {Shankar}\ \emph {et~al.}(2017)\citenamefont {Shankar}, \citenamefont {Cooper}, \citenamefont {Bohnet}, \citenamefont {Bollinger},\ and\ \citenamefont {Holland}}]{Shankar2017}%
  \BibitemOpen
  \bibfield  {author} {\bibinfo {author} {\bibfnamefont {A.}~\bibnamefont {Shankar}}, \bibinfo {author} {\bibfnamefont {J.}~\bibnamefont {Cooper}}, \bibinfo {author} {\bibfnamefont {J.~G.}\ \bibnamefont {Bohnet}}, \bibinfo {author} {\bibfnamefont {J.~J.}\ \bibnamefont {Bollinger}},\ and\ \bibinfo {author} {\bibfnamefont {M.}~\bibnamefont {Holland}},\ }\bibfield  {title} {\bibinfo {title} {Steady-state spin synchronization through the collective motion of trapped ions},\ }\href {https://doi.org/10.1103/PhysRevA.95.033423} {\bibfield  {journal} {\bibinfo  {journal} {Phys. Rev. A}\ }\textbf {\bibinfo {volume} {95}},\ \bibinfo {pages} {033423} (\bibinfo {year} {2017})}\BibitemShut {NoStop}%
\bibitem [{\citenamefont {Degen}\ \emph {et~al.}(2017)\citenamefont {Degen}, \citenamefont {Reinhard},\ and\ \citenamefont {Cappellaro}}]{Degan2017}%
  \BibitemOpen
  \bibfield  {author} {\bibinfo {author} {\bibfnamefont {C.~L.}\ \bibnamefont {Degen}}, \bibinfo {author} {\bibfnamefont {F.}~\bibnamefont {Reinhard}},\ and\ \bibinfo {author} {\bibfnamefont {P.}~\bibnamefont {Cappellaro}},\ }\bibfield  {title} {\bibinfo {title} {Quantum sensing},\ }\href {https://doi.org/10.1103/RevModPhys.89.035002} {\bibfield  {journal} {\bibinfo  {journal} {Rev. Mod. Phys.}\ }\textbf {\bibinfo {volume} {89}},\ \bibinfo {pages} {035002} (\bibinfo {year} {2017})}\BibitemShut {NoStop}%
\bibitem [{\citenamefont {Pezz\`e}\ \emph {et~al.}(2018)\citenamefont {Pezz\`e}, \citenamefont {Smerzi}, \citenamefont {Oberthaler}, \citenamefont {Schmied},\ and\ \citenamefont {Treutlein}}]{Pezze2018}%
  \BibitemOpen
  \bibfield  {author} {\bibinfo {author} {\bibfnamefont {L.}~\bibnamefont {Pezz\`e}}, \bibinfo {author} {\bibfnamefont {A.}~\bibnamefont {Smerzi}}, \bibinfo {author} {\bibfnamefont {M.~K.}\ \bibnamefont {Oberthaler}}, \bibinfo {author} {\bibfnamefont {R.}~\bibnamefont {Schmied}},\ and\ \bibinfo {author} {\bibfnamefont {P.}~\bibnamefont {Treutlein}},\ }\bibfield  {title} {\bibinfo {title} {Quantum metrology with nonclassical states of atomic ensembles},\ }\href {https://doi.org/10.1103/RevModPhys.90.035005} {\bibfield  {journal} {\bibinfo  {journal} {Rev. Mod. Phys.}\ }\textbf {\bibinfo {volume} {90}},\ \bibinfo {pages} {035005} (\bibinfo {year} {2018})}\BibitemShut {NoStop}%
\bibitem [{\citenamefont {Reilly}\ \emph {et~al.}(2023)\citenamefont {Reilly}, \citenamefont {Wilson}, \citenamefont {J\"ager}, \citenamefont {Wilson},\ and\ \citenamefont {Holland}}]{Reilly2023}%
  \BibitemOpen
  \bibfield  {author} {\bibinfo {author} {\bibfnamefont {J.~T.}\ \bibnamefont {Reilly}}, \bibinfo {author} {\bibfnamefont {J.~D.}\ \bibnamefont {Wilson}}, \bibinfo {author} {\bibfnamefont {S.~B.}\ \bibnamefont {J\"ager}}, \bibinfo {author} {\bibfnamefont {C.}~\bibnamefont {Wilson}},\ and\ \bibinfo {author} {\bibfnamefont {M.~J.}\ \bibnamefont {Holland}},\ }\bibfield  {title} {\bibinfo {title} {Optimal generators for quantum sensing},\ }\href {https://doi.org/10.1103/PhysRevLett.131.150802} {\bibfield  {journal} {\bibinfo  {journal} {Phys. Rev. Lett.}\ }\textbf {\bibinfo {volume} {131}},\ \bibinfo {pages} {150802} (\bibinfo {year} {2023})}\BibitemShut {NoStop}%
\bibitem [{\citenamefont {Ye}\ and\ \citenamefont {Zoller}(2024)}]{Ye2024}%
  \BibitemOpen
  \bibfield  {author} {\bibinfo {author} {\bibfnamefont {J.}~\bibnamefont {Ye}}\ and\ \bibinfo {author} {\bibfnamefont {P.}~\bibnamefont {Zoller}},\ }\bibfield  {title} {\bibinfo {title} {Essay: Quantum sensing with atomic, molecular, and optical platforms for fundamental physics},\ }\href {https://doi.org/10.1103/PhysRevLett.132.190001} {\bibfield  {journal} {\bibinfo  {journal} {Phys. Rev. Lett.}\ }\textbf {\bibinfo {volume} {132}},\ \bibinfo {pages} {190001} (\bibinfo {year} {2024})}\BibitemShut {NoStop}%
\bibitem [{\citenamefont {Huang}\ \emph {et~al.}(2024)\citenamefont {Huang}, \citenamefont {Zhuang},\ and\ \citenamefont {Lee}}]{Huang2024}%
  \BibitemOpen
  \bibfield  {author} {\bibinfo {author} {\bibfnamefont {J.}~\bibnamefont {Huang}}, \bibinfo {author} {\bibfnamefont {M.}~\bibnamefont {Zhuang}},\ and\ \bibinfo {author} {\bibfnamefont {C.}~\bibnamefont {Lee}},\ }\bibfield  {title} {\bibinfo {title} {Entanglement-enhanced quantum metrology: From standard quantum limit to heisenberg limit},\ }\href {https://doi.org/10.1063/5.0204102} {\bibfield  {journal} {\bibinfo  {journal} {Applied Physics Reviews}\ }\textbf {\bibinfo {volume} {11}},\ \bibinfo {pages} {031302} (\bibinfo {year} {2024})}\BibitemShut {NoStop}%
\bibitem [{\citenamefont {Steck}(2007)}]{Steck2007}%
  \BibitemOpen
  \bibfield  {author} {\bibinfo {author} {\bibfnamefont {D.~A.}\ \bibnamefont {Steck}},\ }\href {http://steck.us/teaching} {\bibinfo {title} {Quantum and atom optics}} (\bibinfo {year} {2007})\BibitemShut {NoStop}%
\bibitem [{\citenamefont {Silva}\ and\ \citenamefont {Feist}(2022)}]{Silva2022}%
  \BibitemOpen
  \bibfield  {author} {\bibinfo {author} {\bibfnamefont {R.~E.~F.}\ \bibnamefont {Silva}}\ and\ \bibinfo {author} {\bibfnamefont {J.}~\bibnamefont {Feist}},\ }\bibfield  {title} {\bibinfo {title} {Permutational symmetry for identical multilevel systems: A second-quantized approach},\ }\href {https://doi.org/10.1103/PhysRevA.105.043704} {\bibfield  {journal} {\bibinfo  {journal} {Phys. Rev. A}\ }\textbf {\bibinfo {volume} {105}},\ \bibinfo {pages} {043704} (\bibinfo {year} {2022})}\BibitemShut {NoStop}%
\bibitem [{\citenamefont {Mathur}\ \emph {et~al.}(2010)\citenamefont {Mathur}, \citenamefont {Raychowdhury},\ and\ \citenamefont {Anishetty}}]{Mathur2010}%
  \BibitemOpen
  \bibfield  {author} {\bibinfo {author} {\bibfnamefont {M.}~\bibnamefont {Mathur}}, \bibinfo {author} {\bibfnamefont {I.}~\bibnamefont {Raychowdhury}},\ and\ \bibinfo {author} {\bibfnamefont {R.}~\bibnamefont {Anishetty}},\ }\bibfield  {title} {\bibinfo {title} {Su (n) irreducible schwinger bosons},\ }\bibfield  {journal} {\bibinfo  {journal} {Journal of mathematical physics}\ }\textbf {\bibinfo {volume} {51}},\ \href {https://doi.org/10.1063/1.3464267} {10.1063/1.3464267} (\bibinfo {year} {2010})\BibitemShut {NoStop}%
\bibitem [{\citenamefont {Stanley}(2011)}]{Stanley2011}%
  \BibitemOpen
  \bibfield  {author} {\bibinfo {author} {\bibfnamefont {R.~P.}\ \bibnamefont {Stanley}},\ }\href {https://archive.org/details/enumerativecombi0001stan_r3n3} {\emph {\bibinfo {title} {Enumerative Combinatorics: Volume 1}}},\ \bibinfo {edition} {2nd}\ ed.\ (\bibinfo  {publisher} {Cambridge University Press},\ \bibinfo {address} {USA},\ \bibinfo {year} {2011})\BibitemShut {NoStop}%
\bibitem [{\citenamefont {Schwinger}(1952)}]{Schwinger1952}%
  \BibitemOpen
  \bibfield  {author} {\bibinfo {author} {\bibfnamefont {J.}~\bibnamefont {Schwinger}},\ }\href {https://doi.org/10.2172/4389568} {\emph {\bibinfo {title} {On Angular Momentum}}},\ \bibinfo {type} {Tech. Rep.}\ \bibinfo {number} {NYO-3071}\ (\bibinfo  {institution} {US Atomic Energy Commission},\ \bibinfo {year} {1952})\BibitemShut {NoStop}%
\bibitem [{\citenamefont {Liu}\ \emph {et~al.}(2019)\citenamefont {Liu}, \citenamefont {Yuan}, \citenamefont {Lu},\ and\ \citenamefont {Wang}}]{Liu2020}%
  \BibitemOpen
  \bibfield  {author} {\bibinfo {author} {\bibfnamefont {J.}~\bibnamefont {Liu}}, \bibinfo {author} {\bibfnamefont {H.}~\bibnamefont {Yuan}}, \bibinfo {author} {\bibfnamefont {X.-M.}\ \bibnamefont {Lu}},\ and\ \bibinfo {author} {\bibfnamefont {X.}~\bibnamefont {Wang}},\ }\bibfield  {title} {\bibinfo {title} {Quantum fisher information matrix and multiparameter estimation},\ }\href {https://doi.org/10.1088/1751-8121/ab5d4d} {\bibfield  {journal} {\bibinfo  {journal} {Journal of Physics A: Mathematical and Theoretical}\ }\textbf {\bibinfo {volume} {53}},\ \bibinfo {pages} {023001} (\bibinfo {year} {2019})}\BibitemShut {NoStop}%
\bibitem [{\citenamefont {Holland}\ and\ \citenamefont {Burnett}(1993)}]{Holland1993}%
  \BibitemOpen
  \bibfield  {author} {\bibinfo {author} {\bibfnamefont {M.~J.}\ \bibnamefont {Holland}}\ and\ \bibinfo {author} {\bibfnamefont {K.}~\bibnamefont {Burnett}},\ }\bibfield  {title} {\bibinfo {title} {Interferometric detection of optical phase shifts at the heisenberg limit},\ }\href {https://doi.org/10.1103/PhysRevLett.71.1355} {\bibfield  {journal} {\bibinfo  {journal} {Phys. Rev. Lett.}\ }\textbf {\bibinfo {volume} {71}},\ \bibinfo {pages} {1355} (\bibinfo {year} {1993})}\BibitemShut {NoStop}%
\bibitem [{\citenamefont {Jozsa}(1994)}]{Jozsa1994}%
  \BibitemOpen
  \bibfield  {author} {\bibinfo {author} {\bibfnamefont {R.}~\bibnamefont {Jozsa}},\ }\bibfield  {title} {\bibinfo {title} {Fidelity for mixed quantum states},\ }\href {https://doi.org/10.1080/09500349414552171} {\bibfield  {journal} {\bibinfo  {journal} {Journal of Modern Optics}\ }\textbf {\bibinfo {volume} {41}},\ \bibinfo {pages} {2315} (\bibinfo {year} {1994})}\BibitemShut {NoStop}%
\bibitem [{\citenamefont {Cheney}\ and\ \citenamefont {Kincaid}(2012)}]{Cheney2012}%
  \BibitemOpen
  \bibfield  {author} {\bibinfo {author} {\bibfnamefont {W.}~\bibnamefont {Cheney}}\ and\ \bibinfo {author} {\bibfnamefont {D.}~\bibnamefont {Kincaid}},\ }\href {https://books.google.com/books?id=S0imN2tl1qwC} {\emph {\bibinfo {title} {Linear Algebra: Theory and Applications}}},\ The Jones \& Bartlett Learning international series in mathematics\ (\bibinfo  {publisher} {Jones \& Bartlett Learning},\ \bibinfo {address} {Burlington, MA},\ \bibinfo {year} {2012})\BibitemShut {NoStop}%
\bibitem [{\citenamefont {Alex}\ \emph {et~al.}(2011)\citenamefont {Alex}, \citenamefont {Kalus}, \citenamefont {Huckleberry},\ and\ \citenamefont {von Delft}}]{Alex2011}%
  \BibitemOpen
  \bibfield  {author} {\bibinfo {author} {\bibfnamefont {A.}~\bibnamefont {Alex}}, \bibinfo {author} {\bibfnamefont {M.}~\bibnamefont {Kalus}}, \bibinfo {author} {\bibfnamefont {A.}~\bibnamefont {Huckleberry}},\ and\ \bibinfo {author} {\bibfnamefont {J.}~\bibnamefont {von Delft}},\ }\bibfield  {title} {\bibinfo {title} {A numerical algorithm for the explicit calculation of su(n) and $\mbox{SL}(n,\mathbb {C})$sl(n,c) clebsch–gordan coefficients},\ }\bibfield  {journal} {\bibinfo  {journal} {Journal of Mathematical Physics}\ }\textbf {\bibinfo {volume} {52}},\ \href {https://doi.org/10.1063/1.3521562} {10.1063/1.3521562} (\bibinfo {year} {2011})\BibitemShut {NoStop}%
\bibitem [{\citenamefont {Sakurai}\ and\ \citenamefont {Napolitano}(2020)}]{Sakurai2020}%
  \BibitemOpen
  \bibfield  {author} {\bibinfo {author} {\bibfnamefont {J.~J.}\ \bibnamefont {Sakurai}}\ and\ \bibinfo {author} {\bibfnamefont {J.}~\bibnamefont {Napolitano}},\ }\href {https://www.cambridge.org/highereducation/books/modern-quantum-mechanics/DF43277E8AEDF83CC12EA62887C277DC#overview} {\emph {\bibinfo {title} {Modern Quantum Mechanics}}},\ \bibinfo {edition} {3rd}\ ed.\ (\bibinfo  {publisher} {Cambridge University Press},\ \bibinfo {address} {Cambridge, United Kingdom},\ \bibinfo {year} {2020})\BibitemShut {NoStop}%
\bibitem [{\citenamefont {Knill}\ \emph {et~al.}(2000)\citenamefont {Knill}, \citenamefont {Laflamme},\ and\ \citenamefont {Viola}}]{Knill2000}%
  \BibitemOpen
  \bibfield  {author} {\bibinfo {author} {\bibfnamefont {E.}~\bibnamefont {Knill}}, \bibinfo {author} {\bibfnamefont {R.}~\bibnamefont {Laflamme}},\ and\ \bibinfo {author} {\bibfnamefont {L.}~\bibnamefont {Viola}},\ }\bibfield  {title} {\bibinfo {title} {Theory of quantum error correction for general noise},\ }\href {https://doi.org/10.1103/PhysRevLett.84.2525} {\bibfield  {journal} {\bibinfo  {journal} {Phys. Rev. Lett.}\ }\textbf {\bibinfo {volume} {84}},\ \bibinfo {pages} {2525} (\bibinfo {year} {2000})}\BibitemShut {NoStop}%
\bibitem [{\citenamefont {Ostermann}\ \emph {et~al.}(2013)\citenamefont {Ostermann}, \citenamefont {Ritsch},\ and\ \citenamefont {Genes}}]{Ostermann2013}%
  \BibitemOpen
  \bibfield  {author} {\bibinfo {author} {\bibfnamefont {L.}~\bibnamefont {Ostermann}}, \bibinfo {author} {\bibfnamefont {H.}~\bibnamefont {Ritsch}},\ and\ \bibinfo {author} {\bibfnamefont {C.}~\bibnamefont {Genes}},\ }\bibfield  {title} {\bibinfo {title} {Protected state enhanced quantum metrology with interacting two-level ensembles},\ }\href {https://doi.org/10.1103/PhysRevLett.111.123601} {\bibfield  {journal} {\bibinfo  {journal} {Phys. Rev. Lett.}\ }\textbf {\bibinfo {volume} {111}},\ \bibinfo {pages} {123601} (\bibinfo {year} {2013})}\BibitemShut {NoStop}%
\bibitem [{\citenamefont {Roos}\ \emph {et~al.}(2006)\citenamefont {Roos}, \citenamefont {Chwalla}, \citenamefont {Kim}, \citenamefont {Riebe},\ and\ \citenamefont {Blatt}}]{Roos2006}%
  \BibitemOpen
  \bibfield  {author} {\bibinfo {author} {\bibfnamefont {C.~F.}\ \bibnamefont {Roos}}, \bibinfo {author} {\bibfnamefont {M.}~\bibnamefont {Chwalla}}, \bibinfo {author} {\bibfnamefont {K.}~\bibnamefont {Kim}}, \bibinfo {author} {\bibfnamefont {M.}~\bibnamefont {Riebe}},\ and\ \bibinfo {author} {\bibfnamefont {R.}~\bibnamefont {Blatt}},\ }\bibfield  {title} {\bibinfo {title} {`designer atoms' for quantum metrology},\ }\href {https://doi.org/10.1038/nature05101} {\bibfield  {journal} {\bibinfo  {journal} {Nature}\ }\textbf {\bibinfo {volume} {443}},\ \bibinfo {pages} {316} (\bibinfo {year} {2006})}\BibitemShut {NoStop}%
\bibitem [{\citenamefont {Kollath}\ \emph {et~al.}(2016)\citenamefont {Kollath}, \citenamefont {Sheikhan}, \citenamefont {Wolff},\ and\ \citenamefont {Brennecke}}]{Kollath2016}%
  \BibitemOpen
  \bibfield  {author} {\bibinfo {author} {\bibfnamefont {C.}~\bibnamefont {Kollath}}, \bibinfo {author} {\bibfnamefont {A.}~\bibnamefont {Sheikhan}}, \bibinfo {author} {\bibfnamefont {S.}~\bibnamefont {Wolff}},\ and\ \bibinfo {author} {\bibfnamefont {F.}~\bibnamefont {Brennecke}},\ }\bibfield  {title} {\bibinfo {title} {Ultracold fermions in a cavity-induced artificial magnetic field},\ }\href {https://doi.org/10.1103/PhysRevLett.116.060401} {\bibfield  {journal} {\bibinfo  {journal} {Phys. Rev. Lett.}\ }\textbf {\bibinfo {volume} {116}},\ \bibinfo {pages} {060401} (\bibinfo {year} {2016})}\BibitemShut {NoStop}%
\bibitem [{\citenamefont {Colella}\ \emph {et~al.}(2019)\citenamefont {Colella}, \citenamefont {Mivehvar}, \citenamefont {Piazza},\ and\ \citenamefont {Ritsch}}]{Colella2019}%
  \BibitemOpen
  \bibfield  {author} {\bibinfo {author} {\bibfnamefont {E.}~\bibnamefont {Colella}}, \bibinfo {author} {\bibfnamefont {F.}~\bibnamefont {Mivehvar}}, \bibinfo {author} {\bibfnamefont {F.}~\bibnamefont {Piazza}},\ and\ \bibinfo {author} {\bibfnamefont {H.}~\bibnamefont {Ritsch}},\ }\bibfield  {title} {\bibinfo {title} {Hofstadter butterfly in a cavity-induced dynamic synthetic magnetic field},\ }\href {https://doi.org/10.1103/PhysRevB.100.224306} {\bibfield  {journal} {\bibinfo  {journal} {Phys. Rev. B}\ }\textbf {\bibinfo {volume} {100}},\ \bibinfo {pages} {224306} (\bibinfo {year} {2019})}\BibitemShut {NoStop}%
\bibitem [{\citenamefont {Sheikhan}\ \emph {et~al.}(2016)\citenamefont {Sheikhan}, \citenamefont {Brennecke},\ and\ \citenamefont {Kollath}}]{Sheikhan2016}%
  \BibitemOpen
  \bibfield  {author} {\bibinfo {author} {\bibfnamefont {A.}~\bibnamefont {Sheikhan}}, \bibinfo {author} {\bibfnamefont {F.}~\bibnamefont {Brennecke}},\ and\ \bibinfo {author} {\bibfnamefont {C.}~\bibnamefont {Kollath}},\ }\bibfield  {title} {\bibinfo {title} {Cavity-induced generation of nontrivial topological states in a two-dimensional fermi gas},\ }\href {https://doi.org/10.1103/PhysRevA.94.0Error correcting schemes are complicated. 61603} {\bibfield  {journal} {\bibinfo  {journal} {Phys. Rev. A}\ }\textbf {\bibinfo {volume} {94}},\ \bibinfo {pages} {061603} (\bibinfo {year} {2016})}\BibitemShut {NoStop}%
\bibitem [{\citenamefont {LeDesma}\ \emph {et~al.}(2024)\citenamefont {LeDesma}, \citenamefont {Mehling}, \citenamefont {Shao}, \citenamefont {Wilson}, \citenamefont {Axelrad}, \citenamefont {Nicotra}, \citenamefont {Anderson},\ and\ \citenamefont {Holland}}]{Ledesma2024}%
  \BibitemOpen
  \bibfield  {author} {\bibinfo {author} {\bibfnamefont {C.}~\bibnamefont {LeDesma}}, \bibinfo {author} {\bibfnamefont {K.}~\bibnamefont {Mehling}}, \bibinfo {author} {\bibfnamefont {J.}~\bibnamefont {Shao}}, \bibinfo {author} {\bibfnamefont {J.~D.}\ \bibnamefont {Wilson}}, \bibinfo {author} {\bibfnamefont {P.}~\bibnamefont {Axelrad}}, \bibinfo {author} {\bibfnamefont {M.}~\bibnamefont {Nicotra}}, \bibinfo {author} {\bibfnamefont {D.~Z.}\ \bibnamefont {Anderson}},\ and\ \bibinfo {author} {\bibfnamefont {M.}~\bibnamefont {Holland}},\ }\bibfield  {title} {\bibinfo {title} {Demonstration of a programmable optical lattice atom interferometer},\ }\href {https://doi.org/10.1103/PhysRevResearch.6.043120} {\bibfield  {journal} {\bibinfo  {journal} {Phys. Rev. Res.}\ }\textbf {\bibinfo {volume} {6}},\ \bibinfo {pages} {043120} (\bibinfo {year} {2024})}\BibitemShut {NoStop}%
\bibitem [{\citenamefont {LeDesma}\ \emph {et~al.}(2025)\citenamefont {LeDesma}, \citenamefont {Mehling}, \citenamefont {Wilson}, \citenamefont {Nicotra},\ and\ \citenamefont {Holland}}]{Ledesma2024GateSet}%
  \BibitemOpen
  \bibfield  {author} {\bibinfo {author} {\bibfnamefont {C.}~\bibnamefont {LeDesma}}, \bibinfo {author} {\bibfnamefont {K.}~\bibnamefont {Mehling}}, \bibinfo {author} {\bibfnamefont {J.~D.}\ \bibnamefont {Wilson}}, \bibinfo {author} {\bibfnamefont {M.}~\bibnamefont {Nicotra}},\ and\ \bibinfo {author} {\bibfnamefont {M.}~\bibnamefont {Holland}},\ }\bibfield  {title} {\bibinfo {title} {Universal gate set for optical lattice based atom interferometry},\ }\href {https://doi.org/10.1103/PhysRevResearch.7.013246} {\bibfield  {journal} {\bibinfo  {journal} {Phys. Rev. Res.}\ }\textbf {\bibinfo {volume} {7}},\ \bibinfo {pages} {013246} (\bibinfo {year} {2025})}\BibitemShut {NoStop}%
\bibitem [{\citenamefont {J\"ager}\ \emph {et~al.}(2022)\citenamefont {J\"ager}, \citenamefont {Schmit}, \citenamefont {Morigi}, \citenamefont {Holland},\ and\ \citenamefont {Betzholz}}]{Jager2022}%
  \BibitemOpen
  \bibfield  {author} {\bibinfo {author} {\bibfnamefont {S.~B.}\ \bibnamefont {J\"ager}}, \bibinfo {author} {\bibfnamefont {T.}~\bibnamefont {Schmit}}, \bibinfo {author} {\bibfnamefont {G.}~\bibnamefont {Morigi}}, \bibinfo {author} {\bibfnamefont {M.~J.}\ \bibnamefont {Holland}},\ and\ \bibinfo {author} {\bibfnamefont {R.}~\bibnamefont {Betzholz}},\ }\bibfield  {title} {\bibinfo {title} {Lindblad master equations for quantum systems coupled to dissipative bosonic modes},\ }\href {https://doi.org/10.1103/PhysRevLett.129.063601} {\bibfield  {journal} {\bibinfo  {journal} {Phys. Rev. Lett.}\ }\textbf {\bibinfo {volume} {129}},\ \bibinfo {pages} {063601} (\bibinfo {year} {2022})}\BibitemShut {NoStop}%
\bibitem [{\citenamefont {Griffiths}(2017)}]{Griffiths2017}%
  \BibitemOpen
  \bibfield  {author} {\bibinfo {author} {\bibfnamefont {D.}~\bibnamefont {Griffiths}},\ }\href {https://books.google.com/books?id=0h-nDAAAQBAJ} {\emph {\bibinfo {title} {Introduction to Quantum Mechanics}}}\ (\bibinfo  {publisher} {Cambridge University Press},\ \bibinfo {address} {Cambridge, United Kingdom},\ \bibinfo {year} {2017})\BibitemShut {NoStop}%
\end{thebibliography}
%

\end{document}